\documentclass[]{jocg}

\makeatletter
\renewenvironment{abstract}{%
        \small
        \begin{center}%
          {\bfseries Abstract\vspace{-.5em}\vspace{\z@}}%
        \end{center}%
        \quotation
      }      {\endquotation}
\makeatother

\usepackage{graphicx}
\usepackage{microtype}
\usepackage{amsthm,amsfonts,amsmath,amssymb}
\usepackage{cleveref}
\usepackage[ruled,vlined,boxed,linesnumberedhidden]{algorithm2e}
\usepackage{paralist}

\newtheorem{theorem}{Theorem}[section] %
\newtheorem{lemma}[theorem]{Lemma}
\newtheorem{proposition}[theorem]{Proposition}%
\newtheorem{corollary}[theorem]{Corollary}
\theoremstyle{definition}
\newtheorem{definition}[theorem]{Definition}%
\counterwithin{figure}{section} %

\newcommand{\workspace}{free space}
\newcommand{\workkspace}{free-space} %

\newcommand{\epspert}{$\varepsilon$-perturbation} 

\newcommand\defn[1]{{\bf\emph{{#1}}}} 
\newcommand{\emphblack}[1]{\textit{#1}} 

\newcommand{\wind}{\ensuremath{\mathrm{wind}}}
\newcommand{\Sfse}{\mathcal{E}} %
\newcommand{\wwind}[2]{\ensuremath{\mathrm{wind}(#1,#2)}}

\newcommand{\WALG}{\ensuremath{W_{\mathrm{ALG}}}}
\newcommand{\WDP}{\ensuremath{W_{\mathrm{DP}}}}
\newcommand{\cDP} {\ensuremath{c_{\mathrm{DP}}}}
\newcommand{\ws}{weakly simple}
\newcommand{\wsimp}{WSImP}
\newcommand{\simp}{SImP}
\newcommand{\EWP}{\textup{\textsc{Enclosure-with-Penalties}}}
\newcommand{\GrEWP}{\textup{\textsc{Graph-Enclosure-with-Penalties}}}
\newcommand{\GeoEWP}{\textup{\textsc{Geometric-Enclosure-with-Penalties}}}
\newcommand{\OPT}{\mathrm{OPT}}

\title{%
  \MakeUppercase{Finding a Shortest Curve that Separates Few Objects from Many}%
  \thanks
  {This work was initiated at %
\href{https://www.dagstuhl.de/24072}{Dagstuhl Seminar 24072}
  ``Triangulations in Geometry and Topology'' in February 2024.  We %
thank
the organizers, Maike Buchin, Jean Cardinal, Arnaud de Mesmay, and Jonathan Spreer, as well as all participants, for the stimulating atmosphere.  We 
also thank 
Alexander Wolff for significant %
contributions, Saul Schleimer for useful discussions, and the
anonymous reviewers for their constructive feedback.
This paper appeared in the Journal of Computational Geometry \cite{bcflr-fscts-26}.
A preliminary version of this paper appeared in
 the Proceedings of the
 41st International Symposium on Computational Geometry (SoCG 2025),
 Kanazawa, Japan
\cite {bcflr-fscts-25}.
}
}

\author{%
  Therese~Biedl,%
  \thanks{\affil{University of Waterloo, Canada}, 
          \email{\{biedl,alubiw\}@uwaterloo.ca}.   Research supported by NSERC RGPIN-2020-03958, RGPIN-2020-03959.}%
          \orcid{0000-0002-9003-3783}\,
  \'Eric Colin~de~Verdi\`ere,%
  \thanks{\affil{LIGM, CNRS, Univ Gustave Eiffel, F-77454 Marne-la-Vall\'ee,
      France},
          \email{eric.colin-de-verdiere@\goodbreak{}univ-eiffel.fr}.
          This author is partially supported by the ANR project SUGAR (ANR-25-CE40-0416).}%
          \orcid{0009-0007-6951-3698}\,
  Fabrizio~Frati,%
  \thanks{\affil{Roma Tre University, Italy}, 
          \email{fabrizio.frati@uniroma3.it}}%
          \orcid{0000-0001-5987-8713}\,
  Anna~Lubiw,\footnotemark[2]%
          \orcid{0000-0002-2338-361X}\,
  and G\"unter~Rote%
  \thanks{\affil{Freie Universit\"at Berlin, Germany}, 
          \email{rote@inf.fu-berlin.de}}%
          \orcid{0000-0002-0351-5945}
}

\begin{document}

  \def\definitionautorefname{Definition}
  \def\sectionautorefname{Section}
  \def\subsectionautorefname{Section}
  \def\subsubsectionautorefname{Section}
  \def\lemmaautorefname{Lemma}

\maketitle

\begin{abstract}
We present a fixed-parameter tractable (FPT) 
algorithm to 
find a shortest curve that 
separates some polygons from others.   Formally, the
input is  
a set of interior-disjoint simple polygons in the plane,
where $k$ of the polygons are \emphblack{required} to be enclosed  
and the remaining \emphblack{optional} polygons have
non-negative penalties.
The goal is to find a closed curve that is disjoint from the polygon interiors 
and %
encloses the $k$ required polygons, while minimizing the length of the curve plus the penalties of the enclosed optional polygons.  If the penalties are %
high, the output is a shortest curve that separates the required polygons from the others.
The runtime of our %
algorithm is $O(3^k n^3)$, where $n$ is the
number of vertices of the input polygons. The problem is NP-hard if $k$ is not fixed, even in very special cases.

We extend 
the result to a graph version of the problem where the input is a connected plane graph with positive edge weights.  
There are $k$ required faces; the remaining faces are optional and have %
non-negative penalties.
The goal is to find a closed walk in the graph
that encloses the $k$ required faces, while minimizing the weight of the walk plus the penalties of the enclosed optional faces.

We also consider an inverted version of the problem where the required objects must lie outside the curve. 
Our algorithms
solve some other well-studied problems, such as geometric knapsack.
Assuming the Exponential Time Hypothesis, we prove that 
neither the geometric nor the graph version of
our problem can %
be solved in $2^{o(k)}\cdot n^{O(1)}$ time.
\end{abstract}

\section{Introduction}
\label{sec:intro}

\subsection{Problem statement and results}

We investigate the separation problem of finding a shortest %
curve that 
encloses a subset of objects while excluding %
other objects.
A very basic setting is for points in the plane:
given $n$ points in the plane and a subset of size $k$, 
find a minimum-perimeter %
polygon containing the specified~$k$ points and 
excluding the other~$n-k$ points.
This problem is NP-hard when~$k$ may be large,  as proved by Eades and Rappaport~\cite{er-ccmsp-PRL93} for the case $k=n/2$ via a simple reduction from the Travelling Salesman Problem. 

As a special case 
of our main result, 
we give the first algorithm for this problem that is fixed-parameter tractable (FPT) in~$k$.
Our result is far more general and applies 
in two settings, a geometric setting and a graph-theoretic setting.

\paragraph{The Geometric Setting.}
For the \textsc{Geometric-Enclosure-with-Penalties} problem
we generalize from objects that are points to objects that are interior-disjoint simple polygons in the plane, %
and we generalize to a weighted form of exclusion.

{\bf Input.}
The input is a set of simple interior-disjoint polygons partitioned into a set 
$R$ of $k\ge1$ \defn{required polygons} 
and the remaining set $O$ of \defn{optional %
polygons}.
Each optional 
polygon $P \in O$ comes with a non-negative \defn{penalty} $\pi_P$ where we allow $\pi_P = +\infty$.  

{\bf Output.}
The goal is to find a weakly simple polygon $W$ that does not 
intersect the interior of any input polygon
and encloses all 
polygons
of $R$ while minimizing
the \defn{cost} $c(W)$, which 
is defined to be
the Euclidean
length of $W$ plus the penalties of the 
polygons of $O$ that are inside~$W$. 
\begin{figure}
    \centering
  \includegraphics[page=2,scale=1]{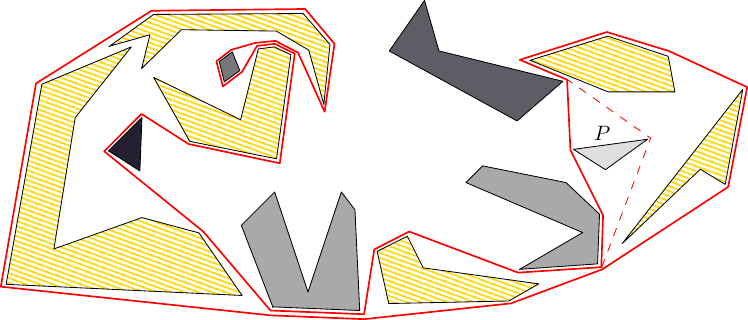}
    \caption{The \textsc{Geometric-Enclosure-with-Penalties} problem. Objects in~$R$ are yellow (hatched%
    ) 
    and objects in~$O$ are gray, %
    darker for objects with larger penalties. A  weakly simple solution polygon~$W$ %
    is shown in red (bold).
  The optimality of $W$ depends on the exact values of the penalties.
    For example, we can compare the penalty~$\pi_P$ of the
    object~$P\in O$, which lies inside~$W$, against %
  the %
  {dashed}{}
  detour 
  that $W$ would have to make
  in order to have~$P$~outside.
  {For visual clarity, $W$ is drawn with
    an offset where it would otherwise touch the objects or itself. }{}%
    }
    \label{fig:example-geom}
\end{figure}
See \Cref{fig:example-geom} for an example.
A polygon with
penalty $+\infty$ 
must 
be excluded. 
A polygon with 
penalty 0
may be included or excluded
without making a difference, so it only acts as an obstacle to the solution curve.
As 
\Cref{fig:example-geom}
illustrates, 
the problem would be ill-defined if we required the solution curve $W$ to be a simple polygon.  
The natural condition is that $W$ should be a 
\defn{weakly simple polygon}, whose boundary may touch or overlap itself but not cross itself
(see Section~\ref{sec:weakly-simple} for details).
An important property is that a weakly simple polygon encloses a well-defined region.

Our first main result is:

\begin{theorem}
\label{thm:main-geom}
   \textup{\textsc{Geometric-Enclosure-with-Penalties}}
   for $k$ required polygons
 can be solved in $O(3^kn^3)$ time and $O(2^kn^2)$ space, if  the input polygons have $n$ vertices in total.
\end{theorem}
If all objects are points, this can be handled by %
approximating each point 
by a small triangle.

\paragraph{The Graph Setting.}
For the \textsc{Graph-Enclosure-with-Penalties} problem,
the objects are faces of a plane graph.

\begin{figure}
    \centering
 \includegraphics[page=3,scale=1]{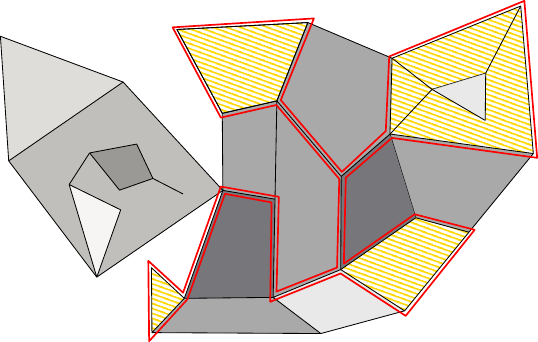}%
    \caption{%
    The \GrEWP\  problem. The colors %
    have the same meaning as in \Cref{fig:example-geom}.
    A weakly simple closed walk $W$ which is a solution for the instance is red (bold).
 The edge weights, which are not shown, are unrelated to the Euclidean lengths.}
    \label{fig:example-graph}
\end{figure}

{\bf Input.} 
The input is a simple connected  plane graph $G$ %
and positive edge weights. 
The bounded faces of $G$ are partitioned into a set $R$ of $k\ge1$ \defn{required faces} and the remaining set $O$ of \defn{optional faces}.  Each optional face $F$ has a penalty $\pi_F$ from $\mathbb R_{\ge 0}\cup\{+\infty\}$.

{\bf Output.} 
The goal is to find a weakly simple closed walk $W$ in $G$ such that faces of
$R$ are inside $W$ while minimizing the cost $c(W)$ which is defined to be the sum of the weights of the edges of $W$ 
plus the penalties of the
faces of $O$ that are inside $W$. See \Cref{fig:example-graph} for an example.
Intuitively, a \defn{weakly simple} closed walk is one without crossings.
For a weakly simple closed walk, the notions of inside and outside are well-defined.

Our second main result is:

\begin{theorem}
\label{thm:main-graph}
   \textup{\textsc{Graph-Enclosure-with-Penalties}}
 can be solved in $O(3^k n^3)$ time and $O(2^k n^2)$ space, where $k$ is the number of required faces and $n$ is the number of vertices of~$G$.
\end{theorem}

\paragraph{A common framework.}

Although the two settings described above seem different, we resolve them into a common geometric framework, which we call \textsc{Enclosure-with-Penalties}.
Our algorithm applies to this 
general problem.
The basic idea is to transform the graph problem into a geometric problem by taking a straight-line embedding of the graph.  The bounded faces of the graph become polygons
slightly more general than simple polygons.  We also consider the outer face as an unbounded polygon.  Then the ``free space'' between the polygons consists only of the graph edges.  This gives us a geometric problem, albeit with arbitrary positive edge weights
defined on edges that have a polygon on each side.
In Section~\ref{sec:framework} we define the \textsc{Enclosure-with-Penalties} problem 
by generalizing the
\textsc{Geometric-Enclosure-with-Penalties} problem
to include these instances.

We remark that 
the resulting algorithm for \Cref{thm:main-graph} makes essential use of 
the 
straight-line embedding of the input graph.  
In particular, the subproblems that we solve depend on the embedding.
This imposition of geometry 
seems artificial, 
but oddly enough, we do not know how to 
formulate our algorithm in a purely combinatorial setting.

\paragraph{Lower bounds.}
To complement our algorithms we prove that, under
the Exponential Time Hypothesis (ETH),
the \textsc{Geometric-} and \textsc{Graph-Enclosure-with-Penalties} problems cannot be solved in $2^{o(k)}\cdot n^{O(1)}$ time, implying that the linear dependence on~$k$ in the exponent of the runtime of our algorithms is the best possible assuming ETH.  The proof is a reduction from unweighted \textsc{Planar Steiner Tree}, which admits a lower bound by a result by Marx, Pilipczuk, and Pilipczuk~\cite[Theorem~1.2]{marx2018subexponential}.
See
Section~\ref{sec:lower-bound}.

\subsection{Our approach}

We use dynamic programming (\Cref{sec:DP}) to build a polygon $W$ that is locally correct---we use segments that 
do not intersect the interior of any object and we account for required objects and 
tally the penalties as we add triangles to $W$.
We will prove that the cost computed by the algorithm is correct, but this is tricky because $W$ itself will not necessarily be weakly simple. 
``Inside'' is no longer well-defined.  Instead, we  
use winding numbers to give a measure of the cost of $W$ 
that matches the cost computed by the algorithm.

In Section~\ref{sec:uncrossing} we give an algorithm to turn %
$W$ to a weakly simple polygon without increasing its cost, which provides our final output. 
Correctness of the whole algorithm is proved in Section~\ref{sec:correctness}.

The runtime for this \emphblack{Reordering Algorithm} is dominated by the runtime for
solving the dynamic programming recursion.  To obtain our claimed
runtime we  speed up the dynamic programming algorithm in \Cref{sec:dijkstra}.  %
This concludes the proofs of Theorems \ref{thm:main-geom} and~\ref{thm:main-graph}.  In the remaining sections, we present extensions, lower bounds, and miscellaneous remarks.

\paragraph{Weakly simple immersed polygons.}
For the proof of our above algorithms, we only need some properties of
the output of the dynamic programming algorithm.  It turns out that the dynamic programming algorithm optimizes over the class of \emphblack{immersed} or \emphblack{self-overlapping} weakly simple polygons.  We provide details in \Cref{sec:immersed}.

\subsection{Extensions}

\paragraph{Point objects.}
\Cref{thm:main-geom} assumes that the input objects are simple polygons.  We extend this result for  an arbitrary mix of point and polygon objects in Section~\ref{sec:points}.

\paragraph{Swapping the inside with the outside.}
We then extend our algorithms to an \defn{inverted} version of 
the \textsc{Enclosure-with-Penalties} problem where the required objects have to be \emphblack{outside} $W$, and the objective is to minimize the length of $W$ plus the penalties of the polygons of $O$ that are \emphblack{outside} $W$. 
The runtime remains the same,
see Section~\ref{sec:inverted}. 
This algorithm provides a new faster solution to the geometric knapsack problem discussed below.

\paragraph{Negative penalties.}
We can allow some number $\ell$ of objects with negative penalties (rewards); in this case, the runtime increases by a factor of~$3^\ell$.  
We present this generalization of Theorems \ref{thm:main-geom} and~\ref{thm:main-graph} in \Cref{sec:negative}.

\subsection{Related work}\label{sec:related}

Cut problems and separator problems in graphs have a long history, and separation problems in geometric settings are a natural and well studied counterpart.%

\paragraph{Geometric knapsack problem.}  
Geometric separation problems were first explored by Eades and Rappaport~\cite{er-ccmsp-PRL93} (as discussed above) and by Arkin, Khuller, and Mitchell, who introduced the \emphblack{geometric knapsack problem}~\cite{arkin1993geometric}, which corresponds to the \emphblack{inverted} version of the \textsc{Geometric-Enclosure-with-Penalties} problem in the special case where there are no required objects.  (In their equivalent formulation, each object has a finite nonnegative value, and the goal is to compute a curve that maximizes the total value of the enclosed objects minus its length.) They gave %
an algorithm with runtime $O(n^4)$~\cite[Theorem~6]{arkin1993geometric}.
In Appendix~\ref{sec:error-AKM},
we note some issues regarding weak simplicity and interior/exterior of their output. 
Since there are no required objects, our algorithm for the inverted problem solves the geometric knapsack problem in time $O(n^3)$.

\paragraph{Relation to homotopy and homology.}  Our problem has a topological flavor and is therefore, in principle, amenable to homotopy and homology techniques.  However, these techniques are unlikely to lead to algorithms that are FPT in~$k$, even assuming only infinite penalties.  
In particular, for the \GrEWP{} problem, enumerating a set of candidate homotopy classes, i.e., the possible ways in which a solution winds around the objects to enclose the required objects and avoid the most undesirable ones, is possible using a technique by Chambers, Colin de Verdière, Erickson, Lazarus, and Whittlesey~\cite{ccelw-scsh-08} (specifically, because the exchange argument~\cite[Proposition~4.2]{ccelw-scsh-08} is also valid for our problem).  However, this results in an algorithm with runtime $K^{O(K)}\cdot n\log n$, where $K$ is the number~$k$ of required objects \emphblack{plus} the number of objects with nonzero penalty.
Similarly, the technique of \emphblack{homology covers}, by Chambers, Erickson, Fox, and Nayyeri~\cite{cefn-mcsg-23}, is applicable, but yields algorithms with runtime $K^{O(K)}\cdot n\log\log n$ or $2^{O(K)}\cdot n\log n$, thus again with an exponential dependence on~$K$.  If there are many objects with nonzero penalty, our algorithm with runtime $O(3^kn^3)$ is faster.

\paragraph{Specifying only the number of objects to be enclosed.} 
Consider the %
related problem
where we are just given a set of $n$ points in general position and a bound $k$ on the \emphblack{number} of points to be enclosed, but not which points must be enclosed.   In this scenario one can show that
a minimum-perimeter polygon enclosing at least $k$ points is convex, contains exactly $k$ points, and can be found in polynomial time by an algorithm of Eppstein, Overmars, Rote, and Woeginger~\cite[Corollary~5.3, Case~3]{eorw-fmakg-DCG92}.  
This algorithm could for example be used to identify an unusual cluster in an otherwise uniformly distributed point set.  However, if the input consists of polygons instead of points, we are not aware of a better method than guessing the $k$ polygons to be enclosed and applying our main result, resulting in an algorithm of runtime $O(\binom Nk3^kn^{3})=O(N^k n^{3})$ if there are $N$ objects.

\paragraph{More variations.} Some related separation and enclosure problems
have been recently studied.
Abrahamsen, Giannopoulos, Löffler, and Rote~\cite{aglr-gmsfs-20} investigated the problem of
separating two or more groups of polygonal objects by a shortest system of fences.
These fences may form an arbitrary plane graph, not necessarily a cycle.
For separating \emphblack{two} groups of polygons by a shortest fence, there is an algorithm with runtime
$O(n^4 \log^3n)$, based on minimum-cut techniques~\cite{aglr-gmsfs-20}.
For separating more than two groups,
only approximation algorithms are known.
While we study a problem in the same spirit, a key difference is that we require a single (weakly) simple cycle, which makes the techniques of this article not applicable for us.

Chan, He, and Xue~\cite{chx-epgo-24}
studied the problem of
choosing a smallest subset of objects from a given set whose union encloses a given point set.
For this problem, there are only
 approximation algorithms.
Klost, van Kreveld, Perz, Rote, and Tkadlec~\cite{kkprt-msbt-25} have recently investigated optimum %
``{blob-trees}'', where the objective is to \emphblack{connect} a set of points by a combination of curves and edges.

\subsection{Overview}

\autoref{sec:prelim} discusses weakly simple polygons.
This is necessary for the precise specification of the output that is expected for
the two versions of our enclosure problem, which are defined in \autoref{sec:framework}.
Sections~\ref{sec:DP}--\ref{sec:dijkstra} describe our main algorithm.
As mentioned above, %
the algorithm
consists of two major steps:
a dynamic programming recursion (\autoref{sec:DP}) and
an algorithm to ensure weak simplicity (\autoref{sec:uncrossing}). Correctness is proved in
\autoref{sec:correctness}.
Finally, \autoref{sec:dijkstra} shows how the runtime can be improved
by a Dijkstra-style evaluation strategy for the dynamic programming recursion.
\autoref{sec:model} discusses the computational model for our
algorithms. In particular, we use the {real RAM} model
for the
 \GeoEWP\ problem.

\autoref{sec:immersed} is optional; it explains our algorithm in terms of immersed (or self-overlapping) polygons.

Sections~\ref{sec:points}--\ref{sec:negative} discuss extensions and variations of our main algorithm: point
obstacles (\autoref{sec:points});
the inverted problem, where the given objects have to remain \emphblack{outside} the curve
(\autoref{sec:inverted});
and negative penalties (\autoref{sec:negative}).

\autoref{sec:lower-bound} proves a conditional lower bound on the complexity of our problem, and the brief concluding \autoref{sec:conclusion} discusses applications, extensions, and open problems.

\section{Preliminaries}\label{sec:prelim}

\subsection{Weakly simple polygons and related concepts
} 
\label{sec:weakly-simple}

Since we have to admit \emphblack{weakly simple} polygons as the
 solution of our problem,
we 
 first 
 discuss
 the notion of weakly simple polygons that we work with, as well as
the related graph-theoretic concept of non-crossing (Euler) tours in
plane graphs.

The intuition is that a weakly simple polygon is one without
crossings, but it is tricky to define crossings, since they need not
be local, see the discussion by Chang, Erickson, and
Xu~\cite{cex-dwsp-15}.
By the standard definition,
a polygon is weakly simple if
for every $\varepsilon > 0$, the vertices can be perturbed by at most $\varepsilon$ to yield a simple polygon~(see below for further details).
For our purposes it is more effective to reformulate weak simplicity combinatorially in terms of plane graphs.

\paragraph{Counterclockwise traversals of non-crossing Euler tours.}
\mbox{}
A graph $G=(V,E)$ is called \defn{plane} if it comes with a drawing in the plane that has no crossing; edges are allowed to use arbitrary curves (not necessarily straight-line segments).  Our graphs will be \emphblack{multigraphs}, i.e., there may be multiple edges connecting a pair of vertices.   
A walk $W$ of length $n$ in a plane graph $G$ is a
sequence $(v_0, v_1, \ldots,v_n )$ of vertices, such
that each $v_iv_{i+1}$  is  an edge of the graph.
A vertex/edge of $G$ may appear multiple times in the sequence.  If $v_0 = v_n$ this is a \defn{closed walk}; otherwise it is an \defn{open walk}.

A \defn{face} of $G$ is a maximal region of the plane not containing vertices or edges; the unbounded face is called \defn{outer face}.
All our graphs will be connected, and as such there is no need to work with the drawing; instead we assume that the multigraph is given
via its combinatorial map (or \emphblack{rotation system}),
which %
specifies the
counterclockwise 
cyclic order of edges around each vertex. %
If there are parallel edges, %
they have distinct identities and must be explicitly ordered in the rotation system. 
The faces (or more precisely, the closed walks that bound faces) can be read off from the combinatorial map.   We assume that one face is 
designated as the outer face.

A \defn{non-crossing Euler tour}
of a plane multigraph is a %
closed walk that traverses each edge exactly once and has no
\defn{vertex crossing}.
A
\defn{vertex crossing}
occurs when %
the tour
visits some vertex $v$ twice, entering once on edge $e$ and leaving on edge $f$, and entering again on edge $g$ and leaving on edge $h$, 
such that $e,f$ and $g,h$ 
interleave %
in the cyclic ordering of edges around $v$, i.e., they appear in the order $e,g,f,h$ 
or $e,h,f,g$. 

Let $G$ be a plane multigraph with a 
non-crossing Euler tour $T$.
Then the vertices of $G$ have even degrees,
and it is well-known that %
the faces of $G$ 
can be 2-colored such that the two faces incident to an edge have
different colors, see
{\cite[p.~200]{Kempe-1879}}
or~\cite[Theorems 34.2 and~34.4]{lw-cc-01}.
Suppose the two colors are grey and white with the outer face colored white. 
We will traverse $T$ so that a grey face lies to the left of the first edge of the tour.  Then, because the tour is non-crossing, every edge of the tour has a grey face to the left.  We call this a \defn{counterclockwise traversal} of~$T$, and we define the \defn{interior} faces of $T$ to be the grey faces. 
Note that the interior faces determine a partition of the edges of $G$.

\paragraph{Weakly simple walks in plane graphs.}

Intuitively, a closed walk $W$ is weakly simple if multiple traversals of an edge of $G$ can be resolved to avoid vertex crossings.  We make this more formal by way of a non-crossing Euler tour that provides a certificate that $W$ is weakly simple.

For edge~$e$ of $G$, define the \defn{multiplicity} $m(e)$ of $e$ in
$W$, to be the number of times $W$ traverses $e$ (in either
direction).  An \defn{expansion} of $W$ is a plane multigraph $M(W,G)$
that replaces each edge $e=ab$ of $G$ by a \defn{bundle} of $m(e)$
parallel edges, each identified with a unique edge of $W$, and
replaces $e$ in the rotation systems of $a$ and $b$ by an ordered
sequence of the edges in the bundle.  Then $W$ corresponds to an Euler
tour in $M(W,G)$.

\begin{figure}
    \centering
    \begin{tabular}{c c}
  \hfill
  \includegraphics[page=1,scale=1]{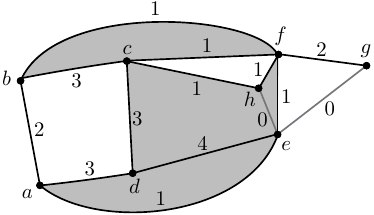} &
  \hfill
  \includegraphics[page=2,scale=1]{Euler.pdf} 
  \hfill
\\
  (a) & (b)
\end{tabular}
    \caption{(a) 
      A %
    closed walk $W= \mathit {abcdcdaededadefgfbcfhcba}$
    in     a simple plane graph $G$.
      The edges of $G$
    are
    labeled with their multiplicities.   %
    (b) A non-crossing Euler tour in an expansion~$M(W,G)$
    of~$W$. Vertices of~$M(W,G)$ are represented as dark disks. The
    Euler tour is shown in red and certifies that the walk $W$ is
    weakly simple.  }
    \label{fig:walks}
\end{figure}

\begin{figure}
    \centering

        \includegraphics[page=1]{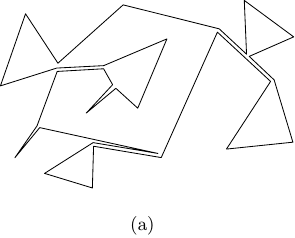}
\hspace*{\fill}
    \includegraphics[page=2]{weakly-ipe-simple-labels.pdf}
\hspace*{\fill}
    \includegraphics[page=3]{weakly-ipe-simple-labels.pdf}

    \caption{(a) A weakly simple polygon drawn via its $\varepsilon$-approximation.
    (b) The edges, after subdividing at forks, are partitioned into
    interior faces.
        Four faces are corridors and five are chambers.  The largest chamber (in yellow) is almost-simple but not simple.
    (c) Non-uniqueness of the faces for a weakly simple polygon that traverses a line segment four times.
    In the top figure the two vertices on the left are transition vertices; this is reversed in the bottom figure.
}
    \label{fig:weakly-simple}
\end{figure}

A closed walk $W$ in a plane graph $G$ is \defn{weakly simple} if it
has an expansion $M(W,G)$ in which $W$ corresponds to a non-crossing
Euler tour.  We call such an $M(W,G)$ (and the correspondence between edges of the multigraph and edges of~$W$) a \defn{certificate} that $W$ is
weakly simple.  Note that certificates are not unique; in particular,
they can have different rotation systems.  For example, see
Figure~\ref{fig:weakly-simple}(c) in which $G$ is a single edge and
$W$ traverses it four times.

Let $M(W,G)$ be a certificate that $W$ is weakly simple. Some faces of
$M(W,G)$ are digons between parallel edges.  Each remaining face is a
union of faces of $G$.  (If an edge of $G$ has multiplicity 0, then
the incident faces are merged in $M(W,G)$.)  Walk $W$
corresponds to a non-crossing Euler tour of $M(W,G)$, which determines
the interior faces of $M(W,G)$.  We say that a face of $G$ is
\defn{interior} to $W$ if it corresponds to an interior face of
$M(W,G)$.  See \Cref{fig:walks}. %
(Observe that the interior faces of $G$ are well-defined independently
of the choice of the certificate because, in a 2-coloring of the faces
of $M(W,G)$, the color of a face of $G$ does not depend on the choice
of rotation system for $M(W,G)$---the two faces incident to edge $e$
of $G$ have the same color if $m(e)$ is even, and opposite colors if
$m(e)$ is odd.)

\paragraph{Weakly simple polygons.}

A polygon  $P$ is a sequence $(p_0, p_1, \ldots , p_{n-1})$  of $n\ge
1$ points %
(the \defn{vertices} of $P$) together with the line segments $p_i p_{i+1 \bmod n}$ (the \defn{edges} of~$P$).
We do not allow edges of 
length~0. 
As a degenerate case, we allow a polygon with a single vertex and no edges.
A polygon is \defn{simple} if the vertices are distinct points and no two edges intersect, except for consecutive edges that intersect at their common vertex. 

For a general non-simple polygon,
a point of the plane may correspond to multiple polygon vertices, and
polygon edges may overlap or cross.
An \defn{interior crossing} of $P$ is a point that is in the relative interiors of two (or more) edges of $P$ that are not collinear.  A \defn{fork} is a vertex of $P$ that lies in the relative interior of an edge of~$P$.
Both interior crossings and forks can be eliminated by subdividing edges, albeit possibly with a quadratic blow-up in the number of vertices of the polygon.

The standard definition~\cite{aaet-rwsp-17,cex-dwsp-15} of a
{weakly simple}
polygon is as follows. For an~$\varepsilon > 0$, we call a polygon that results from moving
  each vertex of a polygon $P$ 
  by at most
$\varepsilon$   
  an \defn{\epspert} of~$P$.
A
polygon %
is then \defn{weakly simple}
if it has fewer than three vertices, or it has at least 
three vertices  and 
for any $\varepsilon > 0$,
it has an {\epspert} that is 
a simple polygon.  We will use an equivalent characterization:  a polygon is weakly simple if and only if it is a limit of
simple polygons
as the Fréchet distance goes to zero~\cite[Theorem 2.1]{cex-dwsp-15}.

In our proofs we find it useful to characterize weakly simple polygons in terms of 
the purely combinatorial notion of 
non-crossing Euler tours in an associated multigraph. 
Let $P$ be a polygon without interior crossings. 
Expanding on definitions from~\cite{aaet-rwsp-17,cex-dwsp-15},
we define 
the \defn{image graph} of %
$P$ to be a plane straight-line graph $G$ formed as follows. First subdivide edges of $P$ at forks.
Next replace every  set of coincident vertices of $P$ by a single vertex of $G$, and
replace every set of %
equal line segments of $P$ 
by a single edge of $G$ 
(%
these are called ``segments'' 
in~\cite{aaet-rwsp-17,cex-dwsp-15}). 
Then $P$ corresponds to a closed walk $W_P$ in the plane graph $G$ and we can apply the concept of a certificate for a weakly simple walk from above. 
In this context we call an expansion $M(W_P,G)$  an \defn{image multigraph} of $P$.

\paragraph{Combinatorial characterization of weakly simple polygons.}  Finally, we can give our combinatorial reformulation of weak simplicity for polygons.

\begin{lemma}
\label{lemma:weakly-simple}
A polygon $P$ 
is weakly simple if and only if it has no interior crossings and
it has an image multigraph in which $P$ corresponds to a non-crossing Euler tour.
\end{lemma}

An image multigraph in which $P$ corresponds to a non-crossing Euler tour is called a \defn{certificate} 
that $P$ is weakly simple.
Again, note that a certificate is in general not unique.

Before turning to the proof of the lemma, we  
discuss equivalent notions of the interior of a weakly simple polygon. 
The definition of the interior of a non-crossing Euler tour gives one definition of the interior of a weakly simple polygon. This is equivalent to the definition of interior in terms of winding numbers, see \Cref{sec:winding}.  
As seen in Figure~\ref{fig:weakly-simple}, the edges of a weakly simple polygon can be partitioned into boundary walks of the interior faces.

The $O(n \log n)$ time algorithm to recognize weakly simple polygons by Akitaya, Aloupis, Erickson, and T\'oth~\cite{aaet-rwsp-17} 
implicitly proves 
Lemma~\ref{lemma:weakly-simple} (as does the earlier algorithm by Chang, Erickson, and Xu~\cite{cex-dwsp-15}).  Akitaya et al.\ use 
\emphblack{strip systems} as their certificates of weak simplicity.  A strip system is more geometric in nature, but has the advantage of having linear size, which is important for their fast algorithm.   
Our image multigraphs have quadratic size, but they more immediately give the properties we need.

We give a direct proof of Lemma~\ref{lemma:weakly-simple} that depends only on the  characterization of a weakly simple polygon as a limit of simple curves as Fréchet distance goes to zero~\cite[Theorem 2.1]{cex-dwsp-15}, which allows adding new vertices to the polygon.

\begin{proof}
Suppose $P$ is weakly simple. Then $P$ has no interior crossings since these cannot be eliminated by $\varepsilon$-perturbations. Let $P'$ be the result of 
subdividing $P$ at forks.  
By definition, for any $\varepsilon>0$ there is a simple
\epspert\ %
of $P$, and this determines  
a simple \epspert\ %
of $P'$, call it~$P'_{\varepsilon}$.  
Any set $U$ of coincident vertices of $P'$
lies in a disc $D$ of radius $\varepsilon$ in $P'_{\varepsilon}$, and the edges incident to $U$  leave $D$ at distinct points. 
From $P'_{\varepsilon}$ we
construct a plane multigraph $M$ by contracting each set $U$ to a single vertex, and 
ordering the incident edges according to their order around~$D$. 
Then $M$ is a plane Eulerian multigraph that expands the image graph of~$P$, and $P$ corresponds to a non-crossing Euler tour of $M$.

For the other direction, suppose $P$ has no interior crossings and suppose $P$ has an image multigraph $M$ in which $P$ corresponds to a non-crossing Euler tour $W$.
As mentioned earlier, a polygon is weakly simple if it is a limit of
simple polygons
as the Fréchet distance goes to zero~\cite[Theorem 2.1]{cex-dwsp-15}.
So it suffices to show that for any
$\varepsilon>0$, %
we can construct 
a simple polygon $P_\varepsilon$ within Fréchet distance
$\varepsilon$ of $P$.
  Note that this is a weaker requirement than for the
  \epspert\ used above
  because %
the polygon $P_\varepsilon$
 is allowed to
have more vertices
than~$P$.

\begin{figure}[htb]
  \centering
  \includegraphics{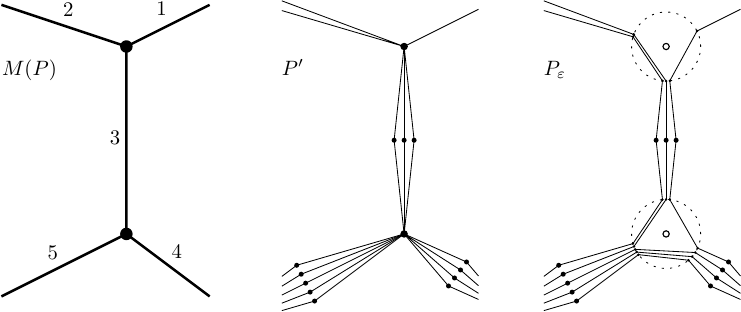}
  \caption{Constructing $P_\varepsilon$ from the image multigraph
    $M$ in two steps.}
  \label{fig:spreading}
\end{figure}

The construction is illustrated for a small section
  of the polygon in \autoref{fig:spreading}.
The coordinates of the vertices of $P$ determine a
straight-line drawing of the image graph $G$ of~$P$ in the
plane.  We expand this to a 1-bend drawing $P'$ of $M$ in which the edges in each bundle are spread apart.  %
Each edge~$e$ of $G$ corresponds to a bundle of $m(e)$ edges in
$M$.  We add a vertex
on %
every edge of the bundle and space these vertices along a small line segment perpendicular to $e$ through its midpoint, using the ordering of the edges in the rotation system of $M$. 

We complete the construction of $P_\varepsilon$ by
altering the drawing of $M$ to spread apart the coincident vertices of $P'$.
For each vertex $v$ of $M$, construct a small disc $D$ of radius~$\varepsilon$ centered at $v$ in the drawing. 
The edges that enter $D$ are incident to $v$, and they cross the boundary of $D$  in the cyclic rotation system ordering. 
Let $D_e$ be the point where edge $e$ enters disc~$D$.
Suppose the Eulerian tour $W$ visits $v$, entering on edge $e$ and leaving on edge~$f$. In the drawing and in $W$ replace segments
$D_e v$ and $v D_f$ by the chord $D_e D_f$. 
If two of these chords of 
$D$ cross, they would correspond to a vertex crossing in $W$.
Thus the result is a simple polygon $P_\varepsilon$.
By making the length of the perpendicular segments and the radius of the
disks smaller than $\varepsilon$, we ensure that the Fréchet distance between
 $P_\varepsilon$ and the original polygon $P$ is smaller than~$\varepsilon$.
\end{proof}

\paragraph{More terminology.}

The above lemma allows us to partition the edges of a weakly simple polygon into boundary walks of 
interior faces, see Figure~\ref{fig:weakly-simple}.  Note that we first subdivide an edge when a vertex lies in its relative interior. 

A vertex of a weakly simple polygon
is a \defn{transition vertex} if its two incident edges
belong to different interior faces.
An interior face of two edges is a \defn{corridor} and an interior face of more than two edges is a \defn{chamber}.  A chamber is not necessarily a simple polygon, but it is 
almost simple (see \Cref{fig:weakly-simple}(b)).
More formally, 
a \defn{bounded  almost-simple polygon} is 
the 
boundary walk of an interior 
face of a connected straight-line graph drawing in the plane. 
We also allow an \defn{unbounded almost-simple polygon} 
by
traversing the boundary of the outer face clockwise.    
An almost-simple polygon has a connected interior and 
a bounded almost-simple polygon
can be triangulated. 
Almost-simple polygons play two roles: the bounded ones arise as chambers; and
our general 
\textsc{Enclosure-with-Penalties} problem 
allows almost-simple input polygons (including a single unbounded one).

We note that the partition of the edges of a weakly simple polygon  into interior faces (and hence the definition of corridors and transition vertices) is not unique, see Figure~\ref{fig:weakly-simple}(c). 
This non-uniqueness, which is inherent in the $\varepsilon$-approximation definition of weakly simple polygons, does not affect our proofs.

\subsection{Winding number and winding parity}
\label{sec:winding}

Our algorithm will construct intermediate polygons that are not necessarily weakly simple, so we will find it useful to generalize %
the notion of ``enclosure'' 
in terms of
winding numbers.
Let $W$ be a polygon and let $x$ be a point not lying on a vertex or edge of~$W$.
The \defn{winding number} $\wwind W x$ of~$x$ with respect to~$W$ (or of~$W$ with respect to~$x$, depending on the viewpoint) is defined as follows.  Take a ray $\rho$ from $x$ that avoids vertices of $W$.  
If an edge of $W$  crosses $\rho$
from right to left,
we count this as~$+1$; a crossing 
from left to right
is counted as~$-1$, and the total
count gives the winding number.
This is well-defined independent of the choice of $\rho$.
The winding number is undefined for points $x$ on $W$. 
Observe that, for a weakly simple polygon $W$ traversed counterclockwise, point $x$ lies in the interior of $W$ if and only if $\wwind W x = 1$.
The \defn{winding parity} of $x$ with respect to~$W$ (or of~$W$ with respect to~$x$) is $\wwind W x \bmod 2$.

\section{Our Common Framework: 
{Enclosure-with-Penalties}}
\label{sec:framework}

In this section we formally define the
\textsc{Enclosure-with-Penalties} problem, which provides a common
framework for both the geometric and graph settings.

\goodbreak

\noindent{\bf Input:}
\begin{itemize}
\item A set of 
interior-disjoint 
almost-simple polygons in the plane. 
We allow a single polygon to be unbounded.
We subdivide polygon edges
to ensure 
that no polygon vertex lies in the interior of an edge of another polygon.

The \defn{\workspace{}} is the plane minus the interiors of the polygons. 

\item A partition of the input polygons into a set $R$ of $k$ \defn{required} polygons and the remaining set $O$ of \defn{optional} polygons.  If there is an unbounded polygon, it must lie in $O$. 
\item For each polygon $P \in O$, a \defn{penalty} $\pi_P \in \mathbb R_{\ge 0}\cup\{+\infty\}$.

\item 
The weight $w_{ab}$ 
of a line segment 
$ab$ 
in the \workspace{} is its Euclidean length,
except for  
\defn{squeezed edges}.
A squeezed edge %
is 
a polygon edge
that is incident
to polygons
on both sides.
We may specify an \emphblack{arbitrary} positive weight for a squeezed edge.
In particular, the weight can be smaller than its Euclidean length,
and the triangle inequality need not hold. %
Subsegments of a squeezed edge get proportional weight, and 
combinations of different squeezed or non-squeezed segments have
their weights added.
\end{itemize}

\noindent{\bf Output:} 
A weakly simple polygon $W$ that lies in the \workspace{}
and contains all 
polygons
of~$R$ while minimizing
the \defn{cost} $c(W)$, which is defined as
\begin{equation}
\label{eq:cost-definition}
c(W) := w(W) + \pi(W)   ,
\end{equation}
where $w(W)$ is the sum of the weights of the edges of $W$, and $\pi(W)$ is the sum of the penalties of the polygons of $O$ that are inside~$W$.
Our main result,
proved in Sections~\ref{sec:DP}--\ref{sec:dijkstra}, is:

\begin{theorem}\label{thm:main}
\textup{\textsc{Enclosure-with-Penalties}}
for $k$ required polygons 
can be solved in $O(3^k n^3)$ time and $O(2^k n^2)$ space,
if the input polygons have $n$ vertices in total.
\end{theorem}

Note that the \textsc{Enclosure-with-Penalties} problem as defined above does not allow point objects. (They are not almost-simple.) 
Section~\ref{sec:points}
shows how to deal with point objects.

\Cref{thm:main-geom} is an immediate consequence of \Cref{thm:main}.
In Section~\ref{sec:intro}, we outlined how \Cref{thm:main-graph} follows from \Cref{thm:main} via a straight-line embedding of the graph.  We now provide the details.

\begin{proof}[Proof of Theorem~\ref{thm:main-graph}, assuming Theorem~\ref{thm:main}]
\label{proof:graph-reduction}
Consider an instance of \textsc{Graph-Enclosure-with-Penalties}, with
simple connected plane graph $G$, required faces $R$, and optional faces~$O$.  
We reduce to an instance of
\textsc{Enclosure-with-Penalties}.

Find a straight-line plane embedding $G'$ of $G$ with the same combinatorial map (i.e., preserving the rotation system).  The faces of $G'$, including the outer face, become the polygons for our new instance. 
Observe that all these polygons are almost-simple.
For the bounded faces of $G'$ we preserve the partition into $R$ and $O$ and the penalties.
The outer face of $G'$ becomes an unbounded polygon.  We put it in the set $O$ with a penalty of $0$ (the penalty is irrelevant, since no weakly simple polygon $W$ can contain the unbounded polygon). 

The \workspace{} of this set of polygons has no interior; it consists only of the edges of $G'$. Each such edge lies between two faces (polygons) so it is a squeezed edge and we assign it the weight of the corresponding edge of $G$. 

This completes the reduction.  
For a graph on $n$ vertices, the reduction produces a set of polygons with $O(n)$ vertices.  The number $k$ of required objects remains the same.
The reduction takes $O(n)$ time.  The runtime claim in Theorem~\ref{thm:main-graph} follows.

There is a one-to-one correspondence between weakly simple polygons in the \workspace{} and weakly simple closed walks in $G$ (as defined in Section~\ref{sec:weakly-simple}), and the interior and the cost are preserved.  Therefore 
a solution to the resulting instance of the \textsc{Enclosure-with-Penalties} problem 
provides a solution to the original graph problem.
\end{proof}

\section{Dynamic Programming Algorithm}
\label{sec:DP}

In this section, we describe the first component of our algorithm for \Cref{thm:main}, which is a dynamic programming algorithm.

The algorithm 
builds a polygon composed of \workkspace{} edges, where 
a \defn{\workkspace{} edge} is an inclusionwise minimal line segment in the \workspace{}
whose endpoints are vertices of the input polygons. 
Equivalently, a \workkspace{} edge is a visibility edge that contains no polygon vertex in its interior.
We prove in Section~\ref{sec:correctness} that this 
restriction to \workkspace{} edges
is valid.
We 
refer to a solution interchangeably as a polygon or as a closed walk in the graph of \workkspace{} edges. 

\subsection{Overview of the algorithm}

The intuition for the algorithm is based on the decomposition of
the region inside a weakly simple polygon $W$
into corridors and chambers joined at ``cutpoints'', see \autoref{fig:walks}.
A cutpoint separates $W$ into subpolygons and partitions the set of enclosed objects.
Our first type of subproblem finds
polygons that enclose a specified subset of $R$ and go through a specified vertex.

A corridor is a digon, and a chamber can be triangulated by adding chords, where a chord may cut through 
polygons.
We therefore use digons and triangles as the basic building blocks to construct our solutions.
A chord cuts off part of the solution. 
Our second type of subproblem %
finds
polygons that use 
 a walk of \workkspace{} edges between two 
given 
 vertices $p$ and $q$ together with the %
 chord
 $pq$ (called the \defn{mouth}) to enclose a specified subset of $R$.

Since a mouth may cut through polygons, we choose a \defn{reference point} $r_P$ in the interior of every input polygon $P$, and aim to enclose $r_P$ for $P \in R$.  Observe that a weakly simple polygon $W$ in the \workspace{} encloses $P$ if and only if $r_P$ lies in the interior of $W$. 
(For convenience, we could insist that the reference point avoids all potential mouth edges,
but in fact, we will be able to properly resolve the ambiguities resulting from such degenerate
choices of reference points.)

The dynamic programming algorithm explicitly keeps track of the subset 
of required objects that are enclosed by partial solutions ($2^k$ possibilities). However,
when combining two partial solutions, 
the algorithm does not have enough information to check whether they cross.
Thus, we allow self-crossing solutions. 
In particular, our use of the word ``enclosing'' is aspirational, and will only be made precise in terms of winding numbers, see Section~\ref{sec:properties_winding}. 
When we state the algorithm, we invite the reader to think of a weakly simple solution without crossings. 

A subproblem of type $C$ (``closed'') is rooted at a vertex $p$, and we build a closed walk 
that goes through $p$ and is composed of %
\workkspace{} edges. 
A subproblem of type $M$
(``mouth'') 
is rooted at 
a {mouth} segment $pq$ between vertices of input polygons,
and we build 
an open walk of \workkspace{} edges from $p$ to $q$; adding segment $qp$ closes the walk.
Note that the mouth need not be a \workkspace{} edge; it may cut through objects.   %

In addition to
the root, each subproblem has two more parameters, $B$ and~$t$:
The set $B\subseteq R$ specifies the precise subset of required
objects that must be %
enclosed,
and the integer $t\ge 0$ is an upper bound on the number of %
edges
of %
the walk.  In each case, the subproblem looks for a \emphblack{minimum-cost}
solution with the required properties.

\subsection{Dynamic programming recursion %
}
\label{sec:DP-recursion}

\begin{figure}[htb]
    \centering
    \includegraphics[page=2]{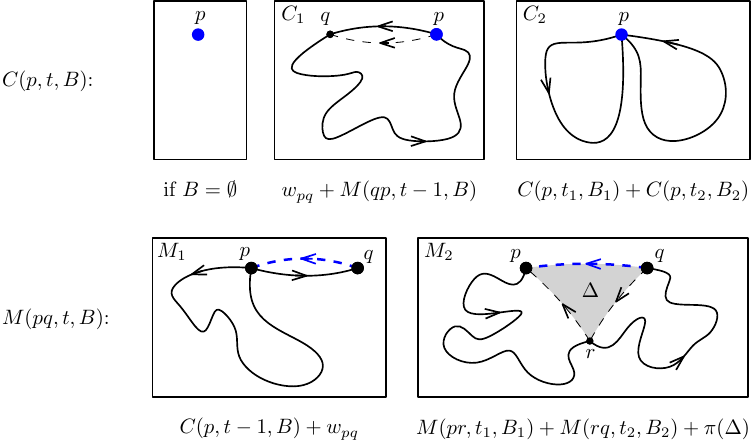}
    \caption{Cases of the recursion. Solid edges are \workkspace{} edges; dashed edges are mouths.%
}
    \label{fig:DP-cases}
\end{figure}

We now give recursive formulas for $C$ and $M$, preceded in each case by an explanation of the formulas.
The formulas with the respective partitions of the walk %
are illustrated in Figure~\ref{fig:DP-cases}.

For $C(p,t,B)$ we have 
two base cases: if $B = \emptyset$, then the %
shortest closed walk is just the point $p$ and its cost is $0$
(Equation~\eqref{eq:C-base-1-new}); and if $B \ne \emptyset$ and $t \le 1$, 
there is no solution, and we set the cost to $\infty$ (Equation~\eqref{eq:C-base-2-new}). 
Otherwise we have two general cases: 
the closed 
walk uses an edge $pq$ of the \workspace{} (for some $q$) plus
a solution $M(qp,t-1,B)$
(Equation
~\eqref{eq:C-expand-new});
or the %
closed walk is composed of two smaller closed walks that both go
through $p$ (Equation~\eqref{eq:C-compose-new}). 
The notation $\sqcup$ means disjoint union: we partition the objects in $B$ into two
sets, each ``enclosed'' 
by exactly one of the two closed walks.

 The base cases define
$C(p,t,B)$ for 
$B = \emptyset$ and for $t\le 1$:
\begin{align}
    C(p,t,\emptyset) &:= 0, \text{ for $t \ge 0$}
\label{eq:C-base-1-new}\\
  C(p,t,B) & :=\infty,
 \text{ for $B\ne\emptyset$ and $t \le 1$}
\label{eq:C-base-2-new}
\end{align}
In the general case, for $B\ne\emptyset$ and $t\ge 2$, we set
\begin{align}
 C(p,t,B) &:=
\min \{C_1, C_2\} \nonumber
\text{, where}
\\   C_1 %
:= {}&
\min \bigl\{\,
    w_{pq} + M(qp,t-1, B)
\bigm|
   pq \text{\ is a \workkspace{} edge\,}
\bigr\} 
\label{eq:C-expand-new}
\\
\label{eq:C-compose-new}
 C_2 %
:= {}&
\min \bigl\{\,
    C(p,t_1, B_1) +
    C(p,t_2, B_2)
\bigm|
t =  t_1 + t_2; \
 B =B_1\sqcup B_2;\
B_1, B_2 \ne\emptyset
\,\bigr\}
\end{align}

For $M(pq,t,B)$ 
there are two possibilities: 
if $pq$
is a \workkspace{} edge, we can use a %
closed walk at~$p$ plus the edge $pq$ (Equation (\ref{eq:M-pq-free-new})); or
we can attach a triangle $\Delta = prq$ to the mouth~$pq$ (Equation (\ref{eq:M-triangle-new})).
In the first case we add the weight of the edge $pq$.
In the triangle case we take into account the polygons with reference points in $\Delta$, where we 
consider $\Delta$ to be closed on $pr$ and $rq$ and open on $pq$.  Define 
$R(\Delta)$ to be the polygons
of $R$ with reference points in~$\Delta$,
and define
$\pi(\Delta)$ to be the sum of 
the penalties of  polygons
of $O$ with reference points in~$\Delta$.

$M(pq,t,B)$ is defined only for $t\geq 1$:
\begin{align}
 M(pq,t,B) &:=
\min \{M_1, M_2\} \nonumber
\text{, where}
\\ 
    M_1 &:= 
\begin{cases}
         C(p,t-1,B) + w_{pq}  &   
    \text{if $pq$ is a \workkspace{} edge}\\
    \infty, &\text{otherwise}
\end{cases}
\label{eq:M-pq-free-new}
\\[-\baselineskip]\nonumber %
\\
 M_2 & := \min
 \label{eq:M-triangle-new}  
  \bigl\{\, 
 M( %
pr,t_1,B_1) + M(rq,t_2, B_2) + 
\pi(\Delta)
\bigm|  %
  \\\nonumber
  &\qquad\qquad\qquad
\Delta = prq\text{ is a counterclockwise\ triangle}, 
   \\\nonumber
  &\qquad\qquad\qquad
t=t_1 + t_2,\, t_1\ge 1, t_2 \ge 1,
\
   B
    =B_1\sqcup B_2
    \sqcup
   R(\Delta) %
    \,\bigr\} %
\end{align}
As we shall see later 
in \Cref{lemma:discrete-existence-new},
the optimal walk has at most $6n$ edges.
Thus, we define the solution to the whole problem as
\begin{align}
\label{eq:final}
\cDP  := \min\{\,
 C(p,6n,R) \mid p\text{ a vertex}
 \,\}   .
\end{align}

In the following sections we prove that $\cDP$ is the correct value.  
Although not required by our proof, we specify for completeness in 
\Cref{sec:immersed}
the class of polygons over which the algorithm optimizes: It is %
the class of \emphblack{immersed} or \emphblack{self-overlapping} 
 \ws\
polygons.
When we allow point objects, the algorithm needs a few refinements,
see
\Cref{sec:points}.

\subsection{Runtime and space}
\label{sec:runtime}
The following lemma provides a crude bound of $O(3^kn^5)$ on the runtime of the dynamic programming algorithm.
A more efficient version of the dynamic programming algorithm, deferred to Section~\ref{sec:dijkstra},
eliminates the parameter $t$ and runs in time $O(3^k n^3)$.

\begin{lemma}\label{lem:runtime-space}
  The runtime of the above dynamic programming algorithm is $O(3^kn^5)$.
\end{lemma}
\begin{proof}
The number of subproblems of type $M$ is $O(n^3 2^k)$: there are $O(n^2)$ choices for the mouth $pq$,
$O(2^k)$ choices for the set $B$, and $O(n)$ choices for the parameter $t$.
The number of subproblems of type $C$ is only $O(n^2 2^k)$, by the same analysis.
Thus, the space requirement is $O(n^3 2^k)$.

The %
recursion that dominates the runtime is \eqref{eq:M-triangle-new} for $M_2$. 
For fixed parameters $pq$, $t$, and~$B$, there are at most $n$ choices for the point $r$,
at most $t=O(n)$ possibilities for $t_1$ and $t_2$, and at most $2^{|B|}$ choices for $B_1$ and $B_2$.
We run through the possible choices of $r$ in an outer loop.
Then, for each triangle $\Delta=prq$, we can determine the reference points that lie in $\Delta$ in a straightforward way in $O(n)$ time, 
and this runtime will be dominated by the inner loop.
This leads to an overall runtime of
\begin{displaymath}
    O(n^3)\times \sum_{B\subseteq R} n \times \bigl(O(n)+ O(n)\times 2^{|B|}  \bigr) =
    O(n^5)\sum_{B\subseteq R} 2^{|B|}=
    O(n^5)\sum_{i=0}^k \binom{k}{i} 2^{i}=
    O(n^5 3^k).
\end{displaymath}
We assume that we can %
access each
of the $O(n^3 2^k)$ entries of the dynamic programming
table in constant time. In particular, a memory word contains at least $k$ bits,
and hence set operations on subsets of $R$ take constant time
if we implement these subsets as bit vectors.

The above computation assumes that we can determine in $O(1)$ time, given two input vertices $p$ and~$q$, whether $pq$ is a \workkspace{} edge.  For this purpose, we precompute a Boolean array of size $n\times n$, with rows and columns indexed by the vertices, storing this information.  The array can be determined in $O(n^2)$ time from the visibility graph of the input polygons, which can be computed in $O(n\log n+e)=O(n^2)$ time for a visibility graph with $e$~edges~\cite{gm-osacv-91}.%
\end{proof}

\subsection{Extracting the solution \texorpdfstring{$\WDP$}{W(DP)}}
\label{sec:extracting}

We now show that with every finite value computed in the  dynamic
programming recursion, we can naturally associate an open or closed walk of
\workkspace{} edges that has this value as its cost, with ``partially enclosed'' objects taken into account according to their reference points.  \Cref{fig:DP-cases} illustrates how each value is computed from the values of subproblems;
the walks corresponding to those subproblems are composed accordingly.  Here are the details.

With each finite value $C(p,t,B)$ 
that is computed in the recursions
\eqref{eq:C-base-1-new}--\eqref{eq:C-compose-new},
we can naturally %
associate a 
polygon $W = W(p,t,B)$, a
closed walk of \workkspace{} edges that goes through~$p$.
Similarly, 
with each finite value $M(pq,t,B)$ computed in the recursions
\eqref{eq:M-pq-free-new}--\eqref{eq:M-triangle-new}, we can associate an open walk of \workkspace{} edges 
$W = W(pq,t,B)$ that goes from $p$ to~$q$.  
For example, in \eqref{eq:M-triangle-new}, where we
form the sum
$M(pr,t_1,B_1) + M(rq,t_2, B_2)$, the open walk $W$ is obtained by
concatenating the open walks associated 
with 
$M(pr,t_1, B_1)$ and $M(rq,t_2,B_2)$.

\begin{definition}
\label{defn:bar-W}
For an open walk $W$ from $p$ to $q$, we define $\overline W$ to be the polygon $W+qp$.
For a closed walk $W$, we define $\overline W$ to be the polygon $W$ itself.
\end{definition}

The recursions
\thetag{\ref{eq:C-base-1-new}--\ref{eq:final}} define only the
\emphblack{values} $C$ and $M$, but we can also recover the
associated walks, as is standard
in dynamic programming algorithms.
We augment the dynamic programming tables with extra information,
remembering for each application of a recursion the
combination of values $t_1,t_2,B_1,B_2$, etc.\ from
which the minimum was obtained. From this information,
we can then recursively reconstruct the
open/closed walks $W$ 
and %
$\overline W$ %
 associated 
 with a given entry of the table
in $O(t)$ time.

We will prove in 
\Cref{lem:cDP-finite-1}
that
$\cDP$ is finite; so the associated closed walk exists:

\begin{definition}
\label{defn:WDP}
$\WDP$ is the polygon
associated with the optimum solution value $\cDP $ in~\eqref{eq:final}.
\end{definition}

\begin{figure}[htb]
    \centering
    \includegraphics{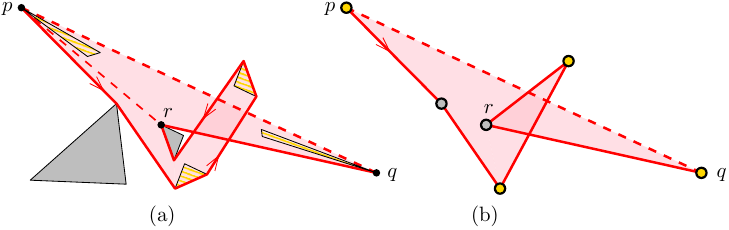}
    \caption{(a) The optimum solution with given mouth $pq$ can 
    self-overlap. 
    The yellow objects are required, and the grey objects have %
    high penalties.
    The solution with mouth~$pr$ is
    a simple polygon. Attaching the triangle $pqr$ to it yields a self-overlapping polygon.  (b) The same example with point objects
    (cf.~\Cref{sec:points}). Observe that the solution covers more than $360^\circ$ around $r$, although this is not apparent locally from looking at the boundary near~$r$.
    }
    \label{fig:overlapping-solution}
\end{figure}

\Cref{fig:overlapping-solution} shows an example of the polygon associated with a {subproblem} of type~$M$. 

\subsection{Properties of \texorpdfstring{$\WDP$}{W(DP)}}
\label{sec:properties_winding}
The polygon $\WDP$ uses \workkspace{} edges, but it
might not be weakly simple,  
and there is no notion of enclosed objects. 
Instead, we use winding numbers: we show that the reference point of any object in $R$ has winding number 1 in $\WDP$, and we  define a cost measure for $\WDP$ in terms of winding numbers %
and prove equality with $\cDP$.  The following definition also applies to an open walk, by considering the winding number of its closure; this will be useful later.%

\begin{definition}
\label{defn:general-cost}
For any open or closed walk $W$ in the \workspace{},
the %
  {associated penalty} %
  is

\begin{equation}\label{eq:gen-pen}
  \pi(W):= %
  \sum
    _{P\in O}
    \wwind 
    {\overline W} {r_P} \cdot\pi_P 
   ,
 \end{equation}
and we define the \defn{cost} as %
 \begin{displaymath}
c(W) = w(W) + \pi(\overline W).
 \end{displaymath}%
\end{definition}
When $W$ is a counterclockwise weakly simple 
closed walk,
this matches the previous definition of
$\pi(W)$ and %
$c(W)$, see Equation~(\ref{eq:cost-definition}).
For a reference point $r_P$ lying on the mouth $qp$ of an open walk $W$ from $p$ to~$q$, we
compute 
$\wwind 
    {\overline W} {r_P}$ as if $r_P$ were slightly moved
    to the right of the segment $qp$, i.e., in the direction where the 
    outside %
    would normally be in case of a %
    counterclockwise simple polygon.
In case of a weakly simple polygon $\overline W$, this means that
points on the mouth are not considered to be enclosed.
For objects with infinite penalty $\pi_P=\infty$, the
  expression $0\cdot\pi_P$ in
  \eqref{eq:gen-pen} is taken as~0. The
penalty $\pi(W)$ and the
  cost $c(W)$ are not always defined, because
the sum in \eqref{eq:gen-pen} may involve terms of the form
$\infty-\infty$ when %
winding numbers are negative.

In this section, we prove that the solution $\WDP$ produced by the algorithm
is a walk in the \workspace{} that has the following properties.
Recall that $\WDP$ is only defined if $\cDP$ is finite.

\begin{lemma}
\label{lemma:WDP-properties}
Assuming $\cDP$ is finite, we have:
\begin{enumerate}[\rm (a)]
\item $\cDP  = c(\WDP)$\textup;
\label{DP-cost}

\item for all $P \in R$, $\wwind \WDP {r_P} =1$\textup;
\label{DP-R}
\item for all points $x$ that do not lie on $\WDP$, $\wwind\WDP x \ge 0$.
\label{DP-nonnegative}
\end{enumerate}    
\end{lemma}

\begin{figure}
    \centering
    \includegraphics%
    {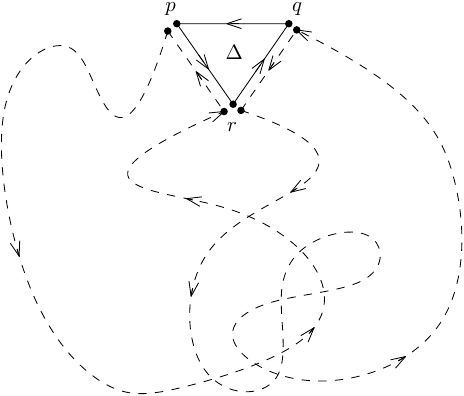}
    \caption{Gluing together closed walks, which may cross each other and are possibly self-crossing.}
    \label{fig:polygon-composition}
\end{figure}

Lemma~\ref{lemma:WDP-properties} is proved  
by induction as the %
    dynamic programming algorithm builds solutions
to subproblems by gluing together open or closed walks. 
The induction must apply also to open walks, and 
we 
use the fact
that winding numbers add when gluing walks together, see Figure~\ref{fig:polygon-composition}.

To prove Lemma~\ref{lemma:WDP-properties}, we need results on winding numbers as walks are glued together.
We first define gluing more precisely.
Two closed walks $W_1$ and $W_2$ can be glued together at a common vertex,
or %
along a common
edge that is traversed in opposite directions
by $W_1$ and~$W_2$. %

More formally:
If $W_1 = (p,q_1,q_2,\ldots ,q_n)$
and
 $W_2 = (p,p_1,p_2,\ldots ,p_m)$,
then the result of gluing the walks along the common point $p$ is
 $W = (p,q_1,q_2,\ldots, q_n,p,p_1,p_2,\ldots, p_m)$.

If $W_1 = (p,q,q_2,\ldots, q_n)$
and
 $W_2 = (q,p,p_2,\ldots, p_m)$
 both use the edge $pq$, but in opposite directions, then
  $W = (q,q_2,\ldots, q_n,p,p_2,\ldots ,p_m)$ is the result.

We have the following easy but key property:
\begin{proposition}[Additivity of winding numbers]
\label{lem:additive}
The winding number is additive with respect to the gluing operation: If $W$ is a closed walk obtained by gluing two closed walks $W_1$ and $W_2$ along a common edge or vertex, then
\begin{displaymath}
\wwind W x = \wwind {W_1} x + \wwind {W_2} x
\end{displaymath}
for all points $x$ that do not lie on $W_1$ or $W_2$.
\end{proposition}

\begin{proof}
  Let $\rho$ be any ray from~$x$ to the unbounded face that avoids the vertices of $W$ and intersects the edges of~$W_1$ and~$W_2$ transversally.

  Assume first that $W$ results from gluing $W_1$ and~$W_2$ at a common vertex; then the multiset of the directed edges of~$W$ is exactly the union of the directed edges of~$W_1$ and of~$W_2$ (counting multiplicities).  Let $r^+$, $r_1^+$, and~$r_2^+$ be the number of times an edge of~$W$, $W_1$, and~$W_2$, respectively, crosses~$\rho$ from right to left; we have $r^+=r_1^++r_2^+$.  Similarly, with the analogous notations $r^-$, $r_1^-$, and $r_2^-$ counting the number of crossings from left to right, we have $r^-=r_1^-+r_2^-$.  Summing up, we obtain the result.

  If $W$ results from gluing $W_1$ and~$W_2$ at a common edge~$pq$,
 the effects of the two oppositely oriented edges $pq$ and $qp$
 cancel out when the winding number is computed. (They contribute 
 to
 $r_1^+$ and $r_2^-$, or 
 to
 $r_1^-$ and $r_2^+$, or not at all.)
 The proof for the first case %
 carries over. %
\end{proof}

We now prove \Cref{lemma:WDP-properties}.

\begin{proof}[Proof of \Cref{lemma:WDP-properties}]
We prove by induction that the properties hold more
generally for all subproblems solved in the dynamic programming algorithm.  To be precise,
consider a finite value $C(p,t,B)$ or $M(pq,t,B)$ computed in the recursions
\eqref{eq:C-base-1-new}--\eqref{eq:M-triangle-new}.  Let $W = W(p,t,B)$ or $W = W(pq,t,B)$ be the closed or open walk associated with the solution and let $\overline W$ be the associated polygon
as in Definition~\ref{defn:bar-W}.
We prove by induction on $t$ that:

\begin{enumerate}[(i)]
\item $C(p,t,B)  = c(W(p,t,B))$, $M(pq,t,B) = c(W(pq,t,B))$;
\label{A}
\item for all $P \in B$, $\wwind {\overline W} {r_P} =1$ and for all $P \in R \setminus B$, $\wwind {\overline W} {r_P} = 0$;
\label{B}
\item for all points $x$ that do not lie on $\overline W$, $\wwind {\overline W} x \ge 0$.
\label{C}
\end{enumerate}

These properties hold in the base case \eqref{eq:C-base-1-new}, where $C(p,t,\emptyset)=0$ and $W$ is the single point~$p$.
For the general formulas we 
rely heavily 
on the additivity
of the winding number with respect to gluing,
 \Cref{lem:additive}.
The cases are as follows, numbered by the equation numbers;
it may help to refer to \Cref{fig:DP-cases}.
\begin{enumerate}
\item[\eqref{eq:C-expand-new}] $C = C_1 = w_{pq} + M(qp,t-1,B)$ where $pq$ is a \workkspace{} edge. 

By induction, the properties hold for 
the open walk $W_0 = W(qp,t-1,B)$.
Let $W$ be the polygon associated with $C$, i.e., $W = pq + W_0$.  
Observe that $W$ is the same polygon as $\overline{W}_0$.  This takes care of properties (\ref{B}) and~(\ref{C}).  For property~(\ref{A}), note that $c(W) = w_{pq} + c(W_0)$.
By induction, $c(W_0) = M(qp,t-1,B)$.  
Thus $c(W) = w_{pq} + M(qp,t-1,B) = C$, which proves property~(\ref{A}).

\item[\eqref{eq:C-compose-new}] $C = C_2 = C(p,t_1,B_1) + C(p,t_2,B_2)$ where $t=t_1 + t_2, B= B_1 \sqcup B_2, B_1, B_2 \ne \emptyset$.

By induction, the properties hold for the polygons $W_1 = W(p,t_1,B_1)$ and $W_2 = W(p,t_2,B_2)$.
The polygon $W$ associated with $C$
is formed by gluing $W_1$ and $W_2$ at the common point $p$.
The weights are additive by definition: $w(W) = w(W_1) + w(W_2)$,
and by 
 additivity of winding numbers, %
 $\wwind W x = \wwind  {W_1} x + \wwind {W_2} x$ for all points $x$ not on $W$. 
Property~(\ref{C}) follows immediately, and
property~(\ref{A}) 
follows by the definition of the cost,  
$c(W)=w(W)+\sum_{P\in O}
\wwind {W} {r_P} \cdot\pi_P$.

Property (\ref{B}) propagates from $B_1$ and $B_2$ to their disjoint union $B$ by the additivity %
of winding numbers. %
More precisely, consider first some $P \in B$.  Since 
$B= B_1 \sqcup B_2$, the polygon %
$P$ is in exactly one of these sets. Suppose without loss of generality that $P \in B_1$.  By induction, $\wwind {W_1} {r_P} = 1$ and $\wwind {W_2} {r_P} = 0$.  Thus $\wwind {W} {r_P} = 1$.  Finally, if $P \in R \setminus B$, then by induction $\wwind {W_1} {r_P} = 0$ and $\wwind {W_2} {r_P} = 0$, so $\wwind {W} {r_P} = 0$, as required.

\item[\eqref{eq:M-pq-free-new}] $M = M_1=  C(p,t-1,B)+ w_{pq}$ where $pq$ is a \workkspace{} edge.

By induction, the properties hold for the polygon $W_0 = W(p,t-1,B)$. 
The open walk $W$ that is associated with $M$ starts at $p$,  traverses 
the polygon $W_0$ and then the edge $pq$, 
ending at $q$. $\overline W$ is formed by gluing the doubled edge $qp$ to the polygon~$W_0$.  
Thus, winding numbers with respect to $\overline W$ are 
the same as for $W_0$, except that they become undefined
for points $x$ on~$pq$. This proves properties (\ref{B}) and~(\ref{C}),
and also that the penalty term
\eqref{eq:gen-pen}
in the cost
for 
$W$ is the same as for~$W_0$. 
 Since $w(W) = w(W_0) + w_{pq} $,
 property~(\ref{A})
 follows. 

\item[\eqref{eq:M-triangle-new}]
$M = M_2= M(pr,t_1,B_1) + M(rq,t_2, B_2) + 
\pi(%
\Delta)%
$
where
$\Delta = prq$ is a counterclockwise  triangle,
$t=t_1 + t_2,t_1\ge 1, t_2 \ge 1$,
$ B  =B_1\sqcup B_2
    \sqcup R(\Delta)
$.

By induction, the properties hold for the open walks $W_1 = W(pr,t_1,B_1)$ and $W_2 = W(rq,t_2,B_2)$. 
Let $W$ be the open walk associated with $M$. 
Then $w(W) = w(W_1) + w(W_2)$.
$\overline W$ is formed by gluing $\overline{W}_1$ and $\overline{W}_2$ to $\Delta$ on the common edges $pr$ and $rq$, respectively.
The argument is analogous to the treatment of
\eqref{eq:C-compose-new} above, except that
we form the combination of \emphblack{three} areas, and
two gluings are performed, along common edges instead of common vertices.

An important point is therefore the treatment of reference points
that lie \emphblack{on} these edges:
By the convention
established in connection with \Cref{defn:general-cost}, %
 the points on the mouths
$pr$ and $rq$ are not considered to be enclosed
by
$\overline{W}_1$ and $\overline{W}_2$, both for
determining 
$R(\overline{W}_i)$ and for computing
$\pi(\overline{W}_i)$.
However, 
when determining 
$R(\Delta)$ and 
$\pi(\Delta)$,
these edges are considered to be part of
$\Delta$, by our conventions of
\Cref{sec:DP-recursion},
in the paragraph before~\eqref{eq:M-pq-free-new}.
Thus the points on the mouth are neither overcounted nor undercounted.

The edge $pq$ is \emphblack{not}
considered as part of
$\Delta$. This is in line with the convention 
of \Cref{defn:general-cost}
that the mouth
$pq$ should not be counted as enclosed by $\overline{W}$.
\qedhere

\end{enumerate}
\end{proof}

\subsection{Optimal walks are found by the algorithm}
\label{sec:best-walks}

In the previous section we showed by induction on subproblems that
 the walk $\WDP$ that is found  
by the dynamic programming algorithm has certain properties (slightly weaker than being weakly simple and enclosing $R$).
In this section we prove a kind of converse:
Every walk %
that has these properties and that  is, moreover, %
weakly simple, is considered as a ``candidate solution'' by the dynamic programming algorithm.
Formally:

\begin{lemma}
\label{lem:cDP-calD}
The value $\cDP$ that the dynamic programming algorithm computes is at most the cost of any weakly simple polygon $W$ that is oriented counterclockwise, encloses $R$, and consists of at most $6n$ free-space edges.
\end{lemma}

By the definition of \cDP, it
suffices to show that for any vertex $p$, $C(p,6n,R)$ is at most the cost of any weakly simple polygon $W$ that goes through $p$, encloses $R$, and consists of at most $6n$ free-space edges.
We give an inductive proof of the more general statement
that $C(p,t,B)$ is at most the cost of any weakly simple polygon $W$ that goes
through~$p$, encloses $B$, and consists of
at most~$t$ \workkspace{} edges. In turn, this requires an analogous inductive statement for $M(pq,t,B)$,
and for this we use the notion of cost of an open walk.
The precise statement is given in
the following technical lemma.  It refers to transition vertices, which were defined in Section~\ref{sec:weakly-simple}.
\begin{lemma}\label{lemma:buildup}
\textup{(A)} Let $W$ be a weakly simple polygon that is oriented counterclockwise and consists of
$\ell 
$ \workkspace{} edges.
Let $p$ be a vertex of $W$ and let $B$ be the objects of $R$
that are enclosed by $W$. 
Then, for all  $t\ge \ell$,
$C(p,t,B) \le c(W)$. 

\textup{(B)}
Let $W_0$ be an open walk with $\ell$ \workkspace{} edges from
vertex
$p$ to
vertex
$q$ such that the
polygon $W 
= W_0 + qp$ is weakly simple
and $q$ is not a transition vertex of $W\!$.
Let $B$ be the objects of $R$ %
whose reference points 
lie inside $W$ and not on $pq$.
Then, for all  $t\ge \ell$,
$M(pq,t,B) \le c(W_0)$.
\end{lemma}

\begin{proof}[Proof of \Cref{lem:cDP-calD}, assuming \Cref{lemma:buildup}]
  Let $W$ be a weakly simple polygon that is oriented counterclockwise, encloses~$R$, and consists of at most $6n$ \workkspace{} edges.  Let $p$ be a vertex of~$W$.  By definition, $\cDP$ is at most $C(p,6n,R)$, which by \Cref{lemma:buildup}(A) is at most~$c(W)$.
\end{proof}

\begin{proof}[Proof of \Cref{lemma:buildup}]
Recall the concepts of an open walk $W_0$, its closure $\overline W_0$ and cost $c(W_0)$ from Definitions~\ref{defn:bar-W} and~\ref{defn:general-cost}.

Let $M$
be a certificate that $W$ is weakly simple, i.e., $M$ is an image multigraph in which $W$ corresponds to a non-crossing Euler tour, see Section~\ref{sec:weakly-simple}.  
Via this correspondence, 
each edge of $W$ has an interior face of $M$ to its left, which provides a 
partition of the edges of $W$ into faces of $M$. We use the cyclic order of edges around faces in the proof.  The reader may find it helpful to refer to Figure~\ref{fig:weakly-simple}.

As in the proof of Lemma~\ref{lemma:WDP-properties}, we go through each 
case of the 
dynamic programming recursion. %
The difference is that in Lemma~\ref{lemma:WDP-properties} we  analyze, in terms of winding numbers, the cost of any %
polygon constructed  by %
the 
dynamic programming algorithm
(potentially not weakly simple),
whereas here we will deconstruct any weakly simple polygon into smaller pieces as defined by the
appropriate recursion formula,
and winding numbers do not come into play.

We prove  claims (A) and (B) simultaneously by induction on $\ell$. 
For part~(A), see
\Cref{fig:structure-A}.

\begin{figure}[htb]
    \centering
    \includegraphics{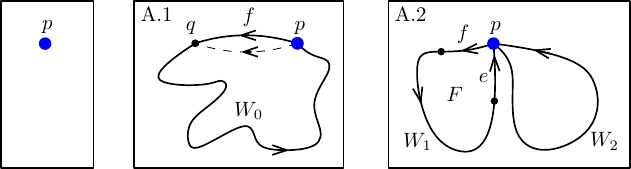}
    \caption{Cases for statement (A) of \Cref{lemma:buildup}.}
    \label{fig:structure-A}
\end{figure}

In the base case, $\ell=0$, polygon $W$ degenerates to a single point,
so only case~(A) applies. For objects with interior, $W$ cannot enclose any objects, so $B = \emptyset$.  By Equation~\eqref{eq:C-base-1-new}, $C(p,t,B) = 0 \le c(W)$.

For part (A), the case $B = \emptyset$ was just dealt with, and for $B \ne \emptyset$ we
distinguish two cases depending whether 
$p$ is a transition vertex in $W$.
Let $f=pq$ be the edge of $W$ that follows~$p$. %

\paragraph{Case A.1. 
$p$ is not a transition vertex.}
Let $W_0$ be the open walk from $q$ to $p$ formed by removing the %
edge $f$ from $W$. Then $\overline W_0 = W$ and $W_0$ has $\ell-1 \le t-1$ \workkspace{} edges. By Equation~\eqref{eq:C-expand-new}, $C(p,t,B) \le  w_{pq}+ M(qp,t-1,B) $, and by induction $M(qp,t-1,B) \le c(W_0)$.  Thus $C(p,t,B) \le   w_{pq}+ c(W_0) = c(W)$, where the last equality comes from the definition of the cost of the open walk $W_0$.

\paragraph{Case A.2. 
$p$ is a transition vertex.}
Let $F$ be the interior face of $M$ incident to edge $f=pq$ and let $e$ be the edge of $W$ that precedes $f$ around $F$.
Suppose edge $e$ enters vertex $r$ of $W$; then vertices $r$ and $p$ are coincident.
Cut $W$ into two polygons, where $W_1$ traverses $W$ from $p$ to $r$ and $W_2$ traverses $W$ from $r$ to $p$.  Observe that $W_1$ and $W_2$ are both weakly simple, and that no point is interior to both.  For $i=1,2$, let $\ell_1$ be the number of edges of $W_i$.  Then $\ell_i > 0$ and $\ell_1 + \ell_2 = \ell$.
Let $B_i = \{ P \in O \mid r_P \text{ lies inside } W_i \}$. 
Then $B = B_1 \sqcup B_2$. 

Suppose that neither $B_1$ nor $B_2$ is empty.  Since $t = \ell_1 + (t-\ell_1)$, Equation~\eqref{eq:C-compose-new}  yields $C(p,t,B) \le C(p,\ell_1,B_1) + C(p,t-\ell_1,B_2)$.  By induction, $C(p,\ell_1,B_1) \le c(W_1)$ and $C(p,t-\ell_1,B_2) \le c(W_2)$.
Thus $C(p,t,B) \le c(W_1) + c(W_2) = c(W)$, where the last equality is because $W_1$ and $W_2$ partition the edges and the interior of $W$.

On the other hand, if some $B_i$, say $B_2$, is empty, then $B_1 = B$. 
By induction, $C(p,t,B) \le c(W_1)$ since $W_1$ has $\ell_1 <%
t$ edges and contains $B$.
Thus $C(p,t,B) \le
c(W_1) \le c(W)$, where the last inequality is because $W_1$'s edges and interior are contained in those of~$W$. 

\medskip

\begin{figure}
    \centering
    \includegraphics[page=2]{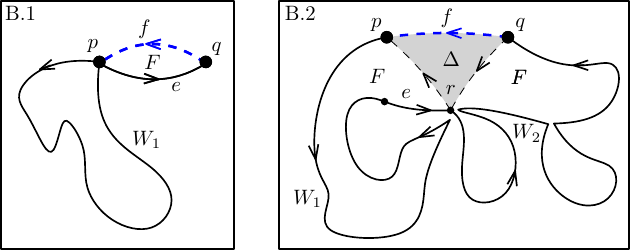}
    \caption{Cases for statement (B) of \Cref{lemma:buildup}.
    }
    \label{fig:structure-B}
\end{figure}

For part (B), we distinguish two cases depending on whether the interior face $F$ of $M$ incident to edge $f=qp$ of $W$ is a corridor or a chamber,
see \Cref{fig:structure-B}.  Note that $B = \emptyset$ is allowed.

\paragraph{Case B.1. $F$ is a corridor.}
Suppose $q$ has incoming edge $e$ and outgoing edge $f$.
By assumption,
$q$ is not a transition vertex.  Thus $e$ is incident to face $F$, and forms the other side of the corridor.  Since $e$ is a \workkspace{} edge, so is $f$.
By~Equation~\eqref{eq:M-pq-free-new}, $M(pq,t,B) \le C(p,t-1,B) + w_{pq} $.  
Let $W_1$ be the polygon formed by deleting 
edges $e$ and $f$ from $W$.  Then $W_1$ is a weakly simple polygon of $\ell-1$ \workkspace{} edges that encloses $B$. By induction, $C(q,t-1,B) \le c(W_1)$.
Thus $M(pq,t,B) \le  c(W_1) +w_{pq} = c(W_0)$, where the last equality is by definition of the cost of the open walk $W_0$.

\paragraph{Case B.2. $F$ is a chamber.}
Take a triangulation of $F$ (which exists because a chamber is an almost-simple polygon)
and 
consider the triangle %
incident to edge $f=pq$.  The triangle lies inside face $F$.  
Let $v$ be the vertex of $M$ that forms the third corner of the triangle.
Vertex $v$ may correspond to more than one polygon vertex, but we choose the ``right'' one as follows.
Let $e$ be the edge of
face $F$ incoming to $v$. 
In $W$, suppose edge $e$ enters vertex $r$.
Name the triangle
$\Delta = pqr$.

Break $W$ into two open walks $W_1$ from $q$ to $r$ and $W_2$ from $r$ to $p$.  Observe that $\overline{W}_1$ and $\overline{W}_2$ are weakly simple polygons and that $r$ is not a transition vertex of $\overline{W}_1$.  
For $i=1,2$, let $\ell_i$ be the number of edges of $W_i$.  Then $\ell_i >0$ and $\ell_1 + \ell_2 = \ell$.  
Let $B_1$
be the set of polygons $P \in R$ with $r_P$
in $\overline W_1$ but not on $qr$, and let $B_2$ be the set of polygons $P \in R$ with $r_P$ in $\overline W_2$ but not on $rp$.  
These sets may be empty.
Let $R(\Delta)$ be the set of  polygons $P \in R$ with $r_P$ inside $\Delta$ where we regard $\Delta$ as being closed on edges $pr$ and $qr$ and open on edge~$pq$.
Then $B = B_1 \sqcup B_2 \sqcup R(\Delta)$.

By Equation~\eqref{eq:M-triangle-new}, $M(pq,t,B) \le M(pr,\ell_1,B_1) + M(rq,t-\ell_1,B_2) + \pi(\Delta)$.
By induction, $M(pr,\ell_1,B_1) \le c(W_1)$ and $M(rq,t-\ell_1,B_2) \le c(W_2)$. 
Thus $M(pq,t,B) \le c(W_1) + c(W_2) + \pi(\Delta) = c(W_0)$ where the last equality is because we have partitioned the edges of $W_0$ and the interior of~$W$.
\end{proof}

\section{Uncrossing,
  Reordering, %
  and Final Output \texorpdfstring{\ensuremath{\WALG}}{W(ALG)}}
\label{sec:uncrossing}

The walk $\WDP$ produced by the dynamic programming algorithm may fail to be weakly simple.  However, 
as we will prove in \Cref{lem:WDP-no-cross5}, $\WDP$
cannot have interior crossings of edges.
As a consequence, it is possible to make $\WDP$ weakly simple just by reordering its edges,
 without increasing its cost,
 turning it into the desired %
weakly simple %
 polygon that encloses~$R$. %
 In this section we present such a \emphblack{Reordering Algorithm},
which forms the final step of our algorithm.

\subsection{Uncrossing Eulerian plane multigraphs}

Our algorithm uses the following known result about taking a plane multigraph (specified via its rotation system) and finding 
a \emphblack{non-crossing Euler tour} in which successive visits to a vertex do not cross each other. (See Section~\ref{sec:weakly-simple}
for more detailed definitions.)

\begin{proposition}[Uncrossing Eulerian plane multigraphs]
\label{prop:eulertour}
  Given a plane connected Eulerian multigraph~$H$ 
  with $m$ edges, specified by its combinatorial map,
we can, in $O(m)$ time, compute a
non-crossing 
  Euler tour 
  of~$H$
  that encloses its interior in counterclockwise direction. 
\end{proposition}
The existence of such a tour in an Eulerian plane multigraph is an easy exercise, see \cite{Singmaster+Grossman-84} or \cite[Lemma 3.1]{tsai-west-11}, and a linear-time algorithm was given by Akitaya and T\'oth~\cite[Corollary~1]{akitaya2018reconstruction}. 
We remark that the %
algorithm of Akitaya and T\'oth %
makes the unstated assumption that the combinatorial map is given.  Their terminology differs from ours, e.g., their input is geometric, and their output is a simple polygon that $\varepsilon$-approximates a non-crossing Euler tour.  For the more general setting of graphs on arbitrary surfaces, a linear-time algorithm  
was recently described by Bulavka, Colin de Verdière, and Fuladi
\cite[Lemma 4]{bcf-csccn-25}, expressed in the framework of cross-metric surfaces.  We outline the proof using our terminology.

\begin{proof}[Idea of the proof of Proposition~\ref{prop:eulertour}.]
Take a 2-coloring (white and grey) of the faces of $H$, with the outer face colored white.  Traverse every grey face counterclockwise to obtain a set of edge-disjoint cycles without vertex crossings.  The plan is to stitch together these cycles to form a non-crossing Euler tour.  Initialize $T$ to one of the cycles.  While there are other cycles, 
find a vertex $v$ where an edge of $T$ and an edge of another cycle $C$
appear consecutively in the cyclic order of edges around $v$, and merge $C$ into $T$ at this point.  This does not create vertex crossings in $T$.  The algorithm can be implemented to run in $O(m)$ time.     
\end{proof}

\subsection{The Reordering %
Algorithm}
\label{sec:uncrossing-geometric}

The Reordering Algorithm will be used not only to convert $\WDP$ to a weakly simple polygon, but also in our correctness proof, so we formulate it with minimal input restrictions.
The algorithm reorders the edges of a walk to make it weakly simple.
Since forks and interior crossings cannot be removed by reordering, we must forbid them as inputs.  
We apply the Reordering Algorithm to $\WDP$ in the
  following \Cref{sec:WALG-defn}. %

{\smallskip\centering\fboxsep=10pt \fbox{\begin{minipage}{0.948\linewidth} %
\noindent\textbf
{Reordering Algorithm.}
\begin{description}
\item[Input:] %
  A walk $W$ in the free space that has no forks or interior crossings.
\item[Algorithm:]
\ \\
The multiset of edges of $W$ forms a plane connected Eulerian graph
$H$.

 Apply \Cref{prop:eulertour}
 to $H$ to find a non-crossing Euler tour.

\item[Output:] 
  The 
  weakly simple polygon $W'$ that corresponds to the Euler tour.  
\end{description}
\end{minipage}}\par}%
\paragraph{Implementation and runtime:} To apply \Cref{prop:eulertour}
we need the combinatorial map of~$H$; so we must sort the incident
edges around each vertex.
If $W$ has $m$ edges, this takes time $O(m \log m)$. After that, \Cref{prop:eulertour} takes $O(m)$ time.  %

\begin{lemma}[Properties of the Reordering Algorithm]
\label{lem:reordering-properties}
\ %
\begin{enumerate}[\rm(a)]
\item\label{wscc} $W'$ is a weakly simple polygon that is oriented counterclockwise.  Thus the winding number of any point not on $W'$ is 0 or 1.
\item $W'$ has the same edges as $W$ and therefore
$w(W') = w(W)$.
\item Every point in the plane that does not lie on $W$ has the same winding parity in $W$ and~$W'$.
\item
\label{wind-kept}
  Suppose $\wind(W,r_P) = 1$ for each $P \in R$, and $\wind(W,r_P) \ge 0$ for each $P \in O$.  Then  $W'$ encloses $R$ and $c(W') \le c(W)$.
\item %
  \label{reduce2}
Suppose that $W$ has finite cost and 
has more than two edges that traverse the same segment 
(in 
any direction).
Then we can delete 
two
copies of the edge before applying \Cref{prop:eulertour}, and the
resulting weakly simple polygon $W''$ satisfies all the above
properties except that
we get strict inequalities
$w(W'') < w(W)$ and
$c(W'') < c(W)$. 
\end{enumerate}
\end{lemma}
\begin{proof}
\begin{enumerate}[\rm(a)]
\item and (b) hold by construction. \stepcounter{enumi}
\item Let $x$ be a point in the plane that does not lie on $W$.  For
  any ray $r$ from $x$ to infinity that avoids the vertices of $W$, the number of edges it crosses is the same for $W$ and $W'$. 
  The direction in which %
  an edge
is traversed
  does not affect the winding parity.
($+1$~and $-1$ have the same parity.)

\item 
From part (a), for any object~$P$,
$\wind(W',r_P)$ is 
  0 or 1, and
by part (c), 
the parity is 
the same as in~$W$. 
Thus our assumption that 
$\wwind W {r_P} \ge 0$ implies that $\wwind {W'} {r_P} \le \wwind W {r_P}$. 

Furthermore, if 
$\wwind W {r_P} = 1$, 
then $\wwind {W'} {r_P} = 1$.
Thus, for $P\in R$, where we have 
$\wwind W {r_P} = 1$
by assumption, we conclude that 
$\wwind {W'} {r_P} = 1$, and thus $W'$ encloses $R$.

It remains to consider costs. 
Recall that, by \Cref{defn:general-cost}, the cost $c(W)$ for an arbitrary polygon $W$ is
\begin{displaymath}
  c(W) = w(W) + \sum_{P \in O} \wwind W {r_P} \cdot \pi_P 
\end{displaymath}
 and similarly for $W'$.
 By part (b), $W$ and $W'$ have the same weight.
We have established above that
$\wwind {W'} {r_P} \le \wwind W {r_P}$ for all $P$,
and this directly gives
$c(W')\le c(W)$.

\item %
  Since at least one copy of the multiple edge remains after deletion,
  $H$ remains connected, and since we delete a pair, it remains Eulerian.
  So we can apply \Cref{prop:eulertour} to obtain $W''$.  Property (a) holds.  For Property~(b), the edges of $W''$ are a proper subset of the edges of~$W$, so $w(W'') < w(W)$.  For (c), consider a ray from a point $x$ not on $W$ to infinity.  The number of edges of $W''$ crossed by the ray may be 2 less than the number of edges of $W$ crossed by the ray, but the parity remains the same.  Finally, for~(d) the only change is the strict decrease in weight, which implies $c(W'') < c(W)$, provided that
$c(W)$ is finite.\qedhere
\end{enumerate}
\end{proof}

\subsection{Definition of the solution walk \texorpdfstring{$\WALG$}{W(ALG)}}
\label{sec:WALG-defn}

We will prove in \Cref{lem:cDP-finite-1}
  that $\cDP$ is finite, so the dynamic programming algorithm
constructs
the walk $\WDP$;
  we will prove in 
 \Cref{lem:WDP-no-cross5}
  that $\WDP$ 
has no interior crossings. %
There are
also %
no forks in $\WDP$ 
because it consists of free-space edges.
Thus, the Reordering Algorithm can be applied to it.

\begin{definition}\label{defn:WALG}
$\WALG$ is the output of the Reordering Algorithm on input $\WDP$.
\end{definition}

\begin{lemma}\label{lem:WALG-properties}
  Assuming %
  $\cDP$ is finite and $\WDP$
  has no interior crossings, $\WALG$ is a weakly simple polygon in the \workspace{}  
that encloses $R$, i.e., it is a feasible solution to
  \EWP. Its cost is bounded by $c(\WALG) \le  %
  \cDP
$.
\end{lemma}
\begin{proof}
$\WALG$ is a weakly simple polygon in the \workspace, by construction (\Cref{lem:reordering-properties}\ref{wscc}).
By Lemma~\ref{lemma:WDP-properties}\ref{DP-R}--\ref{DP-nonnegative},
$ \WDP$ has appropriate winding numbers and thus satisfies the assumptions of
\Cref{lem:reordering-properties}\ref{wind-kept}.
The conclusion of \Cref{lem:reordering-properties}\ref{wind-kept}
about the reordered path $W'=\WALG$
gives us
(i)~the fact that $\WALG$ encloses $R$, and
(ii)~the inequality
$c(\WALG) \le c(\WDP)$.
By the equation $ c(\WDP)=\cDP $
from Lemma~\ref{lemma:WDP-properties}\ref{DP-cost},
this implies the claimed %
inequality
 $c(\WALG) \le %
 \cDP$.    
\end{proof}

\section{Correctness Proof}
\label{sec:correctness}

In the previous section we completed the 
algorithm for the \textsc{Enclosure-with-Penalties} problem.  The algorithm has two major steps: the dynamic programming algorithm to find $\cDP$ and $\WDP$; and the Reordering Algorithm to convert $\WDP$ to the weakly simple polygon $\WALG$. 
The first step relied on the claim that $\cDP$ is finite, and the second step relied on the claim that $\WDP$ has no interior crossings.
In this section, we
prove these two claims and then we prove the main result that $\WALG$ is an optimum solution to the \EWP{} problem.

We defined the \textsc{Enclosure-with-Penalties} problem over the continuous space of all weakly simple polygons, but our algorithm only explores the discrete space of weakly simple polygons consisting of at most $6n$ \workkspace{} edges.
 In fact, we proved in \Cref{sec:best-walks} that the algorithm really does consider all polygons
in this discrete space.  
It remains
to prove that the \EWP\ problem has an optimum solution in
the discrete space.

Thus our task is to convert a hypothetical optimum solution to one consisting of at most $6n$ \workkspace{} edges.  
One main idea is that 
homotopically shortening a walk in the \workspace{} will make it hug object boundaries, and thus convert it to \workkspace{} edges; we achieve this in  \Cref{sec:hom-short}. 
In \Cref{sec:discrete-opt} we address the other conditions, namely the limit of %
$6n$ 
on the number of
edges
and the condition of weak simplicity.  We also deal with two delicate issues: 
objects may have a penalty of $+ \infty$, so we do not know {\it a priori} that there is a solution of finite cost; and
the solution space is continuous, so we do not know that an optimum
solution \emphblack{exists}, as opposed to having an infinite sequence of solutions of ever lower cost. 
Finally \Cref{sec:correctness-conclusion3} completes the correctness proof.

A note about terminology: The walks $W$ that we consider in this section all lie in the 
\workspace{}---they never cut through an object.  This means that for an object $P$, the winding number $\wind(W,r_P)$ is independent of the choice of reference point $r_P$ and we will simply call this the winding number of $W$ with respect to $P$.

\subsection{Homotopic shortening}
\label{sec:hom-short}

In this section we show 
that a polygon 
in the \workspace{} can be ``homotopically shortened'' to hug the
boundaries of the objects.
  Intuitively, homotopic shortening means that
  the walk is considered as a rope and ``pulled taut'' by
  a continuous deformation in the \workspace, 
without moving over any object; the resulting walk will then be composed of \workkspace{} edges.
(The most well-known version of this idea is that a shortest path in a polygon uses visibility edges.)

\begin{lemma}[Homotopic Shortening Lemma]\label{lemma:hom-short}
  Given a polygon~$W$ \textup(not necessarily simple\textup) in the
  \workspace\
  that
 has nonzero winding number for at least one object,
  there is a polygon~$W'$ that is made of free-space edges, has weight
  at most $w(W)$, and has the same winding number as~$W$ with respect
  to every object.  Consequently, $c(W')\le c(W)$.
\end{lemma}

\begin{figure}[htb]
  \centering
  \includegraphics{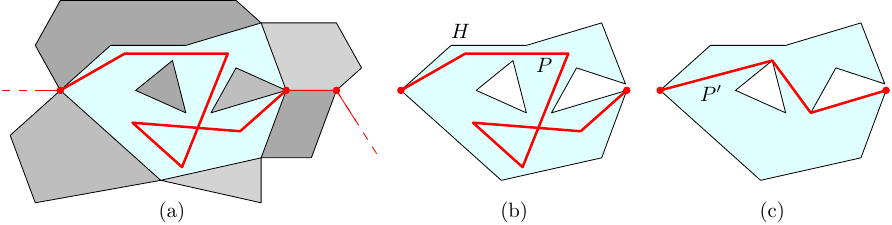}
  \caption{The polygonal region $H$ (shaded blue) inside which a piece $P$ of the walk
    is homotopically shortened. To convert $H$ into a 2-manifold with boundary, as
    required
    in~\cite{hs-cmlpg-94},
    repeated vertices on the boundary must be considered as distinct
    vertices,
    as illustrated symbolically in Part~(b). Part~(c) shows the
    shortened path $P'$.}
    \label{fig:cut-vertex}
\end{figure}

\begin{proof}
  Intuitively, %
 it is obvious that
 shortening $W$ homotopically %
  will result in a walk with
  the desired properties.  %
  For a formal proof, we will invoke
results due to Hershberger and Snoeyink~\cite{hs-cmlpg-94}.
 As stated, these results apply to a polygonal region that is a
2-manifold with boundary; in particular, the region must be bounded and every 
boundary point must have an open neighbourhood homeomorphic to a half-disk.
We make our \workspace{} bounded by adding a sufficiently large box around the input polygons.  
We then cut $W$ into pieces at every vertex of an input polygon and
treat each piece $P$ independently.
(It can happen that there is only one piece $P=W$.)

We identify the component $H$ of the free space in which $P$ lies as follows:
Let $P^0$ denote the piece $P$ after removing its endpoints
(if any),
 and let
$F^0$ 
be the region obtained from the \workspace{} by removing all vertices.
Let $H^0$ be the connected component of $F^0$ that contains~$P^0$, and
let $H$ be that component together with the vertices on its
boundary.

There are several cases:
\begin{itemize}
\item $H$ is a squeezed edge.
   Since some object has non-zero winding number, $P$ is not all of~$W\!$, so the
   endpoints of $P$
   are object vertices. We replace $P$ with the line segment between its endpoints. (It may be reduced to a single vertex if $P$ is a loop.)
 \item
 $H$ is a polygonal region with nonempty interior.
$H$ may fail to be a 2-manifold with boundary, but only 
    at repeated vertices. %
    To remedy this, we ``cut'' $H$ at these 
    repeated vertices, see \Cref{fig:cut-vertex}b.  More precisely, a
    vertex %
    of~$H$ that is locally incident to $p\ge2$ regions
    of~$H$ is replaced by %
    $p$ copies of it.
    The result is a 
    2-manifold with boundary.  
      \begin{itemize}
      \item If $P \ne W$, then the endpoints of~$P$ are vertices of
        $H$.  We replace $P$ with a shortest path $P'$ homotopic,
        within $H$, to~$W$.  %
        This shortest homotopic path $P'$ consists of straight
        line segments whose endpoints are vertices of the polygonal
        region~\cite{hs-cmlpg-94}.
    \item If $P=W$, then $P$ is a closed curve in $H$ that avoids all vertices of the \workspace{}.  We replace $P$ (and $W$) with the shortest closed curve $P'$ that is freely homotopic, within $H$, to~$P$ (where ``freely'' means homotopic to~$P$ without pinning any point).  $P$ cannot collapse to a point since some object has non-zero winding number.  Thus the new curve $P'$, as above, consists of straight line segments whose endpoints are vertices of the polygonal region~\cite[Section~5.1.1]{hs-cmlpg-94}, hence, of \workkspace{} edges.
  \end{itemize}
\end{itemize}

By applying the above procedure to all pieces $P$ of~$W$,
we obtain a walk of \workkspace{} edges~$W'$.  Because $W$ and~$W'$
are homotopic in the \workspace{}, 
for any object the winding numbers of $W$ and $W'$ with respect to the object are the same.
Furthermore
$w(W')\le w(W)$.  It follows that $c(W')\le c(W)$.
\end{proof}

\subsection{Restricting to a discrete space of solutions} %
\label{sec:discrete-opt}

To describe the goal of this section, we introduce notation for three classes of polygons in the free space. 
The class $\cal F$ is the set of feasible solutions to the \textsc{Enclosure-with-Penalties} problem.
The dynamic programming algorithm
in connection with the {Reordering} Algorithm
only explores %
polygons in the  smaller, discrete, 
subclass $\cal D$, where polygons consist of at most $6n$ free-space edges. 
In terms of this notation, our primary objective is to prove that there is a polygon in~$\mathcal{D}$ that is a minimum-cost solution for the class~$\mathcal{F}$.
Since weak simplicity %
is difficult to handle, the proof will take a detour
via a third, larger class $\cal S$,
where weak simplicity is relaxed.
The three classes thus form a hierarchy $\cal D\subseteq\cal F\subseteq\cal S$.
The main result of this section is the stronger statement that there is a polygon
in $\cal D$ that is %
a minimum-cost polygon
even for $\mathcal{S}$, and hence
for all three classes $\cal D$, $\cal F$, and~$\cal S$ simultaneously.
This strengthened version will be used, in \Cref{sec:correctness-conclusion3}, to prove the properties that we had to assume in order to construct $\WDP$ and~$\WALG$.

Although the notation is new, we have encountered these classes before:
the walk~$\WDP$ returned by the dynamic programming algorithm lies in~$\cal S$, by \Cref{lemma:WDP-properties}, and $\cDP$ is at most the cost of an optimum polygon in~$\cal D$, by \Cref{lem:cDP-calD}.
The classes are formally defined as follows: %

\begin{itemize}
\item $\mathcal{S}$ is the set of polygons in the \workspace{} that have finite cost, winding number~$1$ with respect to every object in~$R$, and non-negative winding number with respect to every object in~$O$.   (They need not consist of \workkspace\ edges.)
\item $\mathcal F$ is the set of polygons in
$\mathcal{S}$ that are weakly simple.
It follows that they enclose~$R$ and are oriented counterclockwise.
(Like the polygons in $\mathcal{S}$, they need not consist of \workkspace\ edges.)
\item $\mathcal{D}$ is the set of polygons in
$\mathcal{F}$ that consist of at most $6n$ \workkspace{} edges.
In other words,
$\mathcal{D}$ is the set of counterclockwise-oriented weakly simple polygons that have finite cost, enclose $R$, and consist of at most $6n$ \workkspace{} edges.
\end{itemize}

As mentioned above, we want to prove that there is a polygon in $\cal D$
that is a minimum-cost polygon for $\cal S$, hence also for $\cal D$ and $\cal F$.
We first show that $\mathcal{S}$ is non-empty and has an optimum polygon~$W$.  We want to convert $W$ to a polygon in $\mathcal{D}$ without increasing the cost. Homotopic Shortening converts $W$ to a polygon $W^*$ that consists of \workkspace{} edges.
We suspect that $W^*$ is in fact weakly simple, but this seems hard to prove with current techniques.  So instead, we will prove that $W^*$ has no interior crossings%
, which allows us to apply the Reordering Algorithm and convert $W^*$ to a weakly simple polygon.
Finally, we prove that the result has at most $6n$ edges.

We begin by proving a property of optimum walks in $\mathcal{S}$.

\begin{lemma}\label{lem:nocross5}
   An optimum walk $W\!$ in $\mathcal{S}$ cannot have interior crossings.
\end{lemma}

\begin{proof}
  Our proof procedure is illustrated in \autoref{fig:shortcut}.
  Suppose $W$ has interior crossings.
  First we
  subdivide edges at forks and interior crossings.  This leaves the weight,
  winding numbers, and hence the cost unchanged.
  
  Now we apply the Reordering Algorithm to get a
  weakly simple polygon $W'$.  By \Cref{lem:reordering-properties}\ref{wind-kept},
  $W'$ encloses $R$ and $c(W') \le c(W)$.
 Thus $W'\in \mathcal{S}$.

 \begin{figure}[htb]
   \centering
   \includegraphics[page=2]{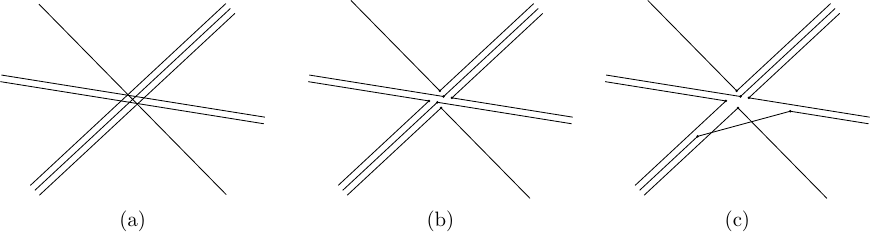}
   \caption{(a) An interior crossing point $x$. (b) 
   The result of subdividing the edges at $x$ and reordering to get a weakly simple polygon $W'$.
    (c) Adding a shortcut.
     (Note 
that coinciding %
parallel
edges
are separated
in the illustration only to make them distinctly visible.%
)
     }
   \label{fig:shortcut}
 \end{figure}

Let $x$ be an interior crossing point of $W$, 
i.e., $x$ is in the relative interiors of at least two non-collinear edges, so it lies in the interior of the \workspace{}. 
We have subdivided all edges that go straight through~$x$.
Thus  $W'$ must have
 a vertex at $x$ 
 whose incident
 edges form an angle different from $180^\circ$,
 because if
$W'$ always continued straight after reaching~$x$, it could not be weakly simple. 
 Now we %
 pick one of %
 these
 corners %
 and shortcut it
 by a %
 small
amount~$\varepsilon$, as shown in
\Cref{fig:shortcut}c:
The 
two %
edges incident to $x$ terminate at distance
$\varepsilon$ before they reach $x$, and we connect their endpoints
by
straight segments.
We choose $\varepsilon$ small enough 
so that {the disk %
of radius~$\varepsilon$   centered at~$x$  is contained in the \workspace{}.
The resulting shortened curve $W^*$ lies in the \workspace{} and has the same winding number as~$W'$ with respect to every object, so it lies in~$\mathcal{S}$ and}
incurs the same penalties
as $W$. The weight is, however, strictly reduced:
$w(W^*)<w(W')=w(W)$,
since segments in the interior of the \workspace{} have Euclidean lengths.
Our assumption that $W$ lies in $\mathcal{S}$ means that it has finite cost, so a strict reduction in weight implies a strict reduction in cost.
Therefore,
$c(W^*)<c(W)$, and $W$ cannot be optimum.
\end{proof}
Here is the announced main result of this section:

\begin{lemma}
\label{lem:simultaneous-opt-new}
  \label{lemma:discrete-existence-new}
     There is a polygon %
     in $\mathcal{D}$ that has minimum cost over all polygons in $\mathcal{S}$.

     Consequently, each class $\cal D$, $\cal F$, and $\cal S$  contains a minimum-cost solution, whose cost is finite by definition.
     Moreover,  the minimum costs in the three classes are equal.

\end{lemma}

\begin{proof}
The second statement follows directly from the first statement,
since $\cal D\subseteq\cal F\subseteq\cal S$.

It remains to prove %
that some polygon in $\cal D$ has minimum cost among all polygons in~$\cal S$.
First, we prove that $\mathcal{S}$ is non-empty by constructing a polygon that
has winding number~1 around each object in~$R$
  and winding number~0 around each object in~$O$,
  and thus %
  finite cost.
  Begin with the counterclockwise boundaries of the polygons in $R$. Then
  join them together as follows. 
  While there is more than one polygon,
find a path in the free space that connects
two vertices of different polygons (possibly a degenerate path
consisting of a single point only),
and combine the two polygons into a single polygon by traversing the path once in each direction. 
The end result is a single polygon that has the required winding numbers.

We now define $\Sfse$ to be the set of polygons in~$\mathcal{S}$
made of \workkspace{} edges.  We prove that, among the minimum-cost
polygons in~$\mathcal{S}$, there is one of minimum cost, $W$, 
that also
lies in~$\Sfse$.  Indeed, since $\mathcal{S}$ is non-empty, %
the Homotopic Shortening
Lemma~\ref{lemma:hom-short} implies that $\Sfse$ is non-empty
as well.
Moreover, for any value~$x$, there are finitely many polygons in~$\Sfse$ with cost at most~$x$, because all \workkspace{} edges have positive weight and all penalties are nonnegative.
So $\Sfse$ must have an element of minimum cost; let $W$ be such a polygon.  By Lemma~\ref{lemma:hom-short} again, $W$ is also a minimum-cost polygon in~$\mathcal{S}$, as desired.

We show that $W$ satisfies the conditions for applying the Reordering Algorithm:
First, $W$ has no forks because it is composed of free-space edges.
By \Cref{lem:nocross5},
since $W$ is a minimum cost solution in $\mathcal{S}$,
it has no interior crossings.
Let $W'$ be the result of applying the Reordering Algorithm to $W$.
By \Cref{lem:reordering-properties}\ref{wind-kept}, $W'$ is a weakly simple polygon that encloses $R$ and has $c(W') \le
c(W)$. Therefore $W'$ is an optimum solution in $\mathcal{S}$.

It remains to prove that $W'$ lies in~$\cal D$, and for this it suffices to show that $W'$ consists of at most $6n$ \workkspace{} edges. 
The edges of $W'$ are the same as the edges of $W$ and thus are \workkspace{} edges.
These edges form
a plane graph with at most~$n$ vertices,
and thus at most~$3n$ edges.  Therefore to prove that $W'$ has at most $6n$ edges it suffices to show that  each edge is used at most twice.

Suppose $W'$ has an edge that is used more than twice.  Then the same
is true of $W$.
Using \Cref{lem:reordering-properties}(\ref{reduce2}),
we would obtain a weakly simple polygon $W''$ that encloses $R$ and has cost
less than $c(W)$,  contradicting the optimality of $W$.
  \end{proof}

   As a side consequence of this lemma,
   we obtain an easy \emphblack{a-priori} bound on the cost of an optimum solution for the \EWP\ problem.
   Such a bound allows us to replace infinite penalties with an explicit large enough number, if desired. 
\begin{corollary}\label{lemma:upper-bound-cost}
    An optimum solution for the \EWP\ problem has cost at most $6n$ times the length $L$ of the longest \workkspace{} edge.
\end{corollary}
\begin{proof}
  Consider an instance of the \EWP\ problem, and temporarily set all penalties to $\infty$.  By \Cref{lemma:discrete-existence-new}, some optimal solution to this
  modified instance lies in~$\cal D$, and thus consists of at most $6n$ \workkspace{} edges and
  has finite cost.
  So it does not enclose any object in~$O$, and thus has cost at most $6nL$.  This is a feasible solution of cost at most~$6nL$ for the original instance.
\end{proof}
 As discussed above in \autoref{sec:runtime}, one can compute all \workkspace\ edges and hence the quantity $L$ in time $O(n^2)$~\cite{gm-osacv-91}.

\subsection{Conclusion of the correctness proof} %
\label{sec:correctness-conclusion3}

In this section we prove our main result that $\WALG$ is an optimum solution to the \EWP\ problem. 
First we justify the assumptions that were made when we defined $\WDP$ and $\WALG$.
The arguments become more transparent if we
use $\OPT(X)$, for $X\in\{\cal D,\cal F,\cal S\}$, to denote the minimum cost of the polygons in~$X$.
When we appeal to a lemma from a previous section, we restate it in terms of this %
notation,  and
to highlight that there is no circular reasoning,
we include %
all its hypotheses.

\begin{lemma}
\label{lem:cDP-finite-1}
$ \cDP \le %
 \OPT(\cal D)=\OPT(\cal F) =\OPT(\cal S)<\infty$.
In particular, $\cDP$ is finite.
\end{lemma}
\begin{proof}
  \Cref{lem:simultaneous-opt-new} says that $\OPT(\cal D)$, $\OPT(\cal F)$, and $\OPT(\cal S)$ are well-defined, finite, and equal.   
  \Cref{lem:cDP-calD} says that $\cDP\le\OPT(\cal D)$. 
Combining these gives the claimed result.
\end{proof}

Since $\cDP$ is finite, the dynamic programming algorithm constructs the corresponding polygon $\WDP$ (\Cref{defn:WDP}). 

\begin{lemma}\label{lem:WDP-no-cross5}
$\WDP$ has no interior crossings.  \end{lemma}
\begin{proof}
\Cref{lemma:WDP-properties} says that, if $\cDP$ is finite, then $\WDP$ lies in~$\cal S$ and has cost~$\cDP$.
By \Cref{lem:cDP-finite-1}, the 
finiteness %
assumption is fulfilled,
and thus $c(\WDP)= \cDP$.
By \Cref{lem:cDP-finite-1},
we also know that $\cDP \le \OPT(\mathcal{S})$, %
and therefore $\WDP$ is optimum in $\cal S$.
\Cref{lem:nocross5} says that   an
  optimum walk in~$\cal S$ cannot have interior crossings.
\end{proof}

As an aside, we note that
while interior crossings are impossible in $\WDP$, 
they can
actually occur in optimal 
solutions of subproblems, as shown in \Cref{fig:overlapping-solution}. 

Since $\WDP$ has no interior crossings, the Reordering Algorithm applies to it, and constructs $\WALG$ (\Cref{defn:WALG}).

\begin{theorem}
$\WALG$
is an optimum solution to the \textup{\textsc{Enclosure-with-Penalties}} problem.     
\end{theorem}
\begin{proof}
  \Cref{lem:WALG-properties} says that if $\cDP$ is finite and $\WDP$ has no interior crossings, then $\WALG$ lies in $\cal F$ and has cost at most~$\cDP$. 
  We have proved above that the assumptions are fulfilled,
  and thus $c(\WALG)\le \cDP$.
  We know that 
$\cDP \le \OPT(\mathcal{F})$ from \Cref{lem:cDP-finite-1}, and therefore $\WALG$ is optimum in $\mathcal{F}$.
In other words, 
 $\WALG$ is an optimum solution to the \EWP\ problem.    
\end{proof}

This theorem implies \Cref{thm:main}, with a weaker bound on the runtime
(\Cref{lem:runtime-space}).  The announced runtime is obtained in \Cref{sec:dijkstra} by speeding up the dynamic programming algorithm.

\section{Reducing the Runtime by a Dijkstra-Style Algorithm}
\label{sec:dijkstra}

In this section, we prove that the runtime of our algorithm to solve the \textsc{Enclosure-with-Penalties} problem can be reduced by a $\Theta(n^2)$ factor,
leading to %
the bound of \Cref{thm:main}.

The dynamic programming algorithm in Section~\ref{sec:DP} is guided
by a parameter $t$, which limits the number of edges of the
walk. %
We have proved (\Cref{lemma:discrete-existence-new}) that there is an optimal solution 
with
at most $6n$ edges.
Hence, the solution %
cannot be improved by 
allowing larger values of $t$;
the iteration stabilizes, and the algorithm can stop when $t$ reaches $6n$. 
The parameter $t$ has been useful for ensuring that
the quantities in the dynamic programming algorithm are well-defined,
and it is essential as an induction variable for the proofs.
We will now eliminate $t$ and solve the recursion in the style of Dijkstra's algorithm for shortest paths. 
In particular, we use a generalization of Dijkstra's algorithm as proposed by Knuth~\cite{Knuth1977}.

More specifically,
we
define $C(p,B):=C(p,6n,B)$ and $M(pq,B):=M(pq,6n,B)$ in terms of the quantities from \Cref{sec:DP}. By 
the above %
observations,
$C(p,B)=C(p,t,B)$ and $M(pq,B)=M(pq,t,B)$ %
for all $t\ge6n$.
Therefore, the limit %
quantities
$C(p,B)$ and $M(pq,B)$ fulfill a variation of the recursions
\thetag{\ref{eq:C-base-1-new}--\ref{eq:M-triangle-new}} where the parameter
          $t$ is eliminated:
\begin{align}
    C(p,\emptyset) &= 0
\label{eq:C-base-1-Dijkstra}
\\
 C(p,B) &=
\min \{C_1(p,B), C_2(p,B)\} %
\text{ for $B\ne\emptyset$, where}
\label{eq:C-general}
\\   &C_1(p,B) 
=
\min \bigl\{\,
    w_{pq} + M(qp, B)
\bigm|
   pq \text{\ is a \workkspace{} edge\,}
\bigr\} 
\label{eq:C-expand-Dijkstra}
\\
\label{eq:C-compose-Dijkstra}
 &C_2(p,B) %
= 
\min \bigl\{\,
    C(p, B') +
    C(p,B'')
\bigm|
 B =B'\sqcup B'';\
B', B'' \ne\emptyset
\,\bigr\}
\\
 M(pq,B) &=
\min \{M_1(pq,B), M_2(pq,B)\} %
\text{, where}
\label{eq:M-general}
\\ 
   & M_1(pq,B) = 
\begin{cases}
        w_{pq}  +  C(q,B), &   
    \text{if $pq$ is a \workkspace{} edge}\\
    \infty, &\text{otherwise}
\end{cases}
\label{eq:M-pq-free-Dijkstra}\\
 &M_2(pq,B)  = \min
 \label{eq:M-triangle-Dijkstra}  
  \bigl\{\, 
    M(pr,B') + M(rq,B'') + \pi(\Delta) 
\bigm|  %
\\ &
\qquad\nonumber
\text{$prq=\Delta$ is a counterclockwise\ triangle%
};
 \\
&\qquad\nonumber B
    =B'\sqcup B''
    \sqcup
     R( \Delta)
    \,\bigr\} %
\end{align}
As in Equation \eqref{eq:final}, we define the solution to the whole problem as
\begin{align}
\label{eq:final-Dijk}
\cDP  := \min\{\,
 C(p,R) \mid p\text{ is a vertex}
 \,\}   .
\end{align}
By \Cref{lemma:discrete-existence-new}
and by the definition of $C(p,B)$ and $M(pq,B)$, the value of $\cDP $ resulting from~\eqref{eq:final-Dijk} is the same as the one resulting from~\eqref{eq:final}. 

\subsection{Setting up a modified system of equations}
\label{app:dij-equations}

The system \thetag{\ref{eq:C-base-1-Dijkstra}--\ref{eq:M-triangle-Dijkstra}}     is a \emphblack{system of equations}
that involves cyclic dependencies. Nevertheless, we show in this section that it has a unique solution. 
Essentially, the reason is that on the right-hand side of the equations, the result of any expression combining some quantities of the form $C(p,B)$ and $M(pq,B)$  is always \emphblack{larger} than these quantities.

For the original
recursions \eqref{eq:C-base-1-new}--\eqref{eq:M-triangle-new}, the absence of a cyclic dependence between the quantities $C(p,t,B)$ and $M(pq,t,B)$ is guaranteed by the second parameter $t$, which is always smaller on the right-hand side
than on the left-hand side. For the system
\eqref{eq:C-base-1-Dijkstra}--\eqref{eq:M-triangle-Dijkstra},
we need to argue differently. In terms of the parameter representing the set of required objects, the parameter $B$, $B'$, or $B''$ on the right-hand side is always a subset of the parameter $B$ on the left-hand side.
 Whenever it is a strict subset, the corresponding equation cannot be part of a cyclic dependence.
 The recursion is more delicate when the same set $B$ appears on the right-hand side.
To separate these cases, we
split $M_2$ into three parts $M_3$, $M_4$ and $M_5$ consisting of
those compositions where $B'=B$, where $B''=B$, and where both $B'$ and $B''$ are strict subsets
of $B$%
. Thus, equation \eqref{eq:M-triangle-Dijkstra} becomes:
\begin{align}
 M_2(pq,B) &=
\min \{M_3(pq,B), M_4(pq,B), M_5(pq,B)\} \label{eq:M-triangle-Dijk-new}
\text{, where %
}\\
 M_3(pq,B) & = \min
 \label{eq:M-equal-Dijk-left}  
  \bigl\{\, 
  \begin{minipage}[t]{9.5cm}\openup 2pt
    $M(pr,B) + M(rq,\emptyset)+ \pi(\Delta) 
\bigm| $ %
\\\null \qquad 
 $\Delta = prq$\text{ is a counterclockwise %
 triangle}, $R(\Delta)=\emptyset\,\bigr\}$ 
  \end{minipage}
    \\
 M_4(pq,B) & = \min
 \label{eq:M-equal-Dijk-right}  
  \bigl\{\, 
  \begin{minipage}[t]{9.5cm}\openup 2pt
    $    M(pr,\emptyset) + M(rq,B) + \pi(\Delta) 
\bigm| $ %
\\\null \qquad 
 $\Delta = prq$\text{ is a counterclockwise %
 triangle}, $R(\Delta)=\emptyset\,\bigr\}$
  \end{minipage}
    \\
 \label{eq:M-subset-Dijkstra}  
     M_5(pq,B)  &= \min
  \bigl\{\, \begin{minipage}[t]{9cm}\openup 2pt
    $M(pr,B') + M(rq,B'') + \pi(\Delta) 
\bigm|$  %
\\
\null\qquad %
$\Delta=prq$ is a counterclockwise\ triangle%
;
 \\
\null\qquad%
$B =B'\sqcup B'' \sqcup
    R(\Delta) %
    $; 
    $B',B''\subsetneq B 
    \,\bigr\}$ %
  \end{minipage}
\end{align}
For $B=\emptyset$, the equations \eqref{eq:M-equal-Dijk-left} and  \eqref{eq:M-equal-Dijk-right}
coincide, but this redundancy is no problem.

Equations \eqref{eq:C-compose-Dijkstra} and \eqref{eq:M-subset-Dijkstra},
defining the quantities $C_2(p,B)$ and $M_5(pq,B)$, have on the
right-hand side quantities whose parameter, $B'$ or $B''$, is a strict
subset of $B$. Thus they cannot be involved in cyclic dependencies.
We distinguish them from the other four quantities
 $C_1(p,B)$, $M_1(pq,B)$, $M_3(pq,B)$, and
$M_4(pq,B)$, where the respective defining
equations \eqref{eq:C-expand-Dijkstra}, \eqref{eq:M-pq-free-Dijkstra}, \eqref{eq:M-equal-Dijk-left}, and \eqref{eq:M-equal-Dijk-right}
have on the right-hand side quantities whose parameter $B$ is the same as the one on the left-hand side.
By inspecting these equations, one can see that
the left-hand quantity that is computed is always strictly bigger
than the ingredients on the right-hand side, and
hence the
 recurrences behave like a \emphblack{superior context-free grammar} which uses {strictly superior functions};
see Knuth \cite[Section 5]{Knuth1977}. Knuth proved that, for such a grammar, the minimum value of a string (representing a composition of functions) derived from each terminal symbol exists, is unique, and can be computed efficiently. In our problem, this translates to the fact that all the values $C(p,B)$ and $M(pq,B)$, where $p$ and $q$ are vertices and $B$ is any subset of the input set of required polygons $R$, exist, are unique, and can be computed efficiently.

Knuth's setup does not directly apply to our problem as far as uniqueness is concerned, because
our functions are not \emphblack{strictly} superior functions.
This is compensated by having positive additive terms
in the recursion. Our uniqueness proof below
(\Cref{le:fast-correctness}) is a straightforward adaptation of Knuth's proof to our situation.

We show that the system of $O(n^2 2^k)$ equations \eqref{eq:C-base-1-Dijkstra}--\eqref{eq:M-pq-free-Dijkstra} and \eqref{eq:M-triangle-Dijk-new}--\eqref{eq:M-subset-Dijkstra} 
has a unique solution~$S=(C,M)$.
(We do not consider the auxiliary quantities $C_1,C_2,M_1,M_2,M_3,M_4,M_5$ as part of the solution~$S$, because they can be directly expressed in terms of $C$ and $M$). This, together with the fact that the solution of equations \eqref{eq:C-base-1-new}--\eqref{eq:M-triangle-new} is a solution to the system, implies that the solution~$S$ is the same as the one coming from equations \eqref{eq:C-base-1-new}--\eqref{eq:M-triangle-new}. 

 \begin{lemma} \label{le:fast-correctness}
     The system of equations 
 \eqref{eq:C-base-1-Dijkstra}--\eqref{eq:M-pq-free-Dijkstra}
 and 
 \eqref{eq:M-triangle-Dijk-new}--\eqref{eq:M-subset-Dijkstra} has a unique solution 
     $S=(C,M)$ with
     $C(p,B)\in \mathbb{R}_{\ge0}\cup\{\infty\}$
     and $M(pq,B)\in \mathbb{R}_{>0}\cup\{\infty\}$.
 \end{lemma}
\begin{proof}
The existence of a solution follows by  substituting the limiting solution of the equations \eqref{eq:C-base-1-new}--\eqref{eq:M-triangle-new} for large enough~$t$.  All quantities $M(pq,t,B)$ in those equations are positive because the quantity $M_1$ (see equation \eqref{eq:M-pq-free-new}) is at least equal to the weight $w_{pq}$ of the mouth, which is positive, and the other term $M_2$ (see equation \eqref{eq:M-triangle-new}) involves the addition of two quantities $M(pr,t_1,B_1)$ and $M(rq,t_2,B_2)$ whose second parameter, $t_1$ or $t_2$, is smaller than~$t$.

We now prove uniqueness. The crucial fact that allows us to exclude a cyclic dependency
    is that in the equations \eqref{eq:C-base-1-Dijkstra}--\eqref{eq:M-pq-free-Dijkstra}
 and 
 \eqref{eq:M-triangle-Dijk-new}--\eqref{eq:M-subset-Dijkstra},  the quantities $M$ and $C$ on the right-hand side that could cause such a cyclic dependency (because they use the same set parameter $B$) must be strictly smaller than the quantities on the left side that are defined through them.
    
 Assume, for contradiction, that there are two  different solutions $S=(C,M)$ and
     $S'=(C',M')$. 
     Among the quantities where the two solutions differ,
     select the ones for which the  parameter $B$ is minimal, and among those,
     consider a pair with the smallest value
     $T = \min\{C(p,B),C'(p,B)\}$
     or $T = \min\{M(pq,B),M'(pq,B)\}$.

 Let us first deal with the case that the smallest difference occurs for~$C(p,B)\ne C'(p,B)$.
Assume without loss of generality that~$T=C(p,B)< C'(p,B)$. By the minimality of~$B$, we have that~$C(p,B')=C'(p,B')$ and~$C(p,B'')=C'(p,B'')$, for any strict subsets~$B'$ and~$B''$ of~$B$. Hence, equation~\eqref{eq:C-compose-Dijkstra} gives us that~$C_2(p,B)=C'_2(p,B)$ and thus~$T=C_1(p,B)< C'_1(p,B)$. By equation~\eqref{eq:C-expand-Dijkstra}, $C_1(p,B)$ is equal to~$w_{pq}+M(qp,B)$ for some \workkspace{} edge~$pq$. Since the weight~$w_{pq}$ is positive,~$M(qp,B)$ is strictly smaller than~$T$, and hence, by the minimality of~$T$, we get $M(qp,B)=M'(qp,B)$. It follows that 
\begin{displaymath}
C_1(p,B)= w_{pq}+M(qp,B) = w_{pq}+M'(qp,B)\ge C'_1(p,B),
\end{displaymath}
 a contradiction.

     The same argument works for the case in which $T = \min\{M(pq,B),M'(pq,B)\}$.
     Here it is necessary to use the fact that all values $M(pq,B)$ are positive.
(Without this assumption, the identically zero solution $M(pq,B)\equiv0$
might
be an alternative solution, for example.) 
\end{proof}

\subsection{The algorithm}
Similar to Dijkstra's shortest-path algorithm, our algorithm maintains tentative values 
$C(p,B)$ and $M(pq,B)$. The smallest of the tentative values
is made permanent, and all right-hand side expressions where this value appears
are evaluated and used to update the corresponding tentative left-hand side values. We now describe the algorithm that carries out this idea: it computes the values $C(p,B)$ and $M(pq,B)$ for all vertices $p$ and $q$ and all subsets $B\subseteq R$ of required polygons. The algorithm has an outer loop that goes through all subsets $B\subseteq R$ %
in order of increasing size $|B|$, or in any other order that is compatible with set inclusion.

When the algorithm needs to compute the values $M(pq,B)$ and $C(p,B)$
for a certain $B$, the values $M(rs,B')$ and $C(r,B')$ have already
been computed for all strict subsets $B'\subset B$, all vertex pairs
$rs$ and all vertices $r$. This allows us to directly compute the values $C_2(p,B)$ for all vertices $p$, via equation \eqref{eq:C-compose-Dijkstra}, and $M_5(pq,B)$ for all vertex pairs $pq$, via equation \eqref{eq:M-subset-Dijkstra}.

The algorithm maintains a set $F_1$ of vertices $p$ for which the value  $C(p,B)$ has been determined, and a set $F_2$ of vertex pairs $pq$ for which the value $M(pq,B)$ has been determined. Initially, $F_1$ and $F_2$ are empty. The algorithm also maintains \emphblack{tentative} values $M(pq,B)$ and $C(p,B)$, which are upper bounds on their final values.
When they become final, the corresponding item $pq$ or $p$ is added to $F_2$ or $F_1$.
Actually, what the algorithm maintains are tentative values for $M_1(pq,B),M_3(pq,B),M_4(pq,B)$ and $C_1(p,B)$, which are initialized to $\infty$. The values of $C(pq,B)$ and $M(pq,B)$ are kept up-to-date via $C(p,B)=\min\{C_1(p,B),C_2(p,B)\}$ and $M(pq,B)=\min\{M_1(pq,B),M_3(pq,B),M_4(pq,B),M_5(pq,B)\}$.

The core of the algorithm consists of making a tentative value final
and adding the corresponding vertex or vertex pair to $F_1$ or $F_2$. The strategy to do so is akin to the strategy of Dijkstra's algorithm for computing shortest paths: We pick the smallest tentative value and make it final. Then we look at all equations where this value appears on
the right-hand side, and update the left-hand side. The pseudo-code in Algorithm~\ref{alg:eff} implements this in a straightforward way. (The only
challenge %
is %
the confusion caused by the necessary renaming of the vertices $p,q,r,s$.)

\newcommand{\codecomment}[1]{\,\textcolor{blue}{\texttt{/\kern-0,7pt/} {\textit{#1}\;\xspace}}}
\setlength{\algomargin}{1em} 
\begin{algorithm*}[ht!]
\caption{{Computation of the values $M(pq,B)$ and $C(p,B)$ for fixed $B$.}}
\label{alg:eff}
\DontPrintSemicolon
\SetKwInOut{Input}{Input}
\SetKwInOut{Output}{Output}
\Input{Set $R$ of required polygons, set $O$ of optional polygons, set $B\subseteq R$}
\Output{Values $M(pq,B)$ for every pair of vertices $pq$ and $C(p,B)$ for every vertex $p$}
\BlankLine
\codecomment{For every strict subset $B'\subset B$, the values $M(rs,B')$ for every pair of vertices $rs$ and $C(r,B')$ for every vertex $r$ have already been computed.}
\For{\rm \textbf{each} vertex $p$}{
Set $C_1(p,B):=\infty$; compute $C_2(p,B)$ by equation \eqref{eq:C-compose-Dijkstra};  set $C(p,B):=C_2(p,B)$ 
}
\For{\rm \textbf{each} vertex pair $pq$}{
Set $M_1(pq,B):=%
M_3(pq,B):=%
M_4(pq,B):=\infty$,  and compute $M_5(pq,B)$ by equation \eqref{eq:M-subset-Dijkstra}; 
set  $M(pq,B) := M_5(pq,B)$ according to \eqref{eq:M-general}
and \eqref{eq:M-triangle-Dijk-new}
}
Set $F_1 :=%
F_2 :=\emptyset$ \codecomment{$F_1$ contains the vertices $p$ for which $C(p,B)$ has been %
computed, and $F_2$ the vertex pairs $pq$ for which $M(pq,B)$ has been %
computed.}
\While{\rm there are vertices not in $F_1$ or vertex pairs not in $F_2$}{
\strut Find the smallest value $D$ among the tentative values $C(p,B)$ with $p\notin F_1$ and the tentative values $M(pq,B)$ with $pq\notin F_2$;
ties are broken arbitrarily. \;
\If{\rm $D$ %
is $C(p,B)$}{ 
Set $F_1 := F_1 \cup \{p\}$ \codecomment{make $C(p,B)$ permanent}
\For{\rm \textbf{each} \workkspace{} edge $sp$ incident to $p$}{
Set $M_1(sp,B) := \min\{M_1(sp,B), w_{sp}  +  C(p,B)\}$ \codecomment{by equation \eqref{eq:M-pq-free-Dijkstra}}
Update $M(sp,B) = \min \{M_1(sp,B),M_3(sp,B),M_4(sp,B),M_5(sp,B)\}$
}
}
\If{\rm $D$ %
is $M(pq,B)$}{ 
Set $F_2 := F_2 \cup \{pq\}$ \codecomment{make $M(pq,B)$ permanent}
\If{\rm $pq$ is a \workkspace{} edge 
}{
Set $C_1(q,B) :=
\min \{C_1(q,B),\,
  w_{qp} + M(pq, B)
\}
$  \codecomment{by equation~\eqref{eq:C-expand-Dijkstra}}
Update $C(q,B) := \min\{C_1(q,B),C_2(q,B)\}$
}
\For{\rm \textbf{each} counterclockwise triangle $\Delta=psq$
with $R(\Delta)=\emptyset$
} {
Set $M_3(ps,B) :=\min\{M_3(ps,B), M(pq,B) +
M(qs,\emptyset)+\pi(\Delta)\}$ \codecomment{by%
  ~\eqref{eq:M-equal-Dijk-left}}
Update $M(ps,B) := \min \{M_1(ps,B),M_3(ps,B),M_4(ps,B),M_5(ps,B)\}$;\\
Set $M_4(sq,B) :=\min\{M_4(sq,B), M(sp,\emptyset) + M(pq,B) + \pi(\Delta) 
\}$ \codecomment{by%
~\eqref{eq:M-equal-Dijk-right}}
Update $M(sq,B) := \min \{M_1(sq,B),M_3(sq,B),M_4(sq,B),M_5(sq,B)\}$}
}}
\end{algorithm*}

 Correctness is established in the same way as for
Dijkstra's algorithm. The smallest tentative value $D$ that is
determined at the beginning of each iteration of the main loop
is simultaneously a lower bound on all tentative values
and an upper bound on all permanent values.
The algorithm ensures that every tentative value always fulfills
its corresponding equation \eqref{eq:C-general} or
\eqref{eq:M-general}. Thus, whenever a value
is finalized, the equation is fulfilled.
The values on the right-hand side on which it depends do not
change any more because they are smaller than~$D$,
and hence they have already been finalized.

Thus, when the algorithm terminates,
the computed values $C(p,B)$ and $M(pq,B)$
fulfill \eqref{eq:C-general}
and
\eqref{eq:M-general}.
The correct values
$C(p,6n,B)$ and $M(pq,6n,B)$ also fulfill 
these equations, and
by \Cref{le:fast-correctness}
the solution of \eqref{eq:C-general}
and
\eqref{eq:M-general} is unique,
and hence the computed values agree
with the correct values.

\subsection{Runtime analysis}
\label{sec:runtime-Dijkstra}

The most numerous quantities are the $O(n^2 2^k)$ values $M(pq,B)$, and hence the space complexity is  $O(n^2 2^k)$.
Compared to the runtime for the dynamic programming algorithm
in \Cref{sec:DP}, we save a factor $n^2$:
The elimination of $t$ reduces the number of recursions
by a factor $\Theta(n)$, and we save another factor
$\Theta(n)$ because we need not go through all decompositions $t=t_1+t_2$ on the right-hand side.  We now give the analysis of the full algorithm.
The
total runtime is $O(n^33^k)$,
as claimed in \Cref{thm:main}.
Clearly, there are $O(n^2 2^k)$ values $M(pq,B)$ and $C(p,B)$, and this defines the space complexity.

We first analyze the computations of the quantities $C_2(p,B)$ and $M_5(pq,B)$, which are computed directly by equations \eqref{eq:C-compose-Dijkstra} and \eqref{eq:M-subset-Dijkstra},
respectively.
The dominating term for the runtime comes from the computation of the $O(n^2 2^k)$ quantities~$M_5(pq,B)$. For each of them, we have to run through all points $r$ and check each counterclockwise triangle $\Delta=prq$:
We have to find the set 
\begin{equation}
    \label{eq:R-region}
R(\Delta) := \{P \in R \mid r_P \in \Delta \},
\end{equation}
and compute the sum $\pi(\Delta)$ of the penalties of the polygons in $O$ whose reference point is in $\Delta$ 
and, in case $R(\Delta)\subseteq B$, run through all partitions of
$B-R(\Delta)$ into two sets~$B'$ and~$B''$.
We describe below a preprocessing step that allows us to obtain the quantity~$\pi(\Delta)$ in constant time.
The set~$R(\Delta)$ can be trivially computed in $O(k)$ time. Thus the total runtime for computing the quantities~$M_5(pq,B)$ is
\begin{equation}
    \label{eq:dominating-part}
    n^2 \sum _{B\subseteq R} 
    \Big( n \times \bigl(O(k) + 2^{|B|} \bigr) \Big) = O(n^32^kk)+O(n^3) \sum _{B\subseteq R} 2^{|B|} = O(n^33^k).
\end{equation}

Let us now look at the runtime of the core of the algorithm, in which, repeatedly, a tentative value is made final. Consider a fixed subset $B\subseteq R$ (there are $2^k$ such sets). We need to maintain a priority queue for the $O(n^2)$ tentative values for the quantities $C(p,B)$ and $M(pq,B)$.
Each of the $O(n^3)$ expressions 
on the right-hand side of any of the equations
 \eqref{eq:C-expand-Dijkstra}, \eqref{eq:C-compose-Dijkstra}, \eqref{eq:M-pq-free-Dijkstra}, and \eqref{eq:M-equal-Dijk-left}--\eqref{eq:M-subset-Dijkstra} is evaluated exactly once (when the corresponding quantity becomes final) or twice (in case $B=\emptyset$, for the expression $M(pq,\emptyset)+M(pq,\emptyset)$).
The evaluation potentially triggers an update to the priority queue, which takes
 $O(1)$ amortized time with Fibonacci heaps \cite{DBLP:books/daglib/0023376,DBLP:journals/jacm/FredmanT87}.
We need to extract the minimum $O(n^2)$ times, at an amortized cost of $O(\log n)$ per operation. 
The overall runtime for the heap operations
is then 
 $O(n^2\log n+n^3)=O(n^3)$.
In summary, the overall runtime for this part of the algorithm is
$2^k \cdot O(n^3)$, which is dominated by \eqref{eq:dominating-part}.

Thus, up to showing how $\pi(\Delta)$ can be determined
in constant time, which we do in the next subsection,
we have established \Cref{thm:main}.

\subsection{Preprocessing for quickly determining the penalty of a triangle}
\label{sec:preprocessing}
One can set up a table with $O(n^2)$ entries
from which, for any triangle $\Delta$ whose vertices are vertices of the input polygons, the sum of the penalties of the polygons whose reference point is in $\Delta$ can be obtained in constant time. 
This is a standard technique in this area, see for example \cite[Section~2]{eorw-fmakg-DCG92}.
We give some %
details.

\begin{figure}[htb]
    \centering
    \includegraphics{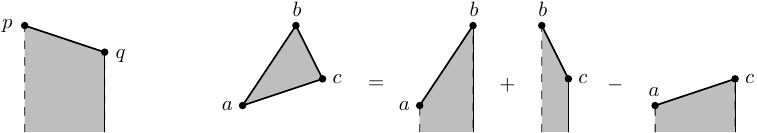}
    \caption{Left: a plank under the segment $pq$. Right: A triangle area $abc$ is obtained by addition
    and subtraction of planks.}
    \label{fig:plank-plus}
\end{figure}

A \defn{plank} is a
region bounded by a line segment $pq$ and two vertical upward rays or
two vertical downward rays
originating at $p$ and $q$. 
We consider the right boundary ray and the open line segment $pq$ to be
part of the plank, but not the left boundary ray including the point~$p$.
\Cref{fig:plank-plus} shows some
     downward planks (``bottomless trapezoids'', so-to-speak).
 Upward planks (``topless trapezoids'') are used in
 \Cref{sec:inverted}.
 
We store for each vertex pair $pq$, the sum of the penalties in
the downward {plank} below the segment $pq$. From this, the same data $\pi(\Delta)$ can then be computed for any triangle
$\Delta=abc$
in constant time by addition and subtraction from three planks,
see \Cref{fig:plank-plus} for an example.
The table can be computed in  $O(n^2)$ time and
$O(n^2)$ space
\cite[Theorem~2.1]{eorw-fmakg-DCG92}.
Even a straightforward $O(n^3)$ preprocessing would
be acceptable for us, as the runtime is dominated by the cost of other parts of the algorithm.

Reference points {of objects} with infinite penalties are handled separately, in the same way: Instead of storing the sum of their penalties, they are merely \emphblack{counted}, to determine whether the triangle in question contains
at least one of them or none.
(An alternative approach is to
replace infinite penalties by sufficiently large finite penalties,
see \Cref{lemma:upper-bound-cost}.)
Finally, recall that $\Delta = prq$ was defined to be open on segment $pq$.  To handle this, we also calculate the sum
the penalties of the reference points \emphblack{on} each segment $pq$ (as well as the number of infinite-penalty points). These have then to be added or subtracted
as appropriate.

\section{The Computational Model}
\label{sec:model}

In this section we discuss our models of computation for the geometric and graph versions of the \EWP{} problem.

\subsection{Geometric-Enclosure-with-Penalties}

{Our algorithm needs to perform orientation tests on input vertices (and reference points).  If vertices
are given by rational coordinates, orientation tests can be carried out with exact arithmetic.   This guarantees the winding number properties of the algorithm.  However, for the geometric version of our problem, we must also compare sums of Euclidean distances to find a solution of minimum length.}
So we assume the real RAM model of computation, in which such square root operations take constant time.
This is the standard model 
in computational geometry
for problems that involve %
{Euclidean} lengths.

One may wonder whether the algorithm would work with a straightforward implementation
that simply
represents edge lengths by floating-point numbers
with machine precision.
There are well-known examples
of geometric algorithms
where a naive floating-point implementation
can lead
to failure or to nonsensical results~\cite{Classroom-examples-08}.
However, in our case, 
the length of the enclosure $W$ is the sum of
at most $6n$ distances, and the straightforward addition produces a result that should be accurate enough for all
practical purposes. %
Indeed, all calculations that the algorithm performs
are additions of sums of at most $6n$ edge lengths, plus penalties.
One might hope that the worst thing that can happen is that some comparisons go %
the wrong way,
resulting in a solution that
is, say, $0.0001\%$ off the optimum.

However, our
argument that $\WDP$ has no interior crossings (\Cref{lem:nocross5}%
)
relies on the triangle inequality in the plane and assumes exact
numbers. It is therefore conceivable that, with rounding, the dynamic programming algorithm produces
a walk $\WDP$ with interior crossings, and the algorithm cannot continue.
There is a way out of this:
To get a valid solution, one can %
eliminate crossings by the method that is implicit
in the proof of \Cref{lem:nocross5}:
Compute the arrangement of the edges, subdivide them at every crossing point, and only then, apply the Reordering Algorithm
to get a weakly simple walk.
(Such a full-fledged \emphblack{Uncrossing Algorithm} is described and analyzed in the conference version  \cite {bcflr-fscts-25} of this article.
We show %
that it
can be implemented to run in time 
$O(t \log t + s)$ for a walk $W$ with $t$ edges, where
$s \in O(t^2)$ 
is the number of interior crossing points of $W$
\cite[Lemma~20 in Appendix~D of the arXiv version] {bcflr-fscts-25}.)
The arrangement computation involves only constant-degree rational constructions and predicates.
Therefore, if the vertices of the input polygon are given, say, as rational numbers,
each computation of the algorithm can be performed exactly in a constant number of arithmetic operations.
In this way, one is guaranteed to find a feasible enclosure that is as close to optimal as
the computation of the cost by floating-point arithmetic allows.

The result of the Uncrossing Algorithm will make bends in the middle of the free space where interior crossings used to be; if these are undesirable, they can be eliminated by Homotopic Shortening (cf.~\autoref{sec:hom-short}),
again requiring constant-degree predicates only and no length comparisons~\cite{hs-cmlpg-94,b-chspp-03,ekl-chspe-06}.
We believe that homotopic shortening of a weakly simple walk introduces no interior crossings,
but if it does, the uncrossing procedure may be repeated.

\subsection{Graph-Enclosure-with-Penalties}
For the graph version of the problem,
we do not need
 the real RAM.
All graph edges are squeezed edges by construction,
and they have some arbitrary weights, given as inputs.
If these weights are integers, the usual
RAM model of computation %
is sufficient.

Note that the algorithm starts by computing a
straight-line embedding in the plane, and
therefore, 
geometry is involved.
In particular, the algorithm needs
orientation tests with respect to the extra
mouth segments, which are not part of the input graph.
However, the length of a mouth edge is never needed.
A straight-line embedding can be constructed on
a grid of size $O(n)\times O(n)$ in $O(n)$ time~\cite{DBLP:conf/soda/Schnyder90,DBLP:journals/combinatorica/FraysseixPP90},
so the size of coordinates is not an issue.

\subsection{The exponential lower bound}
The lower bounds in \autoref{sec:lower-bound} are conditional on 
the Exponential Time Hypothesis (ETH).
  The
  Exponential Time Hypothesis (ETH) is %
  robust against
 polynomial blow-up of the runtime; thus
 the Turing
 machine model or the RAM model are both appropriate in this
 setting.
 We work in the usual RAM model in our lower-bound arguments.
 For the geometric problem,
 \GeoEWP, our reduction will produce a large enough gap
 between YES-instances and NO-instances.
An approximate calculation
of distances with a tolerance of $\pm\frac12$ will be sufficient
to distinguish between them,
and
exact computation is not required.

\section{Weakly Simple Immersed Polygons}
\label{sec:immersed}
\label{SIMP}
The proof of the correctness of our algorithm
relies on some properties of the output of the dynamic programming algorithm.   
In this section, we shed light on the dynamic programming algorithm by defining the precise class of polygons over which
the 
dynamic programming algorithm optimizes. 
(We do this in the notation of the algorithm of
  Section~\ref{sec:DP}; the faster algorithm of Section~\ref{sec:dijkstra}
  computes the same quantities.) The output polygons are more general than
the \ws\ polygons that we want as a solution, because
they can self-cross, but
they are not arbitrary polygons.
While understanding this class of polygons is not
  required for understanding or verifying the correctness of the
  dynamic programming algorithm,
we include the discussion of these polygons to round off the treatment of the problem.

It turns out that
$C(p,t,B)$ 
is
the minimum cost (with
 the extended meaning of \Cref{defn:general-cost})
of a
\emphblack{weakly simple immersed polygon}
 that satisfies the constraints (appropriately modified) regarding
 the number of edges, the set
$B$ of objects whose reference points are enclosed, and the
starting point~$p$.
The same statement holds for $M(pq,t,B)$ with the mouth~$pq$.

 As in Sections \ref{sec:weakly-simple} and~\ref{sec:extracting},
we describe 
such a polygon
as a sequence of vertices forming its boundary cycle, in the form
 $P=(p_1,p_2,\ldots, p_n)$. 
The polygon %
runs counterclockwise around its ``enclosed region'', with the interior to its left.

\paragraph
{Weakly simple immersed polygons (\wsimp).}

A \defn{weakly simple immersed polygon} (\wsimp) is obtained
by gluing together triangles and digons in a tree-like fashion.

There are two base cases:
\begin{itemize}
    \item a counterclockwise nondegenerate triangle 
    $(p,q,r)$
    \item a digon 
    $(p,q)$
\end{itemize}
The two ways of inductively combining two \wsimp s into a larger \wsimp\ 
are the same combinations that we introduced in \Cref{sec:properties_winding} for arbitrary polygons:
\begin{itemize}
    \item Two \wsimp s can be glued together along a common {edge}:
If $P_1 = (p,q,q_2,\ldots, q_n)$
and
 $%
 P_2 = 
 (q,p,p_2,\ldots, p_m)$
 both use the edge $pq$, but in opposite directions, then
 we can form the \wsimp \
  $P = (q,q_2,\ldots, q_n,p,p_2,\ldots, p_m)$.
(This is the same as gluing together two polygons in the plane along a common edge if they lie on different sides
of that edge, except that we do not care whether they overlap.)
\item 
Two \wsimp s $P_1 = (p,q_1,q_2,\ldots, q_n)$ and $P_2 = (p,q_1,q_2,\ldots, q_n)$ can be glued together at a shared {vertex} $p$, forming a new \wsimp  \
$
P = (p,q_1,q_2,\ldots, q_n,p,p_1,p_2,\ldots, p_m)$.
(The vertex $p$ becomes a transition vertex.)
\end{itemize}

By construction, our dynamic programming algorithm computes a \wsimp\ $W$ that has
the correct winding number for all objects in $R$.
Since \wsimp s can be triangulated, the same proof as that of \Cref{lemma:buildup} shows that the cost of~$W$ is optimal. More precisely,
$M(pq,t,B)$ and $C(p,t,B)$ are the minimum cost 
 of a
 {weakly simple immersed polygon} $W$
under
  the following  constraints:
\begin{enumerate}
    \item 
    For $M(pq,t,B)$,
the walk connects the endpoints $p$ and $q$ and is closed by the mouth $qp$;
for  $C(p,t,B)$, it goes through the  startpoint $p$.
\item The number of \workkspace{} edges is at most $t$, not counting the mouth
in case of $M(pq,t,B)$.
\item For each object $P\in B$, $\wwind {%
W} {r_P} =1$,
and for each object $P\in R\setminus B$, $\wwind {%
W} {r_P} =0$.  
\end{enumerate} 
 The cost is interpreted with
 the extended meaning of \Cref{defn:general-cost},
 and the weight $w_{pq}$ is subtracted in case of $M(pq,t,B)$.

If we restrict the base case to triangles and only allow gluings along edges, we arrive at the subclass
of  \defn{(simple) immersed polygons} (\simp{}s).
Here, the construction defines a simply connected surface, which
is obtained by starting with the triangles and performing the gluing as an identification of common points.
The boundary walk of a \simp\ is known as
a \defn{self-overlapping polygon}, %
see for example Evans and Wenk~\cite{evans2023combinatorial}
for a recent discussion.
The boundary walks of
\wsimp s are
related to self-overlapping polygons 
in the same way as {weakly simple} polygons are related to {simple} polygons.

\begin{figure}[htb]
    \centering
    \includegraphics%
    {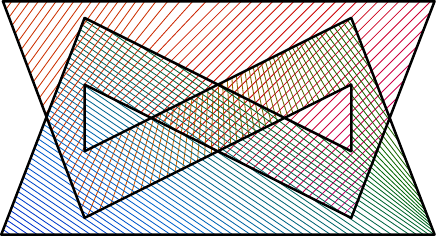}
    \caption{Milnor's doodle. The hatching indicates one of two symmetric ways of viewing this self-%
    overlapping polygon
    as the boundary of an immersed surface. %
    }
    \label{fig:Milnor}
\end{figure}

The relation between a \wsimp\ and its boundary walk is delicate,
just as for self-overlapping polygons:
It is not the case that the \simp  \ as a boundary
determines this surface uniquely.
\Cref{fig:Milnor} shows %
the simplest counterexample, which is
known under the name {Milnor's doodle}.
By construction, a \simp  \ comes with a triangulation, but
the triangulation is obviously not unique.
Shor and Van Wyk~\cite{SHOR-vanWIK-1992} define
a \simp\ as an equivalence class of triangulations
of a self-overlapping polygon, and they
give an algorithm to count the number of \simp s
for a given self-overlapping polygon
\cite[Section~6]{SHOR-vanWIK-1992}.

Thus, for specifying a \wsimp, the boundary walk is not sufficient: We
would have to specify the sequence of gluings describing how the
\wsimp\ was built.
For our purposes, however, the precise \simp\ or \wsimp\ is
irrelevant,
and the boundary walk is all we need:
By \Cref{lem:additive},
we can find out
how often a point in the plane is covered by $P$ by calculating the
winding number.

There is no direct relation between \wsimp s and the set $\mathcal{S}$ of
  solutions introduced in \autoref{sec:discrete-opt}.
  The definition of $\mathcal{S}$ 
  prescribes the winding number of objects in~$R$.
  Even if these conditions were relaxed to 
  allow arbitrary nonnegative winding numbers, %
  not every walk in $\mathcal{S}$ would be a \wsimp, because a \wsimp\
  must have nonnegative winding number
 at every point of the plane %
 (but this is not sufficient, not even for self-overlapping %
 polygons).

\section{Point Objects}
\label{sec:points}

In this section we modify our algorithm to handle input objects that are a mix of points and almost-simple polygons.  The condition that polygons have disjoint interiors is replaced by the condition that the interior of an object may not intersect another object. So a point object is either disjoint from all polygon objects, or it lies on the boundaries of some polygon objects.  We subdivide polygon edges to ensure that no point object lies in the interior of a polygon edge.

Point objects disjoint from polygons could be approximated by tiny
polygons,
but this easy approach fails for point objects that coincide with polygon vertices.
Instead, we show how to modify our algorithm to deal directly with point objects. 

We must clarify the output requirements 
in case a point object $p$ lies on the boundary of a 
solution $W$.
The only reasonable way to decide the matter in this case without
making the problem ill-posed is to consider $p$ to be enclosed or not at our discretion, 
since, by
an arbitrarily small perturbation 
of
$W$ in the vicinity of~$p$, either outcome can
be achieved.
This agrees with the convention used by Eades and Rappaport~\cite{er-ccmsp-PRL93}.
More precisely, we adapt the notion of a feasible solution $W$ as follows:
\begin{quote}
For each %
point object $p\in R$, we require that
$p$ lies in the interior or on the boundary of $W$ (possibly several times).
\end{quote}
The penalty of an optional point object $p\in O$ that lies on the boundary of $W$ is
not counted towards the cost in formula~\eqref{eq:cost-definition}.

We use a reference point $r_P$ for each object $P$.
For a point object,
the reference point must be that point itself. 
A reference point lying on a mouth in a subproblem $M(pq,t,B)$ was already handled by the algorithm. 
What is new is the possibility that a vertex is a point object, either required or optional.

We make two changes to the dynamic programming algorithm:
\begin{enumerate}
\item 
If $p\in R$ is a required point object, %
we 
add an extra possibility to 
equation \eqref{eq:C-base-2-new} for the case
$B=\{p\}$
as follows:
\begin{equation}
    \label{eq:C-base-2-point}
     C(p,t,\{p\}) := 0 \text{ for $t\ge 0$}
\end{equation}

\item For equation \eqref{eq:M-triangle-new} we used a triangle $\Delta = prq$ that was defined to be open on edge $pq$ and closed on edges $pr$ and $qr$.  We
redefine $\Delta$ to 
exclude %
its corners $p,q,r$, i.e., the only boundary points of $\Delta$ that are included are the interiors of edges $pr$ and $qr$.  This affects both $\pi(\Delta)$ and $R(\Delta)$.
\end{enumerate}

We explain the effect of these changes informally, and then outline how our %
proofs of correctness must be modified.

First observe that the changes to $\Delta$ mean that $\pi(\Delta)$ does not count penalties of optional point objects at the corners of $\Delta$ in equation \eqref{eq:M-triangle-new}, which is the correct thing to do. This is the only place in the equations where penalties are added.

Consider now a required point object~$p$. 
At the top level, $p$ lies in $R$, and this is passed to the disjoint sets $B$ used in the recursions.  If ever $p$ is contained in $R(\Delta)$, then this is where $p$ is considered to be enclosed (and it can only be enclosed once in this way). 
At the bottom of the recursion, rule 
\eqref{eq:C-base-2-point} permits us to consider $p$ as enclosed, and rule~\eqref{eq:C-base-1-new} permits us to consider $p$ as not enclosed.  Thus, we can ``catch'' the object $p$ on the boundary if we have not done so already in a triangle.
If the boundary goes through $p$ several times, we can catch
it on one occasion and pass over it on the other occasions. 

To make this more formal, we adapt
\Cref{lemma:buildup} in the following way:

\begin{lemma}%
   \label{lemma:buildup-for-points}
\textup{(A)} Let $W$ be a weakly simple polygon that
goes through some vertex $p$ and consists of  
$\ell 
$ \workkspace{} edges.
Let $B_{\mathrm{in}}$ be the objects of $R$ that are enclosed by $W$,
and let 
 $B_{\mathrm{point}}$ be the point objects of $R$ that
 coincide with vertices of~$W$.
Then, for all  $t\ge \ell$
and for all~$B$ with
$B_{\mathrm{in}} \subseteq B \subseteq
B_{\mathrm{in}} \cup B_{\mathrm{point}}$,
$C(p,t,B) \le c(W)$. 

\textup{(B)}
Let $W_0$ be an open walk with $\ell$ \workkspace{} edges from
{vertex }%
$p$ to
{vertex }%
$q$ such that the
polygon $W 
= W_0 + qp$ is weakly simple.
Let $B_{\mathrm{in}}$ be the objects of $R$ whose reference points 
lie inside $W$ and not on $pq$,
and let 
 $B_{\mathrm{point}}$ 
 be the point objects of $R$ that
 coincide with vertices of~$W$,
 {excluding $p$}. %
In addition,
assume that $q$ is not %
a transition vertex of $W$.
Then, for all  $t\ge \ell$
and for all $B$ with
$B_{\mathrm{in}} \subseteq B \subseteq
B_{\mathrm{in}} \cup B_{\mathrm{point}}$,
$M(pq,t,B) \le c(W_0)$. \qed
\end{lemma}
Note in particular that,
in the statement of part (B), %
we have added the condition that if $p$ is a required point
in $R$, it
cannot be in~$B$
(in addition to requiring that %
$q$ is not a transition vertex).

The induction basis, treating the trivial polygons with $t=0$ edges,
is covered by 
 \eqref{eq:C-base-1-new} and~\eqref{eq:C-base-2-point}.

For the closed walks in statement~(A), when $p$ is a point object in $B$,
we cannot apply the strategy for case A.1, because it would lead to
the subproblem $M(qp,\ldots)$ in which $p$ is a point object,
for which the extra requirement for (B) does not hold.
Thus, in this case, if $p$ is not a transition vertex anyway,
we make a degenerate split %
as in case A.2, with an empty walk $W_2$ and $B_2=\{p\}$,
and consequently $W_1=W\!$, see \Cref{fig:structure-A}.
Equation \eqref{eq:C-compose-new} yields
\begin{displaymath}
    C(p,t,B) \le C(p,t,B\setminus\{p\}) +  C(p,0,\{p\})
    \le C(p,t,B\setminus\{p\}) +  0.
\end{displaymath}
To the subproblem
$C(p,t,B_1)$ with
$B_1=B\setminus\{p\}$,
 case A.1 applies,
since $p$ is no longer an element of $B_1$ for this subproblem.
This leads to %
\begin{displaymath}
    C(p,t,B_1) \le  w_{pq} + M(qp,t-1, B_1) \le  c(W)
\end{displaymath}
and hence to
$
 C(p,t,B)   \le c(W)$.
    
In case B.2, where we split the set $B$ into
$B_1 \sqcup B_2 \sqcup
R(\Delta)$, the splitting is clear:
we have to assign any point objects on $W_1$ to $B_1$ and
any point objects on $W_2$ to $B_2$.
If
the vertex $r$ is a required point and belongs to $B$, we
use the freedom of choice to
put it in~$B_2$ (the subproblem belonging to the mouth $rp$)
and not in $B_1$,
ensuring that the inductive hypothesis can be applied
to the first subproblem $M(pr,t_1,B_1)$.

Case B.1 does not require any changes.

\smallskip

The other part of the correctness proof
is based on
the properties of $\WDP$ proved in Lemma~\ref{lemma:WDP-properties}. 
Lemma~\ref{lemma:WDP-properties}(\ref{DP-R}) claims that for $P \in R$, 
$\wwind {\WDP} {r_P} = 1$.  But for a point object $p \in R$ that lies on $\WDP$, the winding number is undefined. 
Thus, we have to restrict this claim to required point objects that do not lie on $\WDP$ (and ditto in the claims for the subproblems in the inductive proof).
For a point object $p \in R$ lying on $\WDP$, 
we simply observe that it will also lie on $\WALG$ because the Reordering Algorithm does not remove points from the polygon boundary.
Hence $p$ fulfills the adapted requirements of a feasible solution.

\section{The Inverted Problem}
\label{sec:inverted}
In the inverted problem,
the required objects have to be outside the weakly simple polygon~$W$,
and the penalties of the polygons
that are outside $W$ are counted
(see the introduction).
The approach for the original problem has to be adapted. %
To derive a suitable dynamic programming formulation, we decompose the {outside} of $W$ into elementary pieces, as shown in
\Cref{fig:buildup-reversed}:
We form the convex hull of~$W$ and
extend vertical rays upward and downward from the convex hull vertices.
This leads to two additional types of regions:
\begin{figure}
    \centering
    \includegraphics{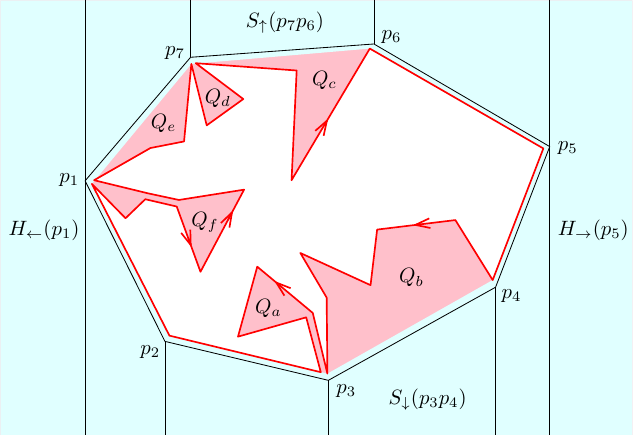}
    \caption{Partition of the outside into pockets $Q_a,\ldots,Q_f$,
    two half-planes, and seven planks.}
    \label{fig:buildup-reversed}
\end{figure}
\begin{itemize}
    \item one left half-plane and one
    right half-plane, each bounded by
    a vertical line through an object vertex;
    \item vertical \emphblack{planks}, that is, regions bounded by a line segment and two vertical upward rays or
     two vertical downward rays.
    We discussed such regions in
         Section~\ref{sec:preprocessing}.
\end{itemize}
\begin{figure}[htb]
    \centering
    \includegraphics{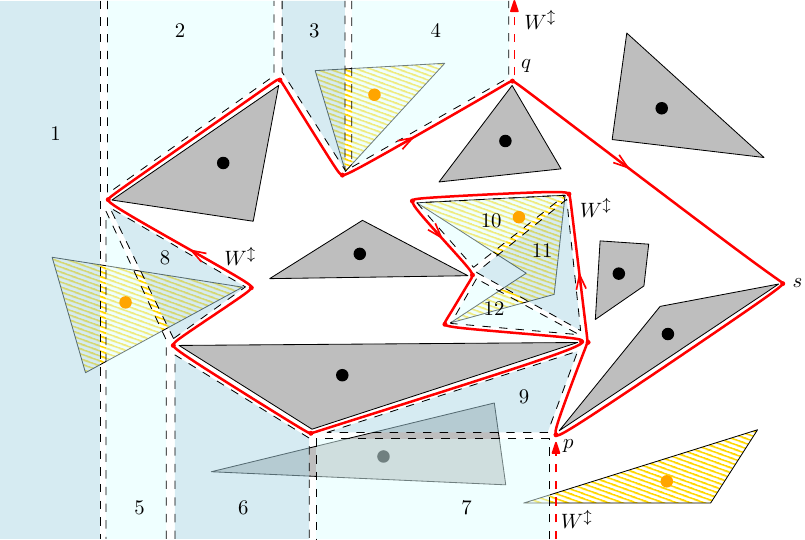}
    \caption{A weakly simple polygon $W$ (red/bold) that has the four required objects (yellow/hatched) on the outside.
   The penalties of two optional (grey) objects is added to the length when the cost is computed.
We grow the outer region triangle by triangle,
considering also ``triangles'' that extend to $-\infty$ or $+\infty$ (vertical planks), or both, like the left half-plane (number~1).
The union of the shaded blue regions is bounded by vertical rays from $p$ downward and from $q$ upward plus
a weakly simple walk %
between $p$ and $q$.
The extended walk $W^\updownarrow$
is a candidate solution considered for the subproblem
$U(p,q,B,t)$, when
$t\ge 13$ and
$B$ consists
of the three leftmost required objects. 
(The fourth required object has its reference point (thick dot) outside
the region.)
Two more planks, spanned by $ps$ and $qs$, plus the right half-plane through 
$s$, would complete the outside region of $W$.
}
    \label{fig:outside}
\end{figure}
We stick to the convention that the interior of the region of interest lies
on the left side of~$W$. %
Accordingly, the solution %
polygon $W$ is now
oriented clockwise.

In the algorithm, we
build the region of interest outside-in, 
see \Cref{fig:outside},
starting from a left half-plane
bounded by two vertical rays through an object vertex.
We add planks from left to right along common rays, and,
as in \Cref{sec:DP}, 
we may also attach
triangles along common edges and digons along common vertices.
In addition to the usual bounded walks, we now 
also consider polygonal walks $W^\updownarrow$ that start from the endpoint of a vertical downward ray
and end at a vertical upward ray.
More precisely, for each pair of vertices $p,q$, we consider
a subproblem of type $U$ (“unbounded”),
which considers regions bounded by
a vertical ray down from $p$,
a walk $W$ from $p$ to $q$, and a vertical ray
up from~$q$,
see \Cref{fig:outside}
for an example;
$p=q$ is allowed.
Accordingly, the algorithm computes  quantities
$U(p,q,t,B)
$
for all $B\subseteq R$ and $t\le6n$.
The two unbounded rays jointly play the role of the mouth.

We now describe how the dynamic programming recursion and the correctness proof of \Cref{sec:correctness} for the non-inverted problem is adapted for the
inverted problem.

\subsection{The dynamic programming recursion}

\def\subminus{\ensuremath{\!\scriptstyle-\!}}
\def\subplus{\ensuremath{\!\scriptstyle+\!}}
\def\subminus{_\leftarrow}
\def\subplus{_\rightarrow}
For simplicity, we assume that distinct vertices and distinct reference points have distinct $x$-coordinates; this can be achieved by a rotation.
We denote by 
$H \subminus(q)$ and $H\subplus (q)$ the left and right half-plane bounded by the vertical
line through~$q$.
$S_\downarrow(pq)$  and
$S_\uparrow(pq)$ are the planks with boundary segment $pq$.
By convention, $p$ is always left of $q$.

\begin{figure}[htb]
    \centering
    \includegraphics[]{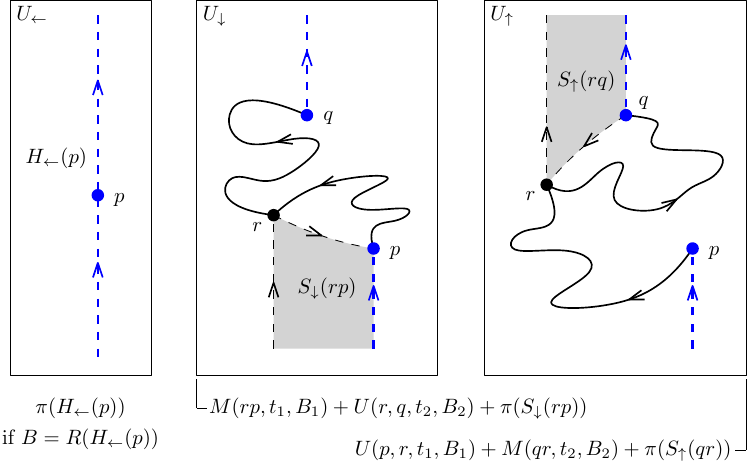}
    \caption{The dynamic programming
    recursion for the inverted problem. Free-space edges are solid;  mouths are dashed.
   }
    \label{fig:DP-cases-reverted}
\end{figure}

The recursion considers three cases, as illustrated in \Cref{fig:DP-cases-reverted}.
The easy case ($U_\leftarrow$) is a left half-plane, which applies only for $p=q$.
The other two cases are symmetric to each other; 
we discuss here only $U_\downarrow$.
This is similar to the term $M_2$ for $M(pq,t,B)$ in recursion~\eqref{eq:M-triangle-new}, except that the plank $S_\downarrow(rp)$ plays the role
of the triangle $\Delta=prq$. One of the subproblems, with mouth $pr$, is an 
``ordinary'' subproblem of type $M$, the other subproblem is of type~$U$.
\begin{align}
  \nonumber %
      U(p,q,t,B) & = \min \{ U \subminus, U_{\downarrow}, U_{\uparrow} \}
      \text{, where}
      \\\label{eq:U-minus}
  U\subminus %
                 &= 
  \begin{cases}
    \pi(H\subminus(p)) %
    ,&\text{if $p=q$ and }
   B= %
   R(H\subminus(p))\\
    \infty, & \text{otherwise}
  \end{cases}
\\[-\baselineskip]\nonumber %
\\\label{eq:U_down}
U_{\downarrow}
 &=
   \min \{\, M(pr,t_1,B_1) + U(r,q,t_2,B_2) +
   \pi(   S_\downarrow(rp) ) \mid
  \\\nonumber
  &\qquad\qquad\qquad
    \text{$r$ left of $p$, } t=t_1+t_2,\
    B =R(S_\downarrow(rp)) \sqcup B_1  \sqcup B_2 
  \ \bigr\}
  \\ \label{eq:U_up} U_{\uparrow}
 &=
   \min \{\, U(p,r,t_1,B_1) +M(rq,t_2,B_2) + 
   \pi(
   S_\uparrow(rq) ) \mid
  \\\nonumber
  &\qquad\qquad\qquad
    \text{$r$ left of $q$, } t=t_1+t_2,
  \ %
    B =R(S_\uparrow(rq)) \sqcup B_1  \sqcup B_2
    \ \bigr\}
\end{align}
$R(\Omega)$ and $\pi(\Omega)$ (for some region $\Omega$) generalize the earlier notations $R(\Delta)$ and $\pi(\Delta)$ and denote the required objects and the sum of penalties of optional objects whose reference point lies inside $\Omega$.
We clarify that
reference
points that lie {on} 
a segment 
$pq$ are treated
as {belonging to} 
$S_\downarrow(pq)$ and $S_\uparrow(pq)$.
No reference points lie on other boundaries of planks and half-planes
since they do not have the same $x$-coordinates as vertices.

The overall solution is
\begin{equation}\label{eq:inverted-opt}
\cDP' := \min\{\,
 U(p,p,%
             R(H\subminus(p)), 6n) + \pi(H\subplus(p)) 
 \mid p\text{ is a vertex}
 \,\}  
\end{equation}
The first term,
with $B=R(H\subminus(p))$,
makes sure that all required objects
with the reference point
to the left of $p$ are covered. The remaining objects
are then automatically covered by the right half-plane $H\subplus(p)$.
For the choice of the leftmost vertex as the vertex $p$ in
\eqref{eq:inverted-opt}, we are considering the degenerate solution
consisting of a path $W$ of length $0$.
This is a valid candidate solution, in case it is more expedient to
pay all penalties than to avoid some optional polygons.

\subsection{Adapting the correctness proof for the inverted problem}

\label{sec:adapt}
To apply the arguments from \Cref{sec:correctness}, we have to
extend the notion of winding number to regions $W^\updownarrow$ whose boundary includes an upward and a downward vertical ray.
We pick an anchor point $X_-$
to the left of all object vertices.
We define $\wwind {W^\updownarrow}{X_-}=1$, and define
the winding number for other points relative to $X_-$ by connecting
them by a curve to $X_-$
and counting signed intersections.
It follows that $\wwind {W^\updownarrow}{x}=0$ for any point $x$ that
is sufficiently far to the right. 
The winding number of planks must also be defined appropriately:
The winding number
of $S_\downarrow(pq)$
or $S_\uparrow(pq)$
is 1 between the two rays, on the ``correct'' side of
the segment $pq$, and 0 otherwise.

When the region is finished off by adding a right halfplane, $W$ becomes a closed clockwise cycle. 
If we were to follow %
the usual conventions, 
the interior
would have %
winding number $-1$ and the exterior
would have %
winding number $0$.
The winding number that we are using has an additive
offset of~1, due to the
stipulation that the point $X_-$ %
has winding number~1.
Thus, according to our convention, the winding number is 1 outside~$W$ and 0 inside~$W$,
which is precisely what we need because
the objects whose presence is checked and whose penalties are added are the objects outside~$W$.

The correctness proof in
\Cref{lemma:buildup} must be adapted as follows.
In the inductive step,
we have a solution $W^\updownarrow$ consisting of
a finite walk $W$ from $p$ to $q$ and two vertical rays.
The case that $p=q$ and the walk $W$ is trivial is handled
by formula \eqref{eq:U-minus} for $U_-$.
Suppose that $p\ne q$ and, w.l.o.g., $p$ is not the leftmost point of $W$.
Then we take the edge $pr$ on the lower convex hull of $W$.
The formula \eqref{eq:U_down} for $U_\downarrow$ shows
that  $W^\updownarrow$ can be reduced to smaller pieces.

There is another
case that requires proper handling,
namely %
the partial solutions of type $C$ (``closed'') that
are incident to the convex hull.
Examples are
the pieces $Q_a,Q_d,Q_f$ in \Cref{fig:buildup-reversed}.
We have to make sure that
the potential solutions that we build can include such pieces.

Before we describe the general approach %
  for handling these
  pieces,
  we look at an example.
Every convex hull edge %
appears exactly once as a
mouth in the recursion,
either
  as the edge $pr$ in
\eqref{eq:U_down}
or as the edge $rq$ in
\eqref{eq:U_up}.
As can be checked in
\Cref{fig:DP-cases-reverted},
it appears always in counterclockwise direction along the boundary.
For example, the edge  $p_6p_7$
in \Cref{fig:buildup-reversed}
appears in a subproblem of the form
$M(p_6p_7,t,B)$, for appropriate parameters $B$ and $t$.
According to \Cref{lemma:buildup}B, this subproblem will
consider partial solutions in which $p_6$ is not a transition vertex.
However, there is no such restriction on $p_7$. Thus we
can
assign all
type-$C$ pieces hanging off $p_7$ to this subproblem. (In the example,
there is only one such piece, $Q_d$.)
Any pieces hanging off $p_6$ would be assigned to the adjacent edge, $p_5p_6$.

The general strategy is as follows. We cut the solution walk $W$ into
pieces at the convex hull vertices
and assign one %
piece to each convex hull edge.
The convex hull edge %
acts as the mouth of
the piece. Pieces of type $C$ that start and end at a hull vertex
$p_i$ are assigned to the hull edge $p_{i-1}p_i$ clockwise from $p_i$.
This rule defines
the
pieces %
unambiguously. %
Moreover, %
we have ensured that
for each mouth $p_ip_{i-1}$, the endpoint $p_{i-1}$ is never a transition vertex of the respective piece. Thus, by \Cref{lemma:buildup}, the weight of the
corresponding piece is an upper bound on 
$M(p_ip_{i-1},t,B)$ with the appropriate parameters $t$ and $B$.

The remainder of the proof, regarding the coverage of outside the convex hull by adding halfspaces and planks, is straightforward. 

The example of \Cref{fig:outside} illustrates
that the region covered by the planks 
need not actually be
the outside of the convex hull of the solution polygon: %
Regions 3 and 4 form an indentation in the convex hull.
 In fact, any ``$x$-monotone hull'' of the solution %
can be taken. %

With these changes, the algorithm of Section~\ref{sec:DP} and
  its speed-up of Section~\ref{sec:dijkstra} carry over to the
  inverted
  version of the general \textsc{Enclosure-with-Penalties} problem.
As for the original  
problem, the algorithm applies to both
the geometric version and the graph version of the
problem, and we obtain the following result.

\begin{theorem}\label{thm:inverted}
  \begin{enumerate}
  \item 
The inverted \GeoEWP\ problem for $k$ required polygons 
can be solved in $O(3^k n^3)$ time and $O(2^k n^2)$ space,
if the input polygons have $n$ vertices in total.
\item The inverted \GrEWP\ problem for $k$ required faces
  in  a %
  plane graph with $n$ vertices %
can be solved in $O(3^k n^3)$ time and $O(2^k n^2)$ space.
\qed
  \end{enumerate}
\end{theorem}

\subparagraph{Graphs on the sphere.}
We mention that for  \GrEWP, it makes sense to state the
problem also for a graph
embedded on the {sphere}, with no distinguished outer face.
In this problem, the desired solution is a
weakly simple closed walk on the sphere with a designation of one of its sides as
the ``interior'' side.
We can solve this version by arbitrarily picking one of the required
faces $F_0\in R$ as the unbounded face and then solving the inverted problem for the resulting plane
graph.

\section{Negative Penalties, or Rewards}
\label{sec:negative}

We now
consider the extension of \textsc{Geometric-Enclosure-with-Penalties} and \textsc{Graph-Enclosure-with-Penalties} in which we allow objects with negative penalties.  In order to have a balanced and general statement, we use a greater variety of types of polygons/faces: 
there is a set~$R^-$ of \defn{required} polygons/faces, a set~$R^+$ of \defn{forbidden} polygons/faces (the notation suggests that these sets correspond to objects with penalty~$-\infty$ and $+\infty$, respectively), and a set~$O^+\sqcup O^0\sqcup O^-$ of \defn{optional} polygons/faces, 
whose penalties are finite and positive for the objects in~$O^+$, zero for the objects in~$O^0$, and finite and negative for the objects in~$O^-$.  The goal is to find a weakly simple closed curve/walk disjoint from the objects, enclosing~$R^-$, excluding~$R^+$, and minimizing the length of the curve plus the penalties of the objects in~$O^+\cup O^0\cup O^-$ that are enclosed by the curve.
\begin{theorem}
  We can solve these generalized problems in time $O(3^kn^3)$ time and $O(2^kn^2)$ space, where $k=\min\{|R^-|+|O^-|, |R^+|+|O^+|\}$.
\end{theorem}
\begin{proof}

Let us first consider the case where $k=|R^-|+|O^-|$.  The idea is to try all subsets of~$O^-$ that can be enclosed in an optimal solution.
  
We run the dynamic programming algorithm with set of required objects $R:=R^-\cup O^-$ and set of optional objects $O:=R^+\cup O^+\cup O^0$, where the objects in~$R^+$ have infinite penalties and those in $O^+\cup O^0$ keep their original nonnegative penalties; this takes $O(3^{|R|}n^3)=O(3^kn^3)$ time.
As part of the %
recursion,
the algorithm determines
$C(p%
,B)$ for all subsets $B\subseteq R$
and all vertices~$p$.
We therefore have available all quantities that enter %
the following formula for the optimum solution:
  \begin{equation}
        \label{eq:final3*}
  C_{\mathrm{final}}^* :=
  \min_{O^-_1\subseteq O^-}
  \left(
    \left(\min_{p\text{ a vertex}} C(p,R^-\cup O^-_1)\right)
   +
    \sum_{P\in O^-_1} \pi_P%
\right)\end{equation}
This formula is justified as follows:
The solutions considered for $\min_pC(p,R^-\cup O^-_1)$ 
are those solutions
that, among the objects in $R=R^-\cup O^-$, enclose precisely
the objects of $R^-\cup O^-_1$ and exclude the remaining objects
of~$O^-$.
In this way, we consider all solutions that
enclose $R$ plus some arbitrary subset $O^-_1\subseteq O^-$
of the negative-weight objects.
The negative weights of the objects in $O^-_1$ are explicitly
added in \eqref{eq:final3*} to get the correct value of the objective function.

The time $O(2^{|O_-|}n)$ for calculating
$C_{\mathrm{final}}^*$
by \eqref{eq:final3*}
is dominated
by the runtime $O(3^kn^3)$ of the dynamic programming algorithm.

  In the other case where $k=|R^+|+|O^+|$, we invoke the algorithm
  of Theorem~\ref{thm:inverted} for the inverted problem
  instead of the algorithm of Theorem~\ref{thm:main-geom} or~\ref{thm:main-graph}.  This swaps $R^+$ with~$R^-$ and~$O^+$ with~$O^-$.  The rest of the argument is identical.
\end{proof}

\section{Exponential Lower Bounds} \label{sec:lower-bound}

We first prove a (conditional) exponential lower bound for the \textsc{Graph-Enclosure-with-Penalties} problem. The proof consists of a simple reduction to our problem from the \textsc{Planar Steiner Tree} problem. 

\begin{theorem}\label{thm:lower-bound}
Assuming the Exponential Time Hypothesis, the \textup{\textsc{Graph-Enclosure-with-Penalties}}
problem cannot be solved in $2^{o(k)}\cdot n^{O(1)}$ time, even when all weights are $1$ and all penalties are $\infty$.
\end{theorem}

\begin{proof}
The proof consists of a reduction from the \textsc{Planar Steiner Tree} problem, whose input is an edge-weighted planar graph $G$ with $n$ vertices and a set $T$ of $k$ vertices of $G$, usually called \defn{terminals}. The problem asks for a minimum-weight tree in $G$ connecting all terminals. Marx, Pilipczuk, and Pilipczuk~\cite[Theorem~1.2]{marx2018subexponential} proved that the \textsc{Planar Steiner Tree} problem cannot be solved in $2^{o(k)}\cdot n^{O(1)}$ time, assuming the Exponential Time Hypothesis, even if the input graph is unweighted (that is, all edge weights are $1$). We now describe the reduction.

\begin{figure}
    \centering
\tabskip = 0pt plus 0.4 fil
\halign to \textwidth{\hfil#\hfil\tabskip = 0pt plus 1 fil
&\hfil#\hfil\tabskip = 0pt plus 0.4 fil
\cr
  \includegraphics[page=1]{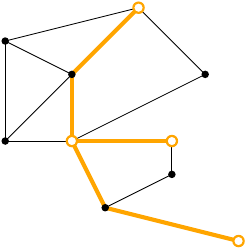}%
& \includegraphics[page=2]{LowerBound.pdf}%
\cr
(a)&(b)
\cr
}
    \caption{(a) An instance of the \textsc{Planar Steiner Tree} problem. Terminals are large empty disks. Edges of a tree $S$ connecting the terminals are thick and yellow. (b) The corresponding instance of the \textsc{Graph-Enclosure-with-Penalties} problem. 
Faces in $R$ are yellow/hatched, faces in $O$ (including the unbounded face) are gray.
    The weakly simple closed walk~$W$ constructed from $S$ is represented by a red 
    curve.}
    \label{fig:lower-bound}
\end{figure}

Given an unweighted planar graph $G$ and a set~$T\subseteq V(G)$, as the one in \Cref{fig:lower-bound}(a), we replace each terminal $v$ in~$T$ with a corresponding \defn{terminal cycle} $C_v$, whose number of edges is equal to $\max\{3,d_G(v)\}$, where~$d_G(v)$ denotes the degree of~$v$ in $G$; each vertex of~$C_v$ is connected to a different neighbor of~$v$, and the interior of~$C_v$ is a face, see \Cref{fig:lower-bound}(b). Denote by~$H$ the obtained plane graph and by $\gamma$ the total number of \defn{terminal edges}, i.e., edges in terminal cycles. Every edge of~$H$ has weight~$1$. 
Let $R$ be the set of faces inside terminal cycles, and let $O$ be the set of all other faces. The faces in $O$ have penalty~$\infty$. This completes the reduction.
We now prove that $G$ contains a tree with weight $\le w$ connecting the terminals in $T$ if and only if $H$ has a weakly simple closed walk
$W$ with weight $\le 2w+\gamma$ that has the faces in $R$ inside (and the faces in $O$ outside).

For the forward implication, given any tree $S$ in $G$ with weight at most $w$ connecting the terminals in~$T$,
one can construct
the desired
walk
as follows. The walk 
traverses each edge in~$S$ twice (once in each direction), and traverses each terminal cycle once,
in 
counter-clockwise~direction.

For the converse, %
let $W$ be a weakly simple closed walk in $H$ with %
weight 
at most $2w+\gamma$ that has the faces in $R$ inside and the faces in $O$ outside.  Walk $W$ contains each terminal edge at least once, because
$W$ must separate $R$ from $O$.
Moreover, 
$W$ uses each non-terminal edge of~$H$ an even number of times, as otherwise one of its incident faces, both of which have penalty $\infty$, would be inside~$W$. 
Since $W$ is connected, it
contains at least twice each edge of a connected subgraph $S_H$  spanning the terminal cycles.
The simple graph $S_G$ in $G$ corresponding to $S_H$ connects all the terminals.
The weight of $S_G$ 
is at most $((2w+\gamma)-\gamma))/2=w$, which is obtained from the weight $2w+\gamma$ of $W$ by subtracting the total weight
of the terminal edges, which is at least $\gamma$, and
by then dividing by two as each edge of $S_G$ is used at least twice in $%
W$. 
We conclude the proof by observing that $S_G$ contains a tree that spans all terminals and has weight at most~$w$. %
\end{proof}

We now present a reduction similar to, and slightly more technical than, the one of \Cref{thm:lower-bound} for the geometric version of our problem.

\begin{theorem}\label{thm:lower-bound-geom}
Assuming the Exponential Time Hypothesis, the \textup{\textsc{Geometric-Enclosure-with-Penalties}} problem cannot be solved in $2^{o(k)}\cdot n^{O(1)}$ time, even when all penalties~are~$\infty$.
\end{theorem}

\begin{proof}
Consider an instance $(G,T)$ of the \textsc{Planar Steiner Tree} problem in which $G$ has~$n$ vertices and edges with weight~$1$. We start by constructing the~$O(n)$-vertex planar graph~$H$ as in the proof of \Cref{thm:lower-bound}. We now construct a sequence of representations of~$H$, and eventually get the desired instance of \textsc{Geometric-Enclosure-with-Penalties}.

First, we construct a {\em visibility representation} $\Gamma$ of $H$ on an $O(n)\times O(n)$ grid~\cite{DBLP:journals/dcg/TamassiaT86}, see~\Cref{fig:lower-bound-geom}(a). In~$\Gamma$, vertices are represented by disjoint horizontal segments lying on grid rows and edges are represented by disjoint vertical segments lying on grid columns. Each vertical segment representing an edge has its endpoints on the horizontal segments representing the end-vertices of the edge and otherwise does not cross any horizontal segment representing a vertex. We scale up the coordinates in the drawing by a factor of $5$. 

\begin{figure}[htb]
    \centering
    \begin{tabular}{c c}
  \includegraphics[page=1]{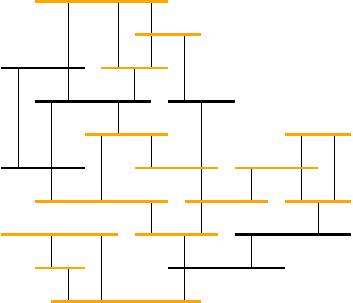}%
& \includegraphics[page=2]{visibility.pdf}\\
(a)&(b)
\end{tabular}
    \caption{(a) A visibility representation $\Gamma$ of the graph $H$ from \Cref{fig:lower-bound}(b). We stress that the relative interior of a vertical segment representing an edge of $H$ does not intersect any horizontal segment representing a vertex of $H$; vertical lines may consist of several vertical segments. (b) A poly-line drawing $\Gamma'$ of $H$ constructed from $\Gamma$. Both representations lie on an $O(n)\times O(n)$ grid.}
    \label{fig:lower-bound-geom}
\end{figure}
Second, we turn $\Gamma$ into a poly-line drawing $\Gamma'$; this can be done by modifying $\Gamma$ only ``close'' to its vertices, see~\cite{DBLP:journals/dcg/Biedl11,DBLP:journals/tcs/BattistaT88} and~\Cref{fig:lower-bound-geom}(b). Specifically, each horizontal segment~$s_v$ representing a vertex~$v$ is replaced by a grid point~$p_v$ on~$s_v$. Also, we shorten each vertical segment representing an edge $uv$ by one unit at the top and at the bottom and connect the endpoints to $p_u$ and $p_v$.

\begin{figure}[htb]
    \centering
\includegraphics[page=3]{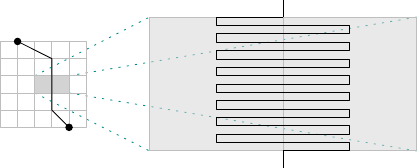}
    \caption{The left part of the figure shows an edge $e$ in the drawing~$\Gamma'$. The right part %
    shows
    an enlarged central section of $e$ in which
    the drawing of~$e$
    is modified in order to transform~$\Gamma'$ into a poly-line drawing~$\Gamma''$ of~$H$ in which $e$ has length between~$n^2$ and~$n^2+1$. The edge~$e$ is only modified in its portion $\sigma_e$ inside the two gray grid cells.}
    \label{fig:lower-bound-geom-2}
\end{figure}
Third, we turn $\Gamma'$ into a poly-line drawing $\Gamma''$ in which all edges have ``almost'' the same length. Intuitively, we are going to modify the representation of each edge by ``orthogonally zig-zagging'' in an intermediate part of the edge, so that the edge has length between $n^2$ and $n^2+1$, see~\Cref{fig:lower-bound-geom-2}. As a consequence of the scaling of $\Gamma$, the representation of each edge $e$ in $\Gamma'$ contains a vertical segment $\sigma_e$ between two grid points $(i,j)$ and $(i,j+1)$ such that $\Gamma'$ has no intersection with the two grid cells incident to $\sigma_e$, other than at $\sigma_e$ itself. Let $\ell_e$ be the length of the polygonal chain representing $e$ in $\Gamma'$ and let $a_e=n^2-\lfloor \ell_e \rfloor$ be the increase of length that we want for $e$. Then we can replace $\sigma_e$ by the orthogonal line passing through points $(i,j), (i+\frac{1}{2},j), (i+\frac{1}{2},j+\frac{1}{a_e}), (i-\frac{1}{2},j+\frac{1}{a_e}),(i-\frac{1}{2},j+\frac{2}{a_e}),(i+\frac{1}{2},j+\frac{2}{a_e}),\dots,(i,j+1)$. The vertical segments of this line have total length $1$, while the horizontal segments have total length $a_e$ (two of them have length $\frac{1}{2}$, while the other $a_e-1$ have length $1$). Hence, the length of $e$ has increased by $n^2-\lfloor \ell_e \rfloor$ and it is now between $n^2$ and $n^2+1$. 

In order to get the instance of \textsc{Geometric-Enclosure-with-Penalties}, we interpret the faces of $\Gamma''$ as polygons: those inside the terminal cycles are in $R$ and those corresponding to faces of $G$ are in $O$ and have penalty $\infty$. Since all edges have approximately the same length, between $n^2$ and $n^2+1$, the same proof as in \Cref{thm:lower-bound} shows that $G$ contains a tree with weight~$\le w$ connecting the terminals in $T$ if and only if there exists a weakly simple closed walk~$W$
with weight~$\le (2w+\gamma)\cdot (n^2+1)$ in the instance of \textsc{Geometric-Enclosure-with-Penalties}. 
Indeed, the ``only if'' part is easy, and the proof for the ``if''
part uses the following argument. From the
weakly simple closed
walk~$W$,
one can extract a
weakly simple closed
walk~$W_H$ in the graph $H$ that uses at most
$\frac{(2w+\gamma)\cdot (n^2+1)}{n^2}=2w+\gamma+\frac{2w+\gamma}{ n^2}$
edges.
 Since~$2w+\gamma =O(n)$, the additional term $\frac{2w+\gamma}{
   n^2}$ is less than 1  provided $n$ is larger than some constant
 $n_0$, which we can obviously assume.
Therefore the
walk~$W_H$ %
uses at most
$2w+\gamma$
edges of~$H$, and as above, this implies
that $G$ has a tree of weight at most $w$ spanning all terminals.
 The described reduction takes polynomial time, since all vertex coordinates in $\Gamma''$ are rational numbers whose numerators and denominators are polynomially bounded.
\end{proof}

The above proof essentially contains a polynomial, parameter-preserving reduction from \textsc{Graph-Enclosure-with-Penalties} to \textsc{Geometric-Enclosure-with-Penalties}.  In passing, we mention that there is also a polynomial-time, parameter-preserving reduction in the other direction:  Given an instance of \textsc{Geometric-Enclosure-with-Penalties}, we know that the output will consist of \workkspace{} edges, so one can compute the graph that is the overlay of all \workkspace{} edges (equivalently, of the visibility graph of the input vertices), assign each subdivided edge a weight that is its Euclidean length, and assign penalty zero to each face of this arrangement that does not come from an input polygon.  This results in an equivalent instance of \textsc{Graph-Enclosure-with-Penalties}.

\section{Applications, Extensions, and Open Problems} \label{sec:conclusion}

\paragraph{Approximation.}
It natural to ask about approximation algorithms. Since the problem
bears a resemblance to the Steiner tree problem (cf. the proof of the lower bound in \Cref{sec:lower-bound}), it is likely that ideas from this problem and
similar problems can be applied. We leave this for future research.

\paragraph{Splitting a surface.}

\begin{figure}[bth]
    \centering %
  \includegraphics[width=.65\linewidth]{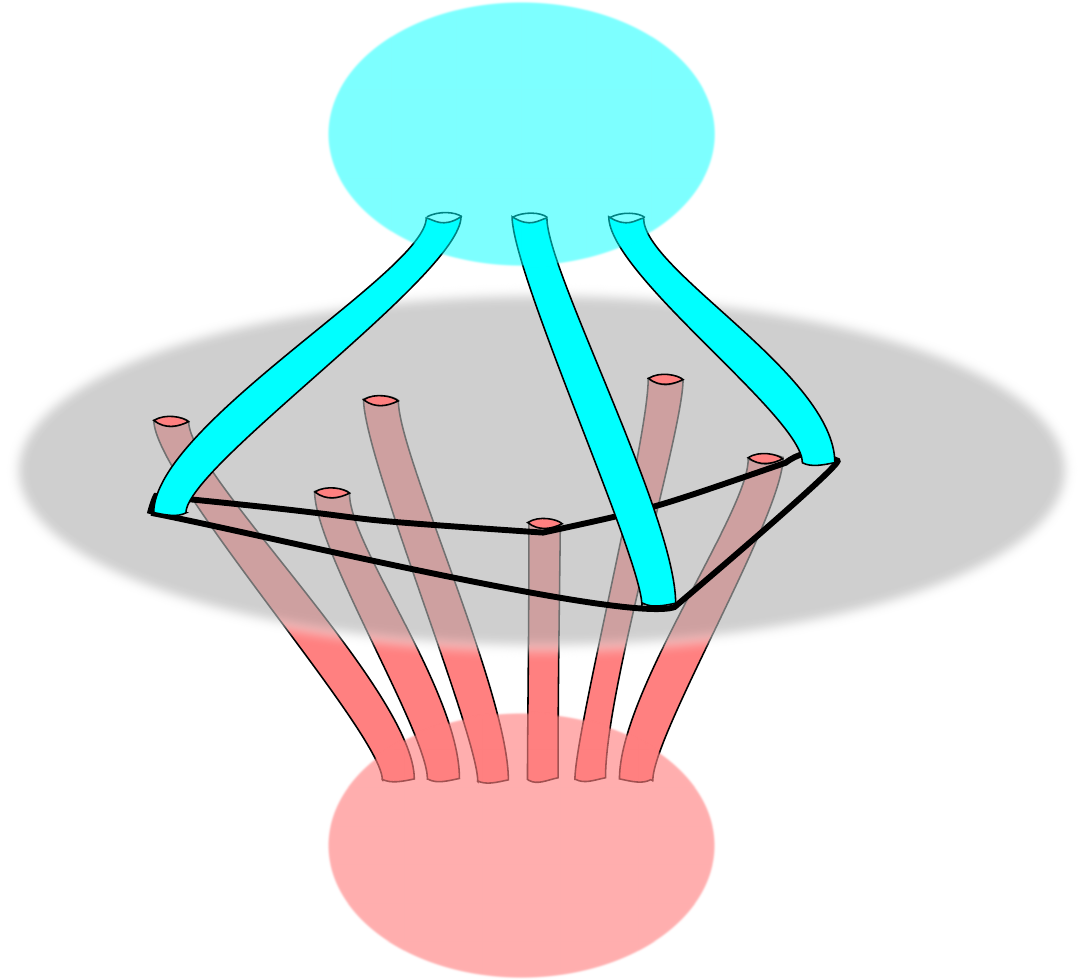}
    \caption{Any instance of \textsc{Graph-Enclosure-with-Penalties} with infinite penalties only, and with parameter~$k$, can be recast as an instance of the problem of splitting off a surface of genus $k-1$.  In this example, the graph~$G$ that is an instance of \textsc{Graph-Enclosure-with-Penalties} is embedded in the middle gray disk; each of the $k=3$ required faces is connected by a tube (in cyan) to the top sphere; each of the 6 optional faces, with infinite penalty, is connected by a tube (in red) to the bottom sphere.  Finally, the gray disk is extended to a sphere (not shown), to obtain a surface~$S$ without boundary.
    Any weakly simple closed walk in~$G$ separating the required faces from the optional ones corresponds to a weakly simple closed walk splitting off a surface of genus~$k-1=2$, and conversely.  (The graph~$G$ is not cellularly embedded on~$S$, but can be made so by adding     edges of large weight.)  Figure inspired by~\cite{ccelw-scsh-08}.}
    \label{fig:splitting}
\end{figure}

Our result may shed some light on the following open problem by Bulavka, Colin de Verdière, and Fuladi~\cite[Conclusion]{bcf-csccn-25}: given an orientable combinatorial surface of genus $g$, and an integer~$g'$, $1\le g'<g$, is it FPT in~$g'$ to compute a shortest weakly simple closed curve that cuts off a surface of genus~$g'$? The problem is FPT in~$g$~\cite[Theorem~6.1]{ccelw-scsh-08}.  Our algorithm for \textsc{Graph-Enclosure-with-Penalties} shows that the answer is yes when restricting to some (admittedly very special) instances, see \Cref{fig:splitting},
 and thus provides some hope for a positive answer in general, although this remains open.

\paragraph{Curved objects and line segments.}
We believe that our approach carries over to more general
objects.
Curved objects that are sufficiently well behaved can be treated by considering all bitangents
as \workkspace{} edges.
We can already handle point objects, as described in 
\Cref{sec:points};
line segments without other vertices in their interior should also be %
doable.
However, extending to weakly simple polygon objects seems difficult.  Even for a path object consisting of two line segments joined at point $p$ it is a challenge to prevent a solution from cutting through the path at $p$.

\paragraph{Recognizing %
weakly simple immersed %
polygons.}
As mentioned, our 
dynamic programming algorithm optimizes over the class of 
weakly simple immersed %
polygons,
see 
\Cref{sec:immersed}.
Weakly simple polygons
can be recognized in  $O(n \log n)$ time~\cite{aaet-rwsp-17},
and  self-overlapping polygons in time $O(n^3)$~\cite{SHOR-vanWIK-1992}.
Weakly simple immersed polygons combine the features of both.
Can they
be recognized efficiently?

\goodbreak
 \phantomsection
   \addcontentsline {toc}{section}{\numberline {--}References}%

\goodbreak
\vskip1cm \goodbreak\vskip-1cm

\appendix

\section{The Geometric Knapsack Algorithm}
\label{sec:error-AKM}

In this section we mention some issues with the algorithm by 
Arkin, Khuller, and Mitchell~\cite{arkin1993geometric} for the geometric knapsack problem. 
The 
problem was described in Section~\ref{sec:related}.
As mentioned in that section,
Arkin et~al.\ %
solve an inverted problem: to
minimize the length of the solution polygon, which they call a ``fence'',
plus the values of the
objects %
outside the 
fence.

We point out two issues: their algorithm does not deal with the %
complications 
that arise due to the possibility of a weakly simple (but not simple) enclosure; and their local approach to 
identifying 
the outside of the 
fence %
can lead to an incorrect output. 

With regard to the first issue, they write, 
``an optimal enclosure in the presence of obstacles will not, in general, be convex. It will, however, be a simple polygon $\ldots$.''
\cite[Section~3, p.~411]{arkin1993geometric}.
Their algorithm relies on identifying the point where a path exits (i.e., crosses) 
the 
fence.
The condition of
being crossed by a path is trickier and less local for weakly simple polygons than for simple polygons, so their algorithm needs to be augmented to specify exactly what happens in degenerate situations.
We believe that this is a technical issue that can be overcome by a
careful treatment of these degeneracies.

To explain the second issue,
which is more severe,
we first
summarize their algorithm.
It relies on the
fact that the solution area can be assumed to be geodesically convex.
The reason is
that 
an obstacle-avoiding shortcut between two points on the
boundary of the 
fence
through the outside
will grow the enclosed area and thus not 
degrade
the solution.
(This argument fails in our more general scenario,
where %
some objects
are required to remain %
outside the 
fence.)

For consistency,
we %
stick with our convention of %
surrounding the enclosed area %
in counterclockwise direction by the fence, while the opposite convention is used
in~\cite{arkin1993geometric}.

The solution consists of edges of $\mathcal{VG}$, the visibility graph of the object vertices.
For each object vertex~$p^*$, the algorithm  
finds an optimal 
fence
that goes through $p^*$, and then optimizes over the choice of $p^*$.  For given $p^*$ they 
find a directed cycle through $p^*$ using edges of $\mathcal{VG}$ with
carefully chosen weights (which they call ``costs''),
which are defined as follows.
Choose a reference vertex on each object.
Consider an object $O$ and a shortest path $\pi_O$ from $p^*$ to the reference vertex of~$O$. 
They prove that there is an optimum solution that, for all~$O$, intersects $\pi_O$ in a connected subpath~\cite[Lemma~3] {arkin1993geometric}.  This implies that 
$O$  is inside  the 
fence
iff the path  $\pi_O$ remains inside the
fence,
and $O$ is outside iff $\pi_O$ exits the fence,
in
which case it exits once at a unique point.  Thus their idea is to
charge the value of an outside object $O$ to the point %
of
$\mathcal{VG}$ where its path $\pi_O$ exits the 
fence.
For an edge of $\mathcal{VG}$ that is crossed by a path $\pi_O$,
 the value of $O$ is simply added to the weight of the edge.
 If $\pi_O$ goes through a vertex~$v$, the question whether
 it crosses the 
 fence
 can only be decided by looking at
 both edges incident to~$v$.
The value of $O$ is added to the cost of any triple $e,v,f$ where the directed edge $e$
enters $v$, the directed edge $f$ leaves $v$,
$\pi_O$ goes through
$v$ and, 
after passing through $v$ enters the interior of the wedge to the
right %
of $e,v,f$.
An~appropriate version of this rule should be used when $v$ is the reference point
of~$O$.

\begin{figure}[htb]
    \centering
    \includegraphics[%
    ]{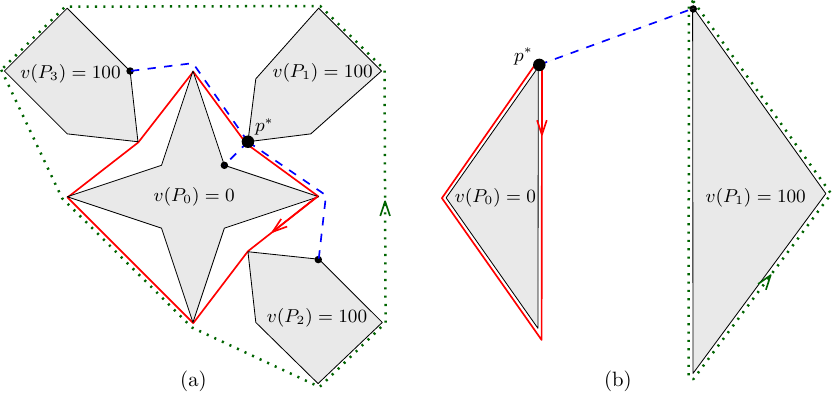}

    \caption{Bad examples for the geometric knapsack algorithm
      of~\cite{arkin1993geometric}.  All edge lengths are very
      small compared to the non-zero object values.  Overlapping paths are slightly displaced for visibility.
    (a) %
    The optimum 
    fence
    is the counterclockwise convex hull of the four polygons (the green dotted curve).
    When the algorithm of~\cite{arkin1993geometric} considers $p^*$
    and the blue paths to the polygon reference points, it finds the
    red cycle with ``interior'' to its left---the length is smaller and the value of ``outside'' polygons is $0$.
    (b) %
    The optimum
    fence (the green dotted curve)
    just hugs $P_1$ in the counterclockwise direction but the algorithm considers $p^*$ and finds the red cycle with ``interior'' to its left.
    The polygon $P_0$, which contains $p^*$, is considered to lie ``outside'' the fence.
}
    \label{fig:geom-knapsack}
\end{figure}

This declares the outside of the directed cycle to be  to its right,
implicitly assuming that the cycle goes in the counterclockwise direction.
However, there is no enforcement that the right side of the cycle will
actually be the unbounded part of the plane. %
 Figure~\ref{fig:geom-knapsack} shows two examples where
 the algorithm mistakenly finds a polygon that counts the values of
 the objects 
 in the bounded region determined by the cycle.
We think that this can be repaired by requiring that $p^*$ is a topmost
vertex of the cycle, i.e., by allowing only visibility edges below $p^*$ when
$p^*$ is considered as the starting point.

This modification does not
 affect the runtime of the algorithm.
 The runtime of the algorithm remains at
 $O(n^2|\mathcal{VG}|)=O(n^4)$,
where $|\mathcal{VG}|$ denotes the number of edges of the visibility
graph~\cite[Theorem~6]{arkin1993geometric}.
Even if the visibility graph is sparse,
i.e., $|\mathcal{VG}|=O(n)$, the algorithm could not improve over our
$O(n^3)$ approach; hence we did not investigate this potential fix of
the algorithm.

\end{document}